%% file: ms.tex
\documentclass[11pt]{article}
\usepackage{caption}
\usepackage{graphicx}
\usepackage[utf8]{inputenc}
\usepackage{apacite}
\usepackage{amsmath,amsthm,amssymb}
\usepackage{tikz,pgf,pgfplots}
\pgfplotsset{compat=1.12}
\usetikzlibrary{arrows,decorations,backgrounds,matrix,automata,
trees,shapes,shadows,plotmarks,calc,positioning,patterns,chains,fit}
\usepackage{fancyvrb}
\usepackage{verbatimbox}
\usepackage{natbib}
\usepackage[vlined,lined,ruled,linesnumbered]{algorithm2e}
\usepackage{gensymb}
\usepackage[normalem]{ulem}
\usepackage{float}
\usepackage{textcomp}
\usepackage{listings}
\usepackage{xcolor}
\definecolor{Pink}{rgb}{255,20,147}
\definecolor{darkgreen}{rgb}{0,0.5,0}
\definecolor{gray}{rgb}{0.4,0.4,0.4}
\definecolor{darkblue}{rgb}{0.0,0.0,0.6}
\definecolor{cyan}{rgb}{0.0,0.6,0.6}
\usepackage{xspace}
\usepackage{hyphenat}
\usepackage[disable]{todonotes} 
\newcommand{\joelComment}[1]{\todo[color=yellow!50!white,fancyline]{#1 -\textit{joel}}\xspace}
\newcommand{\Omit}[1]{}

\usepackage[colorlinks,bookmarksnumbered=true,pagebackref=true]{hyperref}
\usepackage[figure,table]{hypcap}
\hypersetup{
	bookmarks=true,
	bookmarksnumbered=true,
	pdfstartview={FitH},
	citecolor={black},
	linkcolor={black},
	urlcolor={black},
	pdfpagemode={UseOutlines}
}

\lstdefinelanguage{XML}{%
    alsoletter=-,
   morestring=[b]",
    stringstyle=\color{darkgreen},
    morecomment=[s]{>?}{?<},
    morecomment=[s]{<!--}{-->},
    morecomment=[s]{<!}{>},
    commentstyle=\color{cyan},
    moredelim=[s][\color{black}],
    moredelim=*[s][\color{black}]{<}{>},
  morekeywords={population,person, attributes,plan,activity,leg,attribute},
  keywordstyle=\color{magenta}
}
\makeindex
\title{\textsc{Modelling bushfire evacuation \\behaviours}}
\author{\textsc{J. Robertson} \\ Supervisor: \textsc{D.Singh}}
\date{November 2018}

\begin{document}

\maketitle
\input{sec-abstract}
\newpage
\input{sec-acknowledgement}
\newpage
\tableofcontents
\input{sec-intro}

\input{background/sec-background}
\input{chapter-theory/sec-plan-algorithm}

\input{chapter-applications/sec-applications}

\input{sec-conclusion}

\joelComment{TODO: -fix quotes to singles\\
- spellcheck/grammar check everything\\
- add contents with hyperlinks\\
- Header with section title throughout\\
- consistent pronouns\\
- no contractions\\
-consistent dashes\\
- acknowledgements}

\bibliographystyle{apacite}

\bibliography{../thesis.bib}

\end{document}

%% file: sec-abstract.tex
\begin{abstract}
Bushfires pose a significant threat to Australia's regional areas.
To minimise risk and increase resilience, communities need robust evacuation strategies that account for people's likely behaviour both before and during a bushfire.
Agent-based modelling (ABM) offers a practical way to simulate a range of bushfire evacuation scenarios.
However, the ABM should reflect the diversity of possible human responses in a given community.
The Belief-Desire-Intention (BDI) cognitive model captures behaviour in a compact representation that is  understandable by domain experts.
Within a BDI-ABM simulation, individual BDI agents can be assigned profiles that determine their likely behaviour.
Over a population of agents their collective behaviour will characterise the community response.
These profiles are drawn from existing human behaviour research and consultation with emergency services personnel and capture the expected behaviours of identified groups in the population, both prior to and during an evacuation.
A realistic representation of each community can then be formed, and evacuation scenarios within the simulation can be used to explore the possible impact of population structure on outcomes.
It is hoped that this will give an improved understanding of the risks associated with evacuation, and lead to tailored evacuation plans for each community to help them prepare for and respond to bushfire.



\end{abstract}

%% file: sec-acknowledgement.tex
\section*{Acknowledgements}
I would like to thank my supervisor, Dr. Dhirendra Singh, for his help and advice. His guidance has been invaluable and I have enjoyed our collaboration immensely.
Thanks also to Dr. Vincent Lemiale, Dr. Leorey Marquez and Rajesh Subramanian from Data61, who I have worked closely with at times during this year.
Without Data61 I would not have been introduced to bushfire evacuation modelling, and without their support this thesis would not have been possible.

%% file: sec-intro.tex
\section{Introduction}

\defcitealias{CountryFireAuthorityQualitativeReportCFA2009}{CFA \& Sweeney, 2009}

%
%

%

 A key question posed in the aftermath of the 2009 Black Saturday bushfires was whether the official Prepare to Stay and Defend or Leave Early (PSDLE) policy was an effective way to protect under-threat communities \citepalias{CountryFireAuthorityQualitativeReportCFA2009}.
 Whilst it was intended to minimise the number of residents attempting a dangerous late evacuation, there are concerns that the `Stay or Go' approach only leads to confusion over the level of preparation and planning required to stay and defend, and does not properly define how late is `too late' to leave.
Subsequent policies emphasising early evacuation as the only safe option have shown that blanket advisory policy has little effect on people's behaviour \citep{ReidWhereFireCoConstructing2014}.
 This highlights a need for localised bushfire evacuation responses that are co-developed with communities and minimise the risks specific to their townships and localities.

  Realising this goal requires an understanding of the dynamics of an evacuation as well as an insight into how the actions of evacuees and responders influence the process.
 One way to achieve this is to simulate the traffic conditions of a community evacuation using an agent-based model (ABM).
 ABMs are particularly suited to modelling discrete real-world phenomena, where agents represent individuals in the population that can interact with each other and the environment \citep{BonabeauAgentbasedmodelingMethods2002}.
 This method allows emergent or collective behaviours of the community to be captured whilst still preserving the details of the underlying system \citep{ChenAgentbasedmodellingsimulation2008}.
The Belief-Desire-Intention (BDI) architecture \citep{RaoModelingRationalAgents1991} provides a cognitive framework that captures human behaviour within an intuitive structure that is both programmatically efficient and understandable by domain experts \citep{NorlingEnhancingMultiAgentBased2000}. BDI can be used alongside an ABM to add a layer of human reasoning to the actions of agents, allowing them to pro-actively reason about and respond to changing circumstances within the simulation. In this work we will use the BDI-ABM integration outlined by
\citet{SinghIntegratingBDIAgents2016}.

Any evacuation model should take into consideration the expected response of people in the affected community, both to the fire itself and to any evacuation instructions \citep{PelModellingTravellerBehaviour2011}.
One issue is that this behaviour data may be scarce.
Behavioural science research post Black Saturday has focussed on PSDLE decisions \citep{WhittakerCommunitysafety20092013}, but less is known about evacuation travel decisions which also heavily influence an individual's safety. To extract these behaviours, we will use existing data from interviews, surveys and consultation with domain experts to form representative behaviour profiles.
 These profiles will be motivated by bushfire response research that focuses on archetypes in the community \citep{StrahanSelfevacuationarchetypesAustralian2018}, as well as parameter based non-bushfire evacuation response models \citep{LeeIntegratedHumanDecision2010}.

In addition to this, our work will focus on what these behaviour profiles mean for the population \textit{prior} to the evacuation scenario.
The types of activities that people are likely to be engaged in and the dynamics of the background traffic in the region will have an influence on an evacuation should a bushfire threat arise.
We account for these by linking sets of activities to each behaviour profile.
The profile of an agent then determines when it is likely to begin a given activity, and where in the region that activity will occur.
A plan for each agent is generated with details of its activities during the day.
In a given scenario, these plans are interrupted by a specific bushfire threat.
An agent's response depends on its current location and activity as well as its parameterised behaviour profile.

Evaluation focuses on communities in regional Victoria. Behavioural profiles are informally validated by local emergency service personnel and bushfire behaviour experts before being applied to specific communities.
 The simulation is built using the BDI-ABM framework\footnote{\url{https://github.com/agentsoz/bdi-abm-integration}.}, combining the JILL BDI engine\footnote{\url{https://github.com/agentsoz/jill}.} with the MATSim traffic ABM\footnote{\url{https://github.com/matsim-org/matsim}.}.
A range of evacuation scenarios are constructed to experimentally evaluate how sensitive the evacuation outcomes are to the demographic make-up of different communities.

In this thesis we:

\begin{itemize}
  \item Present a review of the literature related to bushfire behaviour, the representation of behaviour in computational models, and the design of social simulations that consider these behaviours.
  \item Outline the design of an algorithm that takes a set of generalised behaviour profiles with associated parameters, and generates a list of agent plans suitable to be run in MATSim.
  \item Describe how these behaviour profiles can be enriched by adding attributes to the plans that dictate how they will respond to a bushfire threat in the BDI-ABM simulation.
  \item Apply the algorithm by generating plans for regions in the Surf Coast Shire of Victoria, describing tangible behaviour profiles drawn from interaction with the local community and its emergency services.
  \item Quantify the bushfire behavioural model through consultation with domain experts and append the BDI attributes to the generated plans.
  \item Run instances of these plans in the BDI-ABM model to illustrate the benefit of conducting bushfire evacuation simulation with the dynamics of the community in mind.
\end{itemize}

The remainder of this document is structured as follows: we present a review of the related literature in Section 2; the design of the algorithm and the BDI behavioural model is given in Section 3; in Section 4 the evacuation simulation application for the Surf Coast Shire is described, and we analyse the algorithm output and the resulting simulation outcomes; lastly, in Section 5 we conclude with a discussion on the current model and directions for future work.

%% file: background/sec-background.tex
\section{Background}

In this section we review  related research into evacuation behaviour, with a particular focus on bushfire and community specific responses.
Following that, we examine social simulation, and specifically, agent-based modelling applied in that context.
Finally, existing models that implement some aspects of our research scope are described.

\input{background/subsec-behaviour-research}

\input{background/subsec-bdi}
\input{background/subsec-social-simulation}
\input{background/subsec-existing-models}
\input{background/subsec-ees}

\Omit{
\subsection{Validation}
***Is this section necessary? Already discussed in ABM context. Either way, it is a mess ATM. Ignore for now. ***\

The BDI framework is easily understandable to domain experts, due to both its compact representation and being intuitive to non-programmers \citep{SinghIntegratingBDIAgents2016}.
This allows behavioural science researchers and emergency personnel to provide informal validation, which \citet{NorlingEnhancingMultiAgentBased2000} emphasises is crucial if behaviours

The BDI-ABM integration is largely motivated by a need to capture a diversity of behaviours within a community of agents under the threat of bushfire, so it is important that these domain experts \citep{LeeIntegratedHumanDecision2010}.

STRUCTURE
*social simulation
*ABM Architecture

*BDI Architecture
*

*HUMAN behaviour
*OFFICIAL inquiries?
*Bushfire

*Traffic and Evacuation -MATSim, Sumo

*Other models

Compare parameter based approach to category based approach
-Shahparvari
Lee
*gaps in research?
}

%% file: background/subsec-behaviour-research.tex
\subsection{Behavioural Research}\label{subsec:behavioural-research}

We proceed by first exploring how behavioural science has understood evacuations in general, and more specifically bushfire behaviours.
Any evacuation simulation needs to take into account the expected human response to environmental cues and received warnings \citep{PelModellingTravellerBehaviour2011}.
Predicting behaviour in an emergency is difficult, and one of the main motivating factors for modelling bushfire evacuations is to gain better insight into expected response to direction and coordination \citep{DashEvacuationDecisionMaking2007}.
The model can only provide this if the underlying human behaviours that the agents are given are realistic.
\

The first aspect to consider is that one should expect systematic
deviations from rational/optimal behaviour in emergency situations  \citep{DashEvacuationDecisionMaking2007}.
Because the situation is likely to be unfamiliar and unusual, people generally do not rely on prior experience and knowledge and instead focus on the information directly available to them, resulting in `myopic' behaviour \citep{PelModellingTravellerBehaviour2011}.
This is exacerbated by the potential for unexpected behaviour in stressful or mentally challenging emergency conditions \citep{KnoopRoadIncidentsNetwork2009}.
People may also fail to comply with directions that they are given, especially if their prepared plan contradicts the instructed action \citep{PelModellingTravellerBehaviour2011}.
Conversely, spontaneous evacuation is known to happen even in places that have not been told to evacuate, so instructed behaviour should not be taken as actual behaviour \citep{LindellCriticalBehavioralAssumptions2007}.

\subsubsection{Bushfire Behaviours}
Whilst there is a body of research into traffic behaviours in evacuations and accidents \citep{LindellCriticalBehavioralAssumptions2007, TuEvacuationplancity2010,PelModellingTravellerBehaviour2011}, it is less available in the context of bushfire disasters.
Behavioural research into bushfires, particularly in an Australian context, has instead tended to focus on the decisions people make between either evacuating or staying and defending property \citep{JohnsonStayGoHuman2011,HaynesAustralianbushfirefatalities2010}.
This is largely a result of the Prepare to Stay and Defend or Leave Early (PSDLE) policy-- colloquially known as `Stay or Go'-- which has been official policy in Australia since the 1990s and explicitly puts the evacuation decision in the hands of individuals rather than the state \citep{ReynoldsHistoryPrepareStay2017}.
In other places where bushfires (wildfires) are common, such as Canada and parts of USA, the norm is that mandatory mass evacuation can and will be enforced by authorities \citep{MoritzLearningcoexistwildfire2014}.
However, related research in North America still tends to focus on the issue of evacuation timing and compliance, and often questions local evacuation policies \citep{PaveglioAlternativesEvacuationProtecting2008,McCaffreyWildfireevacuationits2015}.
This predominant focus on behaviour prior to an evacuation reflects the often rapid and unpredictable nature of a bushfire as compared to other natural disasters, which may allow more calculated and decisive evacuation plans \citep{AdamBdimodellingsimulation2016}.
In a bushfire, the decision between staying or going is most likely to be significant in terms of survival, and so it has become the main focus of behavioural research.
\

Accordingly, we should expect that `Stay or Go' decisions play an important part in an evacuation on a macroscopic level, and that a model of evacuation traffic behaviour will need to account for the circumstances of a person's departure.
In their analysis of community preparedness for the 2009 Victorian `Black Saturday' bushfires, \citet{WhittakerCommunitysafety20092013} highlight the dangers of late evacuation. People who leave late are usually triggered by the sight of heavy smoke or flames, which means that
\begin{quote}`By this time it is likely that driving a vehicle will have become very difficult, with flames, smoke, strong winds, fallen trees, traffic and the urgency of the situation increasing the likelihood of accidents \citep{TibbitsStaydefendleave2007}.'\end{quote}
Late evacuations are associated with a lack of planning and preparedness among those who initially choose to stay and defend, but are not fully committed to that decision. Interviews with Black Saturday survivors establish that there can be a disconnect between intended action and actual behaviour; for instance, 63\% of people who adopted a `wait and see' approach ended up leaving once the fire was visible.
\

A difficulty here is whether to define waiting and seeing as an intended action; as \citet{McLennanHouseholderdecisionmakingimminent2012} note, to `wait and see' is more of an observation that describes a person's mindset rather than a fully formed plan that represents their intentions.
One new approach is to move away from viewing the decision as a binary question of leaving or staying.
\citet{StrahanSelfevacuationarchetypesAustralian2018} form seven self-evacuation archetypes which accommodate a diverse range of responses to a bushfire threat. With foundations in Jung's work on the collective unconscious  \citep{JUNGCollectedWorksJung1969},  archetypes represent `fundamental characteristics of humanity'.
There is also a basis for their use in Australian public policy, where they are generated in a similar way using cluster and discriminant function analysis.

 The results in \citet{StrahanSelfevacuationarchetypesAustralian2018} were derived from interviews with 452 participants who had recently experienced bushfire. The questions focussed on factors like experience, intended and actual responses, self-responsibility, access to information and demographics. The seven archetypes they found are shown in Table~\ref{table:strahan}.

 \begin{table}[!ht]
 \caption{Self-evacuation archetypes according to \citet{StrahanSelfevacuationarchetypesAustralian2018}.}\label{table:strahan}
 \centering
 \begin{tabular}{|p{2.3cm}|p{5cm}|p{4cm}|}
\hline
 Archetype &Key characteristics & Evacuate or Remain \\ [0.5ex]
 \hline
Responsibility Denier& Believe they are not responsible for their personal safety or for their property. &Highly committed evacuators but expect others to direct and assist.\\
\hline
Dependent Evacuator&Expect the emergency services to protect them and their property because they are incapable of taking responsibility for themselves.&Highly committed evacuators but expect others to direct and assist.\\
\hline
Considered Evacuator&Having carefully considered evacuation, are committed to it as soon as they are aware of a bushfire threat.&Committed to self-directed evacuation.\\
\hline
Community Guided&Seek guidance from neighbours, media and members of the community who they see as knowledgeable, well informed and providing reliable advice.&Committed to evacuation on community advice.\\
\hline
\nohyphens{Worried Waverer}&Prepare and equip their property and train to defend it but worry they lack practical experience to fight bushfire putting their personal safety at risk.&Wavering between evacuating and remaining.
\\
\hline
\nohyphens{Threat Denier}&Do not believe that their personal safety or property is threatened by bushfire.
&Committed to remain as perceived lack of threat makes evacuation unnecessary.\\
\hline
 Experienced Independent&Are highly knowledge, competent and experienced and are responsible and self-reliant fighting bushfire.&Highly committed to remaining because they are highly experienced and well prepared.\\ [1ex] 
 \hline 
 \end{tabular}

 \label{table:Arch}
 \end{table}

\citet{ReidWhereFireCoConstructing2014} conducted similar interviews following 11 January 2010, a day where a `Catastrophic' fire warning was issued to residents of Halls Gap.
This warning category was introduced in the aftermath of Black Saturday to indicate that evacuation is highly recommended; despite this, the majority of residents did not leave.
One factor attributed to this response was the nature of risk perception in the community and that `awareness does not always result in a realistic understanding
of how to respond to risk'.
Residents may believe that they have a better understanding of local weather and geographic conditions than a state-wide warning system, and prior experience of similar bushfire events will also strongly influence people's decisions \citep{McCaffreyOutreachProgramsPeer2011}.
Further to this, trust in warning systems can be eroded if they are perceived as a political initiative, or even as an overly `scientific assessment' of risk. Community response resources that incorporate local knowledge are likely to have more impact on people's behaviour.
\

Horse ownership is a good example of a community specific behaviour which might be better understood from a local perspective.
\citet{ThompsonPlannedultimateactions2018} highlights that although pet ownership is generally a factor in people's evacuation plans, horse owners (who are relatively common in rural areas) are presented with unique challenges during a bushfire.
Due to the difficulty of transporting horses pre-emptive or early evacuation is advised by the CFS as a priority action.
However, Thompson et al. find that this blanket strategy may be detrimental if it comes at the expense of other preparatory planning, and could prevent horse owners from forming contingency plans that improve the likelihood of horse survival in the event they are left in place.

%% file: background/subsec-bdi.tex
\subsection{Representing Behaviour in Computational Models}

The complex nature of human behaviour presents challenges for modellers in many different domains. \citet{KulashTraditionalNeighbourhoodDevelopment1990} highlights a particular issue in urban planning, where traffic engineering that fails to take human behaviour into account can result in contradictory or unsafe policies.
Designing hierarchical traffic systems that rigidly place different roads in different classes (often inspired by natural phenomena like river systems) is not effective because it considers humans merely as particles in a flow, rather than autonomous agents who can decide to use a road for reasons separate from its assigned purpose.
Moreover, urban planners cannot assume random distribution or optimal dispersal across a network.
\citet{SongModellingscalingproperties2010} explore patterns of human mobility, and note that traditional continuous-time random-walk models of movement do not reflect the tendency of humans to restrict their exploration to locations near them, nor their propensity to return to frequently visited locations. \joelComment{Not sure about proposed MATSim reference here... you say later to avoid mixing simulation in this section? Also, I probably should investigate the 'co-evolutionary algorithm that converges to a quasi-Nash equilibrium', but how does MATSim address this specific issue... I would've thought that location exploration/frequently visited locations is more determined by the inputted plans for MATSim? Or is this in reference to MATSim's replanning functionality? }
\

This also has relevance in epidemiological modelling, where the distinction between density and frequency-based transmission is important \citep{AntonovicsGeneralizedModelParasitoid1995, Lloyd-SmithFrequencydependentincidence2004}.
The ubiquitous model in epidemiology is the Susceptible-Infected-Recovered (SIR) set of differential equations, and indeed most macroscopic representations of disease are extensions of \citet{Kermackcontributionmathematicaltheory1927}'s model\joelComment{Do you mean inconsistent ref style because two-author citations should have et al as well?}.
\citet{FunkModellinginfluencehuman2010} review the presence of behavioural considerations in disease models, and note that whilst human behaviour is intrinsically linked to disease spread, it is difficult to completely synthesise the pathogen dynamics that the SIR model captures and the complex interplay of `attitudes, belief systems, opinions and awareness of a disease... both in an individual and
in the population on the whole'.
One alternative is to approach disease modelling from a contact network perspective, viewing the population as a complex system \citep{GaleaCausalthinkingcomplex2010,ChristakisSpreadObesityLarge2007}.
Considering individuals within a larger social network and using a bottom-up method allows various different behaviours to be examined; for instance, clustered interaction networks permit the use of percolation thresholds to characterise epidemics \citep{MooreEpidemicspercolationsmallworld2000, Davisabundancethresholdplague2008}.
This is also of interest in economics, where the classical notion of the rational agent is often poorly represented by cumbersome mathematical machinery \citep{Farmereconomyneedsagentbased2009}.
It may not be appropriate to assume that a market will reach a long term equilibrium when complex individual behaviours within the population give rise to sub-optimal emergent behaviours in the collective.
\

Whilst most behavioural research uses surveys and interviews to analyse bushfire responses, there is usually the caveat that both the sample size and the qualitative nature of the data limit the predictive power of
results.
The acknowledgement that intended and actual actions may differ \citep{WhittakerCommunityBushfireSafety2010} means that the interview approach has limited application in capturing behaviour, and actual modelling of bushfire evacuation behaviour has tended to use `the judgement of evacuation planners or experts familiar with the area under study' \citep{BeloglazovSimulationwildfireevacuation2016}.
This is further enforced by the lack of behavioural research into on-road evacuation behaviours in bushfires.
\citet{KennedyModellingHumanBehaviour2012} is critical of the tendency to model behaviour using uniform random variables as it does not leave room for biases and illogical decisions, ignores any notion of memory/learning, and makes it difficult to identify behavioural preferences between different groups of people.
Instead, threshold-based rules are recommended as a simple way of approximating consistent behaviours within an individual agent.

\subsubsection{Belief-Desire-Intention Model}
Any computational model that seeks to use human behaviour to observe emergent outcomes will require a cognitive architecture to represent the human decision making process \citep{GilbertWhendoessocial2006}.
There are several approaches to forming a cognitive model.
A connectionist approach seeks to mirror the design (but not necessarily the complexity) of the brain's neural networks, giving a natural representation of the biological process \citep{McClellandConnectionistmodelspsychological1988}. One shortcoming of a neural network model is that the reasoning process that leads to a particular behaviour is not readily explainable \citep{MedskerDesigndevelopmenthybrid1994}. Other models form a more abstract picture of brain function, with a focus on the storage and deployment of knowledge-- ACT-R and Soar are two well-known examples \citep{AndersonACTsimpletheory1996,LairdExtendingSoarCognitive2007}. Whilst these models are firmly based on experimental evidence and psychological theory \citep{GilbertWhendoessocial2006}, they focus heavily on the individual level of cognition. The Belief-Desire-Intention (BDI) framework provides a way to represent rational decision making that lies more between the individual and social context \citep{
BratmanPlansresourceboundedpractical1988,JarvisHolonicExecutionBDI2008}. Less tied to cognitive science, BDI relies on a `folk psychology' approach that captures how people think about how people think \citep{NorlingEnhancingMultiAgentBased2000}.
\

The BDI framework is formally described by \citet{RaoModelingRationalAgents1991}.
Based on a branching-time possible-worlds model, BDI   seeks to model rational agents that can vary their actions based on changing beliefs, desires/goals and intentions.
\textit{Beliefs} represent an agent's knowledge of the environment, and are not necessarily perfect pictures of the current state of the system \citep{NorlingEnhancingMultiAgentBased2000}.
Combinations of beliefs will leave an agent with certain \textit{desires} that they wish to fulfil.
Desires may be inconsistent with one another, but goals are a subset of desires that are consistent and achievable according to the agent's beliefs.
To achieve these goals, an agent will form \textit{intentions}, or plans of action to execute.
\

\citet{RaoBDIAgentsTheory1995} acknowledge that the abstract formalism of their BDI theory does not lend itself to practical computing, and instead propose a simplified architecture which is more efficient \citep{Ingrandarchitecturerealtimereasoning1992}.
Many modern BDI programming languages are implementations of Rao and Georgeff's abstract interpreter, and some have maintained a link to the formal underpinnings of BDI;
the AgentSpeak(L) language \citep{RaoAgentSpeakBDIagents1996} and its descendent interpreter Jason \citep{BordiniJasonGoldenFleece2005} are useful  in this sense because they allow properties of an empirical BDI implementation to be proven methodically.
GOAL \citep{deBoerAgentProgrammingDeclarative2002} extends this capacity to allow programmable \textit{declarative} goals as part of the propositional logic.
This means that goals, like beliefs, are separate from any required action and can form part of an agent's abstract reasoning.
Other languages include the C++ based dMARS \citep{DInvernodMARSArchitectureSpecification2004} and the Java based JADE \citep{BellifemineDevelopingmultiagentsystems2001}, JACK \citep{BusettaJACKIntelligentAgents1999} and Jill.
Jill is a relatively new Java-based BDI platform  that is lightweight, scalable and geared towards integration with large-scale simulations.
It has been developed specifically to handle a large number of agents.\footnote{See \url{https://github.com/agentsoz/jill/} for details.}\joelComment{Maybe just check my added comments on BDI make enough sense. }

%% file: background/subsec-social-simulation.tex
\subsection{Social Simulation}\label{sec:social-simulation}
The concept of applying computational techniques to social problems has emerged in the last 30 years as a new way to model the complex and often non-linear systems that occur in human societies.
\citet{AxelrodAdvancingArtSimulation1997} proposes in a somewhat extrapolatory manner that social simulation is a `third way of doing science', and that it allows both adaptive (non-optimal) and rational behaviour to be analysed by modellers. \citet{MillerComplexAdaptiveSystems2007} consider the relatively new computational approach as complementary to the more traditional methodologies in social science.
Another major benefit of employing simulation is that very simple and explicit assumptions can be used to generate complex and unpredictable behaviour which would be very difficult to understand analytically \citep{GilbertSimulationSocialScientist2011}.

Drawing ideas from computer science, mathematics, biology, and social science, social simulation provides a forum for cross-discipline interaction, and development has come from a number of domains \citep{AxelrodAdvancingArtSimulation1997}.
The UrbanSim model \citep{WaddellUrbanSimModelingurban2002}
is a polished web-based program that applies simulation to urban planning problems and allows metropolitan planning organisations to forecast the effects of growth in a city's population.
In economics, simulation allows researchers to observe  aggregate rationality emerging from irrational individual actions \citep{GodeAllocativeEfficiencyMarkets1993} and compare the effectiveness of different trading strategies \citep{RustCharacterizingeffectivetrading1994}.
Queuing models are one of the more established applications of social simulation, where discrete events dictate the evolution of a system, and the times between each event are given by some probability distribution \citep{GilbertSimulationSocialScientist2011}.
Schedules generated in this manner generally apply in customer service settings, so such simulations can be used to measure the expected average waiting times and idle times associated with different system states.

\subsubsection{Agent-based Modelling}

One commonly used method in social simulation is agent-based modelling (ABM).
   Through the actions and interactions of autonomous agents within the simulated environment, ABM allows modellers to capture the emergent behaviours of complex systems and has been applied in a range of fields, including public health, urban planning, disaster management and economics \citep{Dawsonagentbasedmodelriskbased2011,AuchinclossBriefintroductoryguide2015}. \citet{MacalTutorialagentbasedmodeling2005}, maintain that `modelling human social behaviour and individual decision-making' is the main purpose of ABM.
They see the possibilities of ABM creating new challenges in social science research, including two central questions:
\begin{itemize}
\item \textit{How much do we know about credibly modelling people’s behaviour?}
\item \textit{How much do we know about modelling human social interaction?}
\end{itemize}
ABMs are particularly suited to modelling discrete real-world phenomena, where agents represent individuals in the population that can interact with each other and the environment \citep{BonabeauAgentbasedmodelingMethods2002}.
The scope of ABM ranges from small-scale academic demonstrations which focus on the key features of the system, to large-scale, validated models with millions of agents that can be used to support policy decisions \citep{MacalTutorialagentbasedmodeling2005}.
Agents are required to be discrete, autonomous components of a system that are able interact with the system's environment (including other agents) and independently adapt behaviour based on these interactions \citep{ChenAgentbasedmodellingsimulation2008}.

The ABM approach allows modellers to:
\begin{itemize}

\item \textbf{Expose emergent behaviours}: These are defined as characteristics of the system that are best observed via the interactions between the individual agents \citep{ChanAgentbasedsimulationtutorial2010}.
\item \textbf{Naturally describe systems}: ABM also allows a natural synthesis of a complex system into a descriptive model \citep{BonabeauAgentbasedmodelingMethods2002}.
\item \textbf{Apply stochasticity}: Complex behaviours are represented in terms of discrete agent activities, which allows randomness to be applied in a targeted and realistic manner \citep{CrooksIntroductionAgentBasedModelling2012}.
\item \textbf{Seek expert validation}: Because agent behaviour is explicitly defined by these activities, domain experts are able to more easily understand, calibrate and validate the model \citep{BonabeauAgentbasedmodelingMethods2002}.
\end{itemize}

ABM is considered by \citet{BonabeauAgentbasedmodelingMethods2002} to be a way of thinking about and analysing systems rather than a specific technological tool.
It provides a microscopic, bottom-up approach to modelling wherein a system is represented by its constituent parts.
 The contrast between ABM and `macroscopic' modelling provides a new perspective on the scientific process.
 For instance, \citet{EpsteinModellingcontainpandemics2009}
  argues that simulating a pandemic with individual agents interacting via social network structures is more effective than classical disease modelling, as it easily allows for heterogeneous factors and non-random mixing \citep{GrimmIndividualbasedModellingEcology2005}.
  \citet{BazzanAgentsTrafficModelling1999} consider the application of microscopic simulation to transportation science.
  They argue that the concept of rationality is very relevant in a traffic simulation, and that maximisation or optimisation should not be the goal.
  Instead it should feature the emotional, non-logical components of decision making, like impulsive lane changing, trying a different route due to impatience, or how a person's internal mental state might be expressed in their driving performance.
  BDI is cited as an appropriate tool for introducing these individual actions within the social framework of the road network.
  \citet{GilbertWhendoessocial2006} makes a similar conclusion, holding that  the micro cognitive level should both influence and be influenced by the macro social level in models of human behaviour.
  This is a crucial element of social simulation and illustrates that an agent-based evacuation model should not focus too heavily on one level of detail.
  The interaction between the agents and their surrounding environment will reveal as much as analysis of standalone social or individual outcomes.

\subsubsection{Synthetic Populations}
Social simulation often requires that the modelled agents to reflect attributes observed in real world data.
A synthetic population is one that is derived from aggregate statistical data to represent the actual population. This is often necessary because the micro-data required to accurately model the population at an individual level is either unavailable or inaccessible \citep{MoeckelCreatingSyntheticPopulation2003}.
There are several established techniques for creating synthetic population datasets, and often these use actual but anonymous individual sample sets attained from census data to generalise incrementally to create a full population \citep{HarlandCreatingrealisticsynthetic2012}.
These methods, including deterministic re-weighting \citep{BallasSimBritainspatialmicrosimulation2005} and the conditional probabilities model \citep{BirkinSynthesisSyntheticSpatial1988}, aim to match the distribution of persons and households to demographic data \citep{JainCreatingSyntheticPopulation2015}.
However, these processes require a stable and realistic set of actual sample data \citep{WickramasingheHeuristicDataMerging2017}.
In regions that feature a large transient or fluctuating non-resident population then this information may not exist in a reliable form, as any existing individual or household data may not generalise to the period that the simulation is aiming to model.

%% file: background/subsec-existing-models.tex
\subsection{Existing Evacuation Models}
In this section, several models  related to the aims of this thesis are described.
They help guide strategies in the development of our own model, as well as reveal approaches that might be counter-productive.
\citet{ShahparvariMultiObjectiveDecisionAnalytics2015} approach the problem of late evacuation in a bushfire from an optimisation perspective with a focus on limited resources and constraining time windows, road disruptions and shelter capacities.
This process is useful in determining which factors have the most impact on an evacuation but has little potential to accurately predict the outcome of an evacuation as human behaviour in an unpredictable situation is likely to be sub-optimal, although there have been attempts to validate optimised plans via agent-based simulation \citep{Pillacconflictbasedpathgenerationheuristic2016}.
\

One way that other models have sought to capture suboptimal behaviour has been to emulate anxiety and human interaction using sensory parameters.
This has been a focus in crowd-based evacuation simulation, where agents will speed up due to an awareness of danger or follow another agent if they are related to them (i.e. parent-child pairs) \citep{OkayaBDIAgentModel2011}.
\citet{LeeIntegratedHumanDecision2010} expand on this idea with their leader-follower dynamic, which they define via a `confidence index'.
Agents who perceive sounds and smoke from a bomb blast will update their confidence index and then, based on some threshold, transition through different states that affect their decision options (using the BDI framework).
If their confidence index is low, they are likely to follow other agents rather than assess other cues for exit points.
Agents also have a predefined type-- `Novice' or `Commuter'-- which changes their knowledge of escape routes and their initial likelihood to be a follower or leader.
\

This concept of following or leading can be translated to traffic evacuations and on road behaviour.
\citet{YuanTrafficevacuationsimulation2017} propose a modular driving decision framework that allows for herding, congestion avoidance, panic-affected driving and variable shelter selection.
They incorporate a number of different routing behaviour options-- e.g. following shortest path, following a leading car or using a GPS-- and then use a BDI cognitive layer to determine the course of action to achieve their goal.
One important aspect of their model is that it allows for heterogeneity of route choices rather than focussing on the optimal solution, highlighting \citet{SadriArifMohaiminHowEvacuateModel2014} findings that `evacuees may not follow a recommended route but may take a usual or familiar one instead'.
Further, they claim that the route and destination choices of drivers have the most affect on clearance time in a large scale evacuation.
\

\citet{ScerriBushfireBLOCKSModular2010} use this same modular approach to explore bushfire response strategies.
The decision on when and where to leave uses a cell-based ABM and focuses on a number of attributes like age, gender, panic level and known information.
A separate traffic model uses the lightweight coordinate-based Repast program, and has agents choose a route that takes the shortest path they believe to be safe.
This belief is updated dynamically as the traffic simulation progresses.
The separation of the cognitive and traffic models, with a parameterised trigger leading to the dynamic traffic simulation, allows the outcomes of each decision model to be more easily analysed.
\

A somewhat unique bushfire evacuation model is provided by \citet{AdamModellingHumanBehaviours2017}, who directly use survivor testimonies to formulate agent behaviour.
They argue that this allows them to simulate actual behaviours, rather than behaviours prescribed by experts.
A tension between perceived and actual danger was highlighted in their review of post Black-Saturday interviews, where people would quickly shift their perception from passive to hyper-alert once the risk became undeniable.
Using a finite-state machine, survivor responses were translated into several stage states that an agent may pass through-- `Unaware', `Aware indecisive',`Preparing to escape', `Escaping' and `Preparing to defend'.
These stages are linked by various triggers that allow them to move into a new state.
Triggers are changes in person attributes in response to changes in the environment, which is limited to a simplistic grid model lacking geospatial attributes.
However, their focus on reproducing realistic behaviours allows them to closely match their results with observed death causes in a fire, validating their intensive, qualitative approach.

%% file: background/subsec-ees.tex
\subsection{The Emergency Evacuation Simulator}
As we have already seen, capturing irrational behaviours is especially important in an evacuation simulation \citep{BarrettDevelopingDynamicTraffic2000}.
Time pressures and limited route choice, along with competing and overwhelming new layers of information (much of which may be unknown to an individual agent) should result in a wide array of responses which diverge from a hypothetical optimal solution.
\citet{PadghamIntegratingBDIAgents2014} provide one such model which employs BDI to augment an ABM in the context of bushfire evacuation.
They describe a method where an BDI agent acts as the cognitive `brain' to the active `body' of a corresponding ABM agent.
The ABM will act based upon instructions from the BDI agent in addition to responding directly to interactions within the ABM system, and the BDI agent will reason and make decisions based on information the ABM agent encounters.
These coupled agents can communicate via

\begin{itemize}
\item\textbf{percepts}: where an ABM agent notifies the BDI agent of an event in the system.
\item\textbf{queries}: where a BDI agent requests information from the ABM agent.
\item\textbf{actions}: where the BDI agent, having reached a decision, informs the ABM agent of what action to take.
\item\textbf{action states}: where an agent conveys to its counterpart that the status of an action has been updated.

\end{itemize}

This model has been refined into the Emergency Evacuation Simulator (EES) tool designed to assist emergency services in Australia respond to natural disasters \citep{SinghEmergencyEvacuationSimulator2017}.
The tool features a web interface which allows domain experts to build and test community evacuation scenarios first-hand, with a view towards it being both a useful planning tool and a decision support system during an emergency.
As a currently active project with a number of different stakeholders, the EES provides a flexible and open platform in which to further test BDI bushfire behaviours as they are developed in this thesis.
\

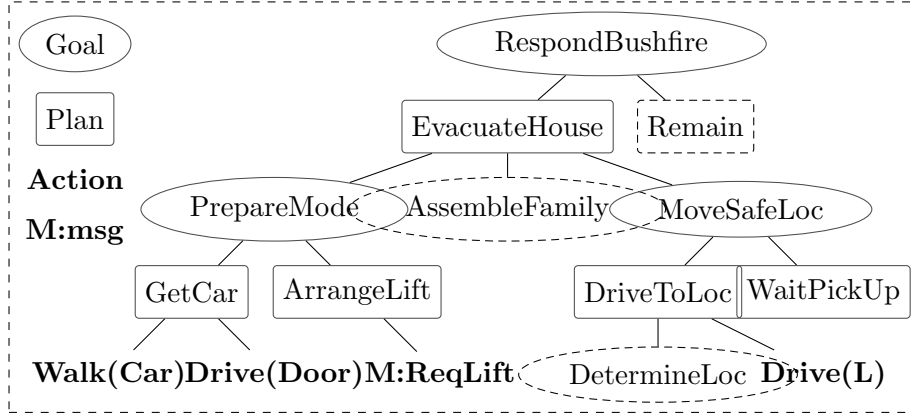
\begin{figure}
  \centering
  \caption{An example of a goal-plan tree in a bushfire context \citep{SinghIntegratingBDIAgents2016}. }\label{fig:gp-tree-ex}
  \input{background/fig-gp-bushfire.tex}

\end{figure}

Development of the Jill BDI engine has recently focussed on expanding the underlying capabilities of the BDI-ABM integration, and much of Jill's architecture is constructed to interact directly with an ABM \citep{SinghIntegratingBDIAgents2016}.
Taken as its own modular component in the integration, Jill allows BDI agents to make decisions based on a goal-plan hierarchy.
An agent holds a number of belief states, which may be altered by information received from the ABM environment.
High-level goals are formed based on changes to these belief states, and these are reasoned through via a set of conditional plan options.
A selected plan may involve a further set of more specific goals, and this process continues until eventually a plan lands upon some decision on how the counterpart agent should act and respond within the ABM.
This process is easily described via a goal-plan tree, which maps the decision-making process into a flowchart that visualises and explicitly orders the various goal and plan options that a BDI agent may have.
Apart from being useful in the process of designing a behavioural model, these goal-plan trees are beneficial as explanatory tools that domain experts and stakeholders can understand.
An example goal-tree is provided in Figure~\ref{fig:gp-tree-ex}.
\
\

For its ABM component the EES utilises MATSim, a multi-agent traffic simulation framework that is implemented in Java \citep{HorniIntroducingMATSim2016}.
The traffic flow model is designed for large-scale simulations and uses a queue-based approach to move agents through a network graph made up of weighted nodes and links.
Storage and flow capacities in the network constrain and shape the movement of an agent, but the main impetus in the model is an agent's desire to complete an individual plan containing a list of activities with corresponding locations with end times, along with a travel mode between each activity.
This plan is initially provided as a per-agent input to the simulation, and is iteratively updated by a co-evolutionary algorithm \citep{PopoviciCoevolutionaryPrinciples2012} that optimises the plan's `score' an individual level rather than aiming for system equilibrium.
These plans are collectively stored in the \texttt{population.xml} input file.
\

MATSim is a long-running and extendable open source project that now features a large number of contributions augmenting its core function as a traffic simulator.
To facilitate these augmentations, MATSim plans can have any number of contribution-specific attributes appended to them, which allow additional parameters and inputs to be associated with each agent as needed. See Figure~\ref{fig:MATSim-plan} for an example of a MATSim \texttt{population.xml} file with attributes added to each agent's plan.
\
\begin{figure}
  \centering
  \caption{A MATSim \texttt{population.xml} file with attributes added to agent plans. }\label{fig:MATSim-plan}
  \input{background/matsim-pop-plan.tex}

\end{figure}
\

BDI-MATSim contributions using BDI engines such as JACK and JADE have included  applications to taxi administration and bushfire evacuation \citep{PadghamMakingMATSimAgents2016}.
In an emergency context, one of the distinguishing features of a BDI-MATSim simulation is that plans should not be iteratively improved; all decision-making must be performed `on the fly' by the BDI agent.
This represents the idea that people in an emergency cannot rely on their experience or prior learned behaviours, and must react to new situations as they are presented.
With this arrangement in place, the initial plan inputs become more significant, because they will directly dictate where agents are prior to the emergency situation.
A large part of the contribution in this thesis is providing a method to generate these plans, including the BDI-specific parameters that are required for each agent.
Together, these input procedures will shape the resultant simulation, so establishing this process is an essential design step for the EES project.

%% file: background/fig-gp-bushfire.tex
\begin{tikzpicture}[level distance=1.1cm]

\tikzstyle{txt}=[scale=1]
\tikzstyle{color1}=[black]
\tikzstyle{plan}=[scale=1,draw=black!70,rounded corners=0.05cm,minimum height=0.7cm]
\tikzstyle{plan1}=[plan,black!30]
\tikzstyle{plan2}=[plan,draw=orange!0,top color=orange!20, bottom color=orange!60, shading angle=45]
\tikzstyle{plan3}=[plan2, opacity=0.5]
\tikzstyle{goal}=[scale=1,draw=black!70,ellipse,minimum height=0.65cm,minimum width=0.75cm]
\tikzstyle{goal1}=[goal,black!30]
\tikzstyle{goal2}=[goal,draw=blue!0,top color=blue!10, bottom color=blue!30, shading angle=45]
\tikzstyle{goal3}=[goal2, opacity=0.5]
\tikzstyle{trace}=[scale=1,minimum size=0.7cm,font={\bfseries},solid,style=circle,fill=blue,text=white,yshift=2em,opacity=0.5]
\tikzstyle{trace1}=[trace,opacity=1.0]
\tikzstyle{percept}=[draw,star,star points=7,minimum height=0.65cm,minimum width=0.75cm]
\tikzstyle{level 1}=[sibling distance=2.5cm] 
\tikzstyle{level 2}=[sibling distance=3.1cm] 
\tikzstyle{level 3}=[sibling distance=2.2cm]
\tikzstyle{level 4}=[sibling distance=2.2cm]

\node[goal] (BR) {RespondBushfire}
	child[solid] {node[plan] (EH) {EvacuateHouse}
		child {node[goal,sibling distance=0.5cm] (OT) {PrepareMode}
			child {node[plan] (GC) {GetCar}
				child {node[txt] (WC) {\textbf{Walk(Car)}}}
				child {node[txt] (DD) {\textbf{Drive(Door)}}}
			}
			child {node[plan] {ArrangeLift}
				child[missing] { node {} }
				child {node[txt] {\textbf{M:ReqLift}}}
			}
		}
 		child {node[goal,densely dashed,black] (AM) {AssembleFamily}
 		}
		child {node[goal] (SL) {MoveSafeLoc}
			child {node[plan] {DriveToLoc} 
				child[missing] { node {} }
				child {node[goal,densely dashed,black] (DL) {DetermineLoc}
				}
				child {node[txt]{\textbf{Drive(L)}}}
			}
			child {node[plan] (WPU) {WaitPickUp}}
		}
	}
	child {node[plan,densely dashed,black] (SD)
	{Remain}} ;
\begin{pgfonlayer}{background}
\node[goal, xshift=-7cm,yshift=0.0cm] (LG) {Goal};
\node[plan, xshift=-7cm,yshift=-1.0cm]{Plan};
\node[txt,xshift=-7cm,yshift=-1.8cm]{\textbf{Action}};
\node[txt,xshift=-7cm,yshift=-2.5cm]{\textbf{M:msg}};
\node [fit=(WPU) (WC) (DL) (BR) (LG)][rectangle, minimum height=4cm, minimum
width=2.5cm, thin, draw, dashed]{};
\end{pgfonlayer}
\end{tikzpicture}

%% file: background/matsim-pop-plan.tex
\begin{lstlisting}[%
language=XML,
numbers=none]

<population>

 <person id= "1" >
    <attributes>
      <attribute name="BDIAgentType" class="java.lang.String" >io.github.agentsoz.ees.agents.bushfire.Resident</attribute>
    </attributes>
    <plan selected="yes" score="143.50345971">
      <activity type="home" x="766728.617380239" y="5754449.11681867" end_time="09:40:00" />
      <leg mode="car" />
      <activity type="work" x="789872.47261438" y="5752811.26023417" end_time="16:34:00" />
      <leg mode="car" />
      <activity type="home" x="766728.617380239" y="5754449.11681867" />
    </plan>
  </person>

 <person id= "2" >
    <attributes>
      <attribute name="BDIAgentType" class="java.lang.String" >io.github.agentsoz.ees.agents.bushfire.Resident</attribute>
    </attributes>
    <plan selected="yes" score="93.2987721">
      <activity type="home" x="786831.511691175" y="5770083.81316717" end_time="08:16:00" />
      <activity type="shops" x="759299.781695201" y="5729901.27980373" end_time="10:34:00" />
      <leg mode="car" />
      <activity type="beach" x="791120.803641559" y="5752820.35296198" end_time="13:07:00" />
      <leg mode="car" />
      <activity type="home" x="786831.511691175" y="5770083.81316717" />
      </plan>
  </person>

<population>
\end{lstlisting}

%% file: chapter-theory/sec-plan-algorithm.tex
\section{Population Generation Algorithm}
A person's response to a bushfire will inevitably be influenced by their plan for the day, including what they are doing when they become aware of the threat and what they plan to do next.
In this section, we describe a plan generation algorithm that allows us to easily generate the planned activities for every individual in a population on a  given day and apply attributes that reflect how they will respond to a bushfire.
If we want to represent these individuals as agents in an EES simulation then their plans need to follow the same semantics that MATSim requires.
The algorithm output for an agent plan should consist of a set of sequenced activities, each with a location and start-time, along with travel legs linking them. Any additional parameters should be included as attributes per agent plan, with the collection of these plans forming a \texttt{population.xml} file as in Figure~\ref{fig:MATSim-plan}.
\

Although there are a number of established methods for synthesising populations from census data \citep{HarlandCreatingrealisticsynthetic2012}, we are interested in generating populations in a range of diverse settings that may differ fundamentally in their underlying demography.
 The make-up of a population and its vulnerability to a bushfire threat will vary depending on the weather conditions, the time of year, which day of the week and even the time of day that the fire strikes \citep{ReidWhereFireCoConstructing2014}.
 These variations are exacerbated when the considered population has a significant transient element, or in a scenario where a special event is occurring in the region that skews both the location and movement patterns for a large portion of the population.
 We therefore require a flexible set of inputs that can be manipulated to reflect the population in a specific scenario. Additionally, inputs should be simple enough for emergency personnel to understand and validate against observed data.
\

This section will proceed by first outlining in a generalised form the inputs that the algorithm requires to generate agent plans.
We then describe the algorithmic process for creating an agent plan, which includes assigning activities to an agent, sequencing these activities into a day plan, and giving each activity in the sequence an appropriate location.
Once plan generation is established, a bushfire behaviour Belief-Desire-Intention model is formulated.
The necessary BDI attribute parameters alongside the plan generation inputs constitute a set of behaviour profiles that broadly define how a given agent will react in a bushfire emergency scenario.

 \input{chapter-theory/subsec-inputs}
 \input{chapter-theory/subsec-activities}
\input{chapter-theory/subsec-locations}
 \input{chapter-theory/subsec-behaviours}

%% file: chapter-theory/subsec-inputs.tex
\subsection{Inputs}

We first consider the inputs that the algorithm requires for a given bushfire scenario with region $R$ and population of agents $P$ over a time period $T$:

\begin{itemize}
  \item A set of subgroups $S$ and a size $P_s$ for each $s\in S$ that partitions $P$ up into categories based on an agent's relationship to $R$.
  \item A set of activities $A$ (of size $K$) that an agent can be assigned to at any time $t\in T$.
  \item A choice of time-step size $\frac{T}{N}$, where $N$ denotes the number of time-steps.
  \item An activity distribution $\Delta_s$ for each subgroup $s$ that defines the expected proportion of $s$-agents engaged in activity $\alpha_k\in A$ at each time-step $t_n$ for $n=1,2,...,N$.
  \item A duration weighting $d_{s,\alpha_k}$ for each $(s,\alpha_k)$ pair that dictates how long an $s$-agent will take to complete activity $\alpha_k$.
  \item A set of locations $\mathcal{L}$ in $R$ defined by a suitable coordinate reference system $W$.
  \item A location mapping $M_{s,\alpha_k}$ for each $(s,\alpha_k)$ pair that defines the set of possible locations in $\mathcal{L}$ where an $s$-agent can complete the activity $\alpha_k$.
  \item An allocation number $a_\ell$ for each location $\ell\in\mathcal{L}$ that defines the expected number of agents attending $\ell$ during the given scenario.
  \item a family $\{L_i\}$ (indexed by $I$) of localities that partitions $\mathcal{L}$ into neighbourhoods.
  \item A travel factor $g_s$ for each $s$ that governs how likely an $s$-agent is to choose a new location outside of its current locality.
\end{itemize}


\subsubsection{Subgroups}
We partition the population of agents $P$ into subgroups that broadly reflect the diversity of people in the area and where they have come from.
These will differ depending on $R$ and the given bushfire scenario, but broadly there should be a focus on knowledge of the area, intended duration of stay in the area, and connection to the community through friendships, relationships and assets.
The most basic input is thus a description of how the population should be broken up into distinct groups. Let $S$ be the set of all subgroups $s$.
We also denote
$$
\{P_{s}|s\in S\}
$$
as the partition of the population of agents $P$ into subgroups.
One simple example of a subgroup partition would be breaking the population up into residents and visitors.

\subsubsection{Activity Distributions}\label{Dist}
Subgroups are mainly formed around expected behaviours during a bushfire, but they are also useful in defining differing intentions for people in their daily activities.
For a given bushfire scenario, a set of activities that a significant number of the population is expected to participate in is determined.
These could include going to the beach, shopping, eating at a restaurant or attending a large event.
Any desired granularity is allowable at this step, but it is better to tend towards more general activity labels here, with more specific sub-activities distinguished by location (see Section~\ref{sec:locmaps}).
Additionally, activities that are only undertaken by a minority can be grouped together into an all encompassing \texttt{other} activity.
The only enforced activity condition is that $\alpha_1$ should specify a \texttt{home} location for each agent.
This plays a fundamental role in how an agent's plan develops, as we shall see.

\

We denote

$$
A= \{\alpha_1,\alpha_2,...,\alpha_k,...,\alpha_K\}
$$

as the set of $K$ possible activities $\alpha_k$ that an agent can be assigned to at time $t\in T$.

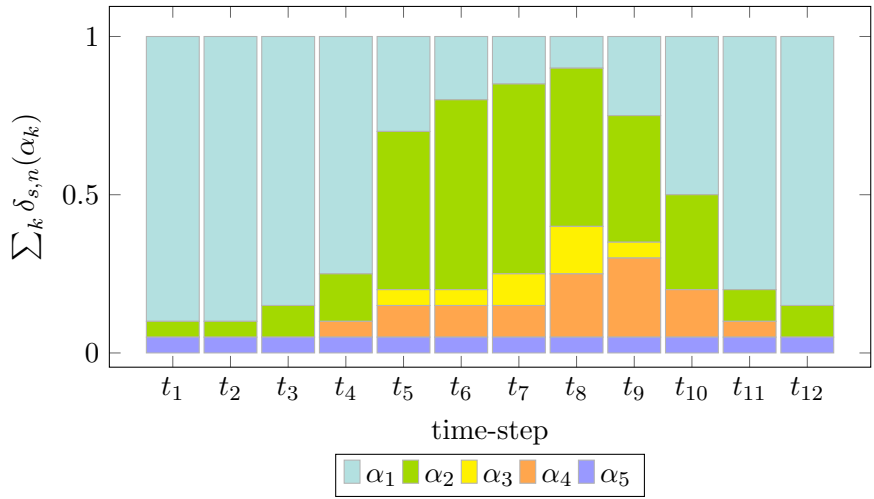
\begin{figure}[ht]
  \begin{center}
\input{figs/activity-distribution.tex}
\caption{Activity distribution table for subgroup $s$ with $K=5$ and $N=12$. }\label{fig:actdist}
\end{center}
\end{figure}

The time period $T$ is then discretised into $N$ steps, and for each time-step $t_n$, for $n=1,2,...,N$,  and each subgroup $s$, a map
$$
\delta_{s,t_n}: A \longrightarrow [0,1] \,\,\,\,\,
$$
$$
\text{s.t.} \sum_{k=0}^K \delta_{s,t_n}(\alpha_k)=1
$$
defines the expected proportion of subgroup $s$ to be engaged in activity $\alpha_k$ during time-step $t_n$.
The set
$$\Delta_s=\{\delta_{s,t_n}|n=1,2,...,N\}$$ of $N$ mappings for each subgroup then defines the distribution of that subgroup over the day.
The activity distribution input is most easily understood in terms of the chart in Figure~\ref{fig:actdist}, with each activity being assigned to a proportion of $P_s$ at every $t_n$.

In Section~\ref{sec:probmat} we will derive start times from these input distributions using each activity's duration.
We do not impose the start times as an input because the concept of describing what a population is doing across various time-steps is simpler to input than specifying exactly when a proportion of people begin and end activities.
Certainly the former is more intuitive for the user, and seems a lot more feasible from a validation and measurement perspective.

\subsubsection{Location Mapping}\label{sec:locmaps}

We have now defined the types of activities available to an agent, and the activity distributions which dictate when they can occur.
It remains to specify for all activities their possible locations in the region.
Again, this mapping is dependent on the subgroup an agent belongs to, which allows variations within activities to be expressed according to who is  performing them.
For instance, we may have visitor agents who have their \texttt{other} activity mapped to popular tourist destinations in the region, whilst resident agents would have schools, golf clubs etc. as  \texttt{other} locations.
Location mapping is particularly important in distinguishing the choice of \texttt{home} location for each $s$-agent.
\\

With a set of coordinates $W$ (in a suitable coordinate reference system\footnote{We assume a Cartesian coordinate system e.g. Universal Transverse Mercator (UTM).}) that cover $R$, we define a location as $\ell = (x,y)\in W$, and let

$$
\mathcal{L}=\{\ell\, | \,\ell\in R\cup W\}
$$
denote the set of all locations in the region. Then for each subgroup $s$ and each activity $\alpha_k$,

$$M_{s,\alpha_k}\subset \mathcal{L}  $$

defines the allowable locations for $s$ and $\alpha_k$.
Note that it we may have the case

$$M_{s,\alpha_k}=\varnothing \iff \delta_{s,n}(\alpha_k)=0 \,\,\,\forall n$$
implying that that an agent of type $s$ will never perform activity $\alpha_k$ at any point during the time period $T$.
\\

In summary for each subgroup $s$ there are three main inputs that must be specified by the user: the distribution of activities, the locations those activities can occur and the number of $s$-agents.
There are other more specific input parameters that will be introduced at the the relevant step of the algorithm explained below.

%% file: figs/activity-distribution.tex
\pgfplotstableread{
hour home work beach shops other
01 .90 .05 .00 .00 .05
03 .90 .05 .00 .00 .05
05 .85 .10 .00 .00 .05
07 .75 .15 .00 .05 .05
09 .30 .50 .05 .10 .05
11 .20 .60 .05 .10 .05
13 .15 .60 .10 .10 .05
15 .10 .50 .15 .20 .05
17 .25 .40 .05 .25 .05
19 .50 .30 .00 .15 .05
21 .80 .10 .00 .05 .05
23 .85 .10 .00 .00 .05

}\loadedtable

\begin{tikzpicture}
  \begin{axis}
    [
      scale only axis, 
      height=4.8cm,
      ylabel=$\sum_k \delta_{s,n}(\alpha_k)$,
      ylabel near ticks,
      xlabel=time-step,
      width=0.8\textwidth,
      bar width = 20pt,
      ybar stacked,
      xtick=data,
      xticklabels={$t_1$,$t_2$,$t_3$,$t_4$,$t_5$,$t_6$,$t_7$,$t_8$,$t_9$,$t_{10}$,$t_{11}$,$t_{12}$,},
      reverse legend,
      legend columns=-1,
      legend style={at={(0.5,-0.24)}, anchor=north}
    ]
    \addplot+[black!30,fill=blue!40] table[x=hour, y=other] {\loadedtable}; \addlegendentry{$\alpha_5$};
    \addplot+[black!30,fill=orange!70] table[x=hour, y=shops] {\loadedtable}; \addlegendentry{$\alpha_4$};
    \addplot+[black!30,fill=yellow!100!black] table[x=hour, y=beach] {\loadedtable}; \addlegendentry{$\alpha_3$};
    \addplot+[black!30, fill=lime!85!black] table[x=hour, y=work] {\loadedtable}; \addlegendentry{$\alpha_2$};
    \addplot+[black!30,fill=cyan!30] table[x=hour, y=home] {\loadedtable}; \addlegendentry{$\alpha_1$};
  \end{axis}
\end{tikzpicture}

%% file: chapter-theory/subsec-activities.tex
\subsection{Activity Assignment}
We now explain the algorithm's process per agent plan.
 The first step in constructing an $s$-agent plan is to decide which activities will be performed at each time-step.
 These activities are drawn from the corresponding distribution table and then sequenced according to their assigned duration.

\subsubsection{Selecting a Feasible Activity Set}\label{sec:probmat}
For each subgroup $s$, the input distributions must be converted into a cumulative probability matrix.
The distributions only tell us where the agents should be at a given time, and not when they begin that activity.
Therefore the start times must be derived, and to do this we must know the expected duration of each activity.
These durations will be some multiple of the length of each time-step
$\frac{T}{N}$. We denote a duration weight
$$d_{s,\alpha_k}\in \{1,2,...,N\}  $$
such that $\frac{T}{N}d_{s,\alpha_k}$ gives the corresponding duration for subgroup $s$ performing activity $\alpha_k$.
Then at each time-step, we can determine the number of $s$-agents who are expected to \textit{start} activity $\alpha_k$ by subtracting the proportion who started $\alpha_k$ during any of the $d_{s,\alpha_k}-1$ previous time-steps. We form the recursive function

\begin{align}
\xi_{s,t_n}(\alpha_k)=\delta_{s,t_n}(\alpha_k)-\sum_{j}\xi_{s,t_{n-j}}(\alpha_k) \,\,\,\,\,\,\,\,\,j\in (0,d_{s,\alpha_k}-1]\cap\mathbb{N}  \label{eq:startprop}
\end{align}

Since for $n\leq 0$, $\xi_{s,t_n}=0$ (we assume no agent has started an activity outside of the time period), $\xi_{s,t_1}=\delta_{s,t_1}$ for all $\alpha_k$ and thus we can recursively define the starting proportion for each activity at each time period using \eqref{eq:startprop}.
\

Now, for each $s$ we can view the collection of $\xi_{s,t_n}(\alpha_k)$ as a $K\times N$ matrix $\Xi_{s}$ that describes all start time proportions for that subgroup:
$$
\Xi_s=\begin{pmatrix}
 \xi_{s,t_1}(\alpha_1) & \xi_{s,t_2}(\alpha_1)&\cdots& \xi_{s,t_N}(\alpha_1) \\
  \xi_{s,t_1}(\alpha_2) & \xi_{s,t_2}(\alpha_2)&\cdots& \xi_{s,t_N}(\alpha_2) \\
\vdots & \vdots &\ddots &\vdots \\
  \xi_{s,t_1}(\alpha_K) & \xi_{s,t_2}(\alpha_K)&\cdots& \xi_{s,t_N}(\alpha_K)
\end{pmatrix}
$$

Note that the sum of the $n^{th}$ column in $\Xi_s$,
$$\sum_{j=1}^k\xi_{s,t_n}(\alpha_k)\leq 1
$$
(the remaining proportion represents $s$-agents who are currently engaged in an activity at $t_n$), and so we normalise $\Xi_s$

\begin{align*}
\Xi_s^*&=\begin{pmatrix}
 \frac{\xi_{s,t_1}(\alpha_1)}{\sum_{j=1}^k\xi_{s,t_1}(\alpha_j)} & \frac{\xi_{s,t_2}(\alpha_1)}{\sum_{j=1}^k\xi_{s,t_2}(\alpha_j)}&\cdots& \frac{\xi_{s,t_N}(\alpha_1)}{\sum_{j=1}^k\xi_{s,t_N}(\alpha_j)} \\
 \frac{\xi_{s,t_1}(\alpha_2)}{\sum_{j=1}^k\xi_{s,t_1}(\alpha_j)} & \frac{\xi_{s,t_2}(\alpha_2)}{\sum_{j=1}^k\xi_{s,t_2}(\alpha_j)}&\cdots& \frac{\xi_{s,t_N}(\alpha_2)}{\sum_{j=1}^k\xi_{s,t_N}(\alpha_j)} \\
\vdots & \vdots &\ddots &\vdots \\
\frac{\xi_{s,t_1}(\alpha_k)}{\sum_{j=1}^k\xi_{s,t_1}(\alpha_j)} & \frac{\xi_{s,t_2}(\alpha_k)}{\sum_{j=1}^k\xi_{s,t_2}(\alpha_j)}&\cdots& \frac{\xi_{s,t_N}(\alpha_k)}{\sum_{j=1}^k\xi_{s,t_N}(\alpha_j)} \\
\end{pmatrix} \\
&=\begin{pmatrix}
 \xi_{s,t_1,1}^* &
 \xi_{s,t_2,1}^* &
 \cdots        &
 \xi_{s,t_N,1}^* &\\
 \xi_{s,t_1,2}^* &
 \xi_{s,t_2,2}^* &
 \cdots        &
 \xi_{s,t_N,2}^* & \\
\vdots & \vdots &\ddots &\vdots \\
\xi_{s,t_1,K}^* &
\xi_{s,t_2,K}^* &
\cdots        &
\xi_{s,t_N,K}^*
\end{pmatrix}
\end{align*}
where each $(k,n)$ entry now represents the probability of an $s$-agent who has  \textit{ended} any activity at time-step $t_N$ beginning new activity $\alpha_k$.
Finally, we take the cumulative sum of each column to attain a cumulative probability matrix:
$$
C_s=\begin{pmatrix}
 \xi_{s,t_1,1}^* &
 \xi_{s,t_2,1}^* &
 \cdots        &
 \xi_{s,t_N,1}^* &\\
 \sum_{k=1}^2\xi_{s,t_1,k}^* &
 \sum_{k=1}^2\xi_{s,t_2,k}^* &
 \cdots        &
 \sum_{k=1}^2\xi_{s,t_N,k}^* & \\
\vdots & \vdots &\ddots &\vdots \\
\sum_{k=1}^{K-1}\xi_{s,t_1,k}^* &
\sum_{k=1}^{K-1}\xi_{s,t_2,k}^* &
\cdots        &
\sum_{k=1}^{K-1}\xi_{s,t_N,k}^*\\
1 & 1 & \cdots & 1
\end{pmatrix}
$$

Now that we have a well-defined probabilistic description of when an agent will perform an activity, we can begin to construct the plan by selecting activities from $C_s$.

For a given $s$-agent, we generate a $1\times N$ vector $U$ of values drawn from the uniform distribution on $(0,1)$.
Then, conditionally checking the $n^{th}$ value of $U$, $u_n$ against the $n^{th}$ column of $C_s$, we select one activity for each $t_n$.
Form $B$, a $K\times N$ matrix with each $b_{k,n}$ entry defined by
\begin{align*}
  b_{1,n}&=\begin{cases}
    1 &  u_n<c_{1,n}\\
    0 &  u_n \geq c_{1,n}
  \end{cases} & k=1, 1\leq n\leq N\\
  b_{k,n}&=\begin{cases}
    1 & c_{k-1,n}\leq u_n<c_{k,n}\\
    0 &  u_n \geq c_{k,n}
  \end{cases} & 2\leq k\leq K, 1\leq n\leq N
\end{align*}

For each column, we only have one non-zero value, so $B$ is a Boolean matrix representing an activity at each time-step $t_n$.
These pre-selections can be thought of as potential activities at that time, dependent on the duration of previous activities.

To actualise these activities, we inject the duration weights into $B$ by defining $B'$:

\begin{align*}
b'_{k,n}=\begin{cases}
  d_{s,\alpha_k} &  \text{when }b_{k,n}=1\\
  0 &   \text{when } b_{k,n}=0
\end{cases} && 1\leq k\leq K, 1\leq n\leq N
\end{align*}

We now cycle through each time-step to determine its affect on the next. Algorithm \ref{alg:activity-select} illustrates this process whereby the duration weight $d_{s,\alpha_k}$ blocks out subsequent time-steps until the full duration of $\alpha_k$ has been performed.
Note that we define
$$
\boldsymbol{b}_{\cdot,n}
$$
to be the $n^{th}$ column of $B$.
\begin{algorithm}
\caption{Activity selection for an $s$-agent}
\label{alg:activity-select}
    Form $B$, $B'$ for $s$-agent \\
    Set \verb|block_flag|=0 \\
      \For{$n \in\{ 1,2,...,N\}$}
      {
        \If{$\tt block\_flag>0 $}
        {
          $\boldsymbol{b}_{\cdot,n}=\boldsymbol{b}_{\cdot,n-1}$\\
          \tt block\_flag=block\_flag $-1$
        }
        \Else
        {
          \tt block\_flag$=\sum_{k=1}^K b'_{k,n}-1$
        }
      }
return $B$
\end{algorithm}

At the end of this process, the resultant Boolean matrix $B$ will contain a feasible plan which describes what the $s$-agent will be doing at each time-step, in line with the format of the initial distributions (see Figure~\ref{fig:B-ex}).

\begin{figure}[ht]
  \caption{An example output from Algorithm~\ref{alg:activity-select} with $K=5$ and $N=12$.}\label{fig:B-ex}
\setcounter{MaxMatrixCols}{20}
$$B=\begin{pmatrix}
   1 & 1 & 1 & 0 & 0 & 0 & 0 & 0 & 0& 0 & 1 & 1 &\\
   0 & 0 & 0 & 1 & 1 & 1 & 1 & 0 & 0& 0 & 0 & 0 &\\
   0 & 0 & 0 & 0 & 0 & 0 & 0 & 1 & 1& 0 & 0 & 0 &\\
   0 & 0 & 0 & 0 & 0 & 0 & 0 & 0 & 0& 1 & 0 & 0 &\\
   0 & 0 & 0 & 0 & 0 & 0 & 0 & 0 & 0& 0 & 0 & 0 &
  \end{pmatrix}$$
\end{figure}

\subsubsection{Sequencing the Daily Plan}
It remains to translate the $t_n$ at which the agent changes activity into meaningful time and provide an ordered list of start times for each activity in the plan.
We define a recursive function on the columns of $B$

\begin{align*}
f(\boldsymbol{b}_{\cdot,n})=\boldsymbol{b}_{\cdot,n}-\boldsymbol{b}_{\cdot,n-1} && 2\leq n\leq N
\end{align*}

with $f(\boldsymbol{b}_{\cdot,1})=\boldsymbol{b}_{\cdot,1}$.

We then set

$$
F=f(B)=f([\boldsymbol{b}_{\cdot,1},\boldsymbol{b}_{\cdot,2},\ldots ,\boldsymbol{b}_{\cdot,N}])=[f(\boldsymbol{b}_{\cdot,1}),f(\boldsymbol{b}_{\cdot,2}),\ldots ,f(\boldsymbol{b}_{\cdot,N})]
$$
which will have entries from $\{-1,0,1\}$. The only elements of $F$ that will be greater than $0$ i.e. unaffected are those where the agent has changed plan from $t_{n-1}$ to $t_n$. Therefore we can form $F'$ such that each entry is defined
\begin{align*}
f'_{k,n}=\begin{cases}
  1 &  f_{k,n}>0\\
  0 &  f_{k,n} \leq 0
\end{cases} && 1\leq k\leq K, 1\leq n\leq N
\end{align*}

Then, the sum of the $n^{th}$ column in $F'$ defines a Boolean \texttt{true} or \texttt{false} as to whether an activity begins at time-step $t_n$, and the row $k$ containing the non-zero entry in column $n$ determines the activity $\alpha_k$ that begins then. This process is illustrated in Algorithm~\ref{alg:time-select}.

\begin{algorithm}
\caption{Algorithm for selecting times that activities begin}
\label{alg:time-select}
Form $F'$ for $s$-agent \\
create empty list \texttt{plan}\\
$\texttt{activity}=\alpha_1$\\
$\texttt{time}=0$ \\
add $(\texttt{time},\texttt{activity})$ to \texttt{plan}\\
\For{$n \in\{ 1,2,...,N\}$}
{
  \If{$\sum_{k=1}^{K}f'_{k,n}=1$}
  {
    \For{$k \in\{ 1,2,...,K\}$}
    {
      \If{$k=1$}
      {
      $\texttt{activity}=\alpha_k$
      }
    }
    $\texttt{time}=t_n+\texttt{runif(-1,1)}\frac{T}{2n}$\\ \label{algline:fuzz}
    add $(\texttt{time},\texttt{activity})$ to \texttt{plan}\\
  }
}
$\texttt{activity}=\alpha_1$\\
$\texttt{time}=T$ \\
add $(\texttt{time},\texttt{activity})$ to \texttt{plan}\\
return \texttt{plan}
\end{algorithm}

Line~\ref{algline:fuzz}  gives the activity a start time at a random point within the length of the time-step $t_n$.
This is important because it ensures an even distribution of start times for activities over the population; if the times were fixed to each $t_n$, then we would have unnatural `waves' of trips occurring periodically. Note that \texttt{runif(-1,1)} on this line denotes a uniformly distributed random variable in the interval $(-1,1)$.
Another purpose served in Algorithm~\ref{alg:time-select} is the allocation of an additional activity $\alpha_1$ at the start and end of the plan.
This is a requirement of MATSim's re-planning features, and it ensures that agents always return to their \texttt{home} the end of the time period.

%% file: chapter-theory/subsec-locations.tex
\subsection{Assigning Locations to Activities}\label{Locs}
Once an $s$-agent's time-ordered list of activities is determined, locations must be assigned to each activity.
Activity locations are chosen based on a number of factors, including the distance required to travel, and the relative popularity of the location.
In this section we will often introduce probabilities that are (inversely) proportionate to distances. In the algorithm, these probabilities are properly formed by normalising the inverse distances over the sum of all inverse distances (in much the same way that we normalised $\Xi_s$ in Section~\ref{sec:probmat}).
To notate this at each step would be cumbersome and distracting (see the end of Section~\ref{sec:locale} for evidence of this), so we maintain the proportionate notation throughout.

\subsubsection{Localities}\label{sec:locale}
If we first assume that an $s$-agent is at a certain location $\ell^0$, then the question is how to choose the location $\ell^*$ of the next activity $\alpha_k$ in the list.
This could be done based on inverse distance:
$$\text{Pr}(\alpha_k \text{ occurs at }\ell)\propto \frac{1}{\text{dist}(\ell^0,\ell)}$$

where dist$(\ell^0,\ell)$ denotes the Euclidean distance from $\ell^0$ to $\ell$ for some $$\ell \in M_{s,\alpha_k}.$$
 However, when there are a large number of location options for an activity, it becomes costly to calculate the distance for every agent to every potential destination, and scalability to larger areas or increasing the detail of location data becomes problematic.
Further, a purely distance-based metric tends to result in trips that cluster around the agent's home location; for each new trip, the agent will generally take the closest option to their current location.
This is amplified in built-up areas, where an agent is likely to have more options close to them.
Whilst intuitively this makes sense and agrees with many of the assessments made by \citet{SongModellingscalingproperties2010}, over a population it does result in movement patterns that are almost \textit{too} optimal, with agents either commuting between communities or not at all.
It is desirable to encourage some intra-community trips which disperse traffic away from major arterial routes, and reflect the possibility (also discussed by Song et al.) that people within a neighbourhood may have preferences that do not depend entirely on distance.
\

Both of the above issues can be addressed if we group locations by some locality area that they lie within.
These localities can be based on the existing suburban/postal districts in the area, by ABS statistical measures, or by any other grouping appropriate to the community in question.
The \textbf{localities} $\{L_i\}$ of a region are defined as a family of $I$ sets such that

\begin{align*}
L_i\subset \mathcal{L}
\end{align*}

$\forall i\in I$, and
\begin{align*}
\bigcup_{i\in I} L_i= \mathcal{L},&& \bigcap_{i\in I} L_i= \varnothing
\end{align*}

i.e. $\{L_i\}$  is a partition of $\mathcal{L}$.
\

\begin{figure}[ht]
  \begin{center}
\includegraphics[scale=0.3]{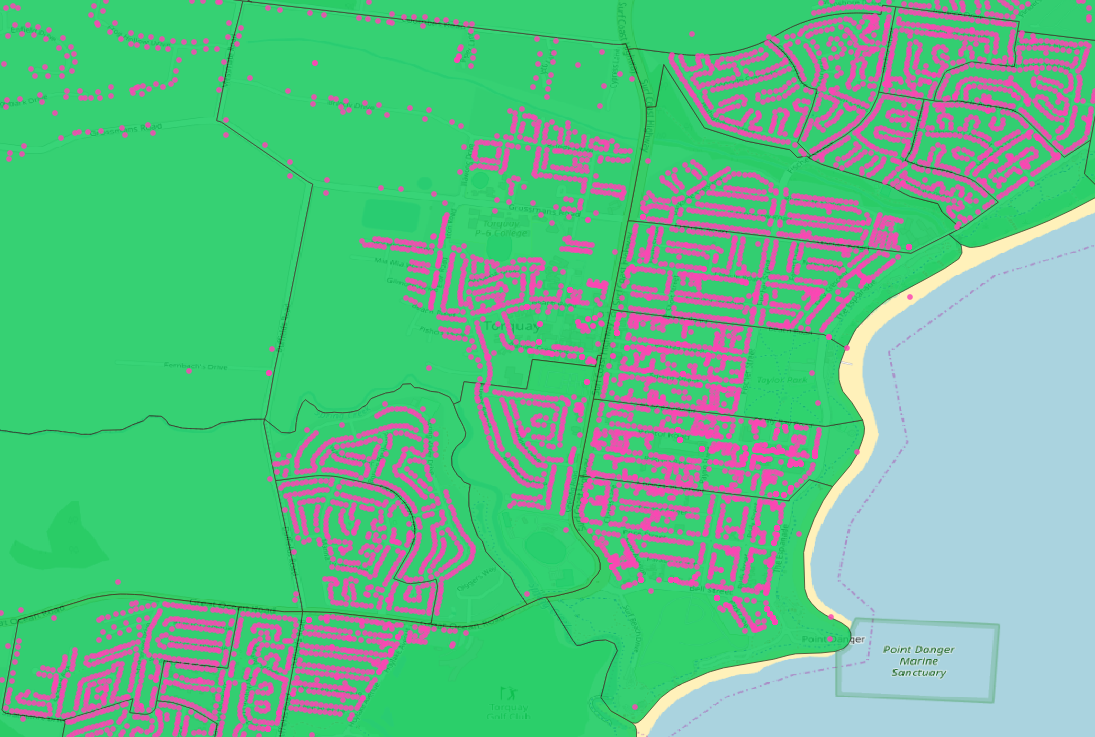}
\caption{Example of location points (pink) being grouped by locality polygons (green).}\label{fig:locale}
\end{center}
\end{figure}

In practical terms, we draw a polygon cover of $R$ and denote all locations under polygon $L_i$ as belonging to locality $L_i$ (illustrated in Figure~\ref{fig:locale}).

For each $L_i$, we define a centroid location $\ell_i$ as

$$
\ell_i=(\frac{1}{|L_i|}\sum_{\ell \in L_i}x_{\ell},\frac{1}{|L_i|}\sum_{\ell \in L_i}y_{\ell})
$$
where $|L_i|$ is the number of locations in  $L_i$ and $x_\ell$ and $y_\ell$ denote the $x$ and $y$ co-ordinates of $\ell\in W\cup L_i$.
Now if we take the current locality to be $L_{i^0}\ni \ell^0$, we can choose a new locality $L_i$ with probability proportional to the inverse distance from centroid $\ell_{i^0}$ to centroid $\ell_i$:

$$\text{Pr}(\alpha_k \text{ occurs in }L_i)\propto \frac{1}{\text{dist}(\ell_{i^0},\ell_i)}$$

It is then a simple matter to randomly choose from the selected locality $L_{i^*}$ a location
$$\ell^*\in L_{i}^*\cap M_{s,\alpha_k}$$

for $\alpha_k$, the next activity in the plan list.
A major benefit of this process is that the distances between each centroid are fixed and only need to be calculated once during the entire algorithm runtime.
Thus we can drastically cut down on distance calculation requirements whilst also gaining the ability probabilistically encourage mildly suboptimal choices.
\

One problem this presents is that the most likely choices for $\ell^*$ should be located within $L_{i^0}$, but the distance to them in terms of locality centroids is

$$\text{dist}(\ell_{i^0},\ell_{i^0})=0$$
so the probability of selecting from the current locality is undefined. However, this provides an opportunity to introduce a new parameter to the algorithm.
For each subgroup $s$ we define a travel factor $g_s$,
which determines how likely an $s$-agent is to leave its current locality.
This allows for agents in some subgroups to be more likely to remain within their neighbourhood, and alternatively can encourage other subgroups to move around the region throughout the day.
Under pure locality-based selection, $g_s$ directly represents the probability of an $s$-agent moving to another locality for $\alpha_k$.
This definition for $g_s$ implies that
\begin{align}
\text{Pr}(\alpha_k \text{ occurs in }L_{i^0})= 1-g_s\label{probbo}
\end{align}

and logically,
$$g_s=\sum_{i\in I\setminus\{i^0\}}{\text{Pr}(\alpha_k \text{ occurs in }L_{i})}\propto \sum_{i\in X}\frac{1}{\text{dist}(\ell_{i^0},\ell_i)} $$
where $X=\{i\in I\setminus\{i^0\}|L_i\cap M_{s,\alpha_k}\neq \varnothing\}$ i.e. $X$ only consists of $i$ such that locality $L_i$ contains valid locations at which activity $\alpha_k$ can be performed by an $s$-agent.
So we can write \eqref{probbo} as
\begin{align*}
\text{Pr}(\alpha_k \text{ occurs in }L_{i^0})=\frac{1-g_s}{g_s}\sum_{i\in I\setminus\{i^0\}}{\text{Pr}(\alpha_k \text{ occurs in }L_{i})}
\end{align*}

and this allows us to define a pseudo `distance'
\begin{align}
  \text{dist}^0(\ell_{i^0},\ell_{i^0})=\frac{g_s}{{(1-g_s)}\sum_{i\in X}\frac{1}{\text{dist}(\ell_{i^0},\ell_i)}}\label{pseud}
\end{align}
relative to all other locality distances that contain locations in $M_{s,\alpha_k}$. Observe that this definition recovers \eqref{probbo} after normalisation:
\begin{align*}
\text{Pr}(\alpha_k \text{ occurs in }L_{i^0})&=\frac{\frac{1}{\text{dist}^0(\ell_{i^0},\ell_{i^0})}}{\frac{1}{\text{dist}^0(\ell_{i^0},\ell_{i^0})}+\sum_{i\in X}\frac{1}{\text{dist}(\ell_{i^0},\ell_i)}}\\
&=\frac{\frac{1-g_s}{g_s}\sum_{i\in X}\frac{1}{\text{dist}(\ell_{i^0},\ell_i)}}{ \frac{1-g_s}{g_s}\sum_{i\in X}\frac{1}{\text{dist}(\ell_{i^0},\ell_i)}+\sum_{i\in X}\frac{1}{\text{dist}(\ell_{i^0},\ell_i)}}\\
&=\frac{\frac{1-g_s}{g_s}\sum_{i\in X}\frac{1}{\text{dist}(\ell_{i^0},\ell_i)}}{ \frac{1-g_s}{g_s}\sum_{i\in X}\frac{1}{\text{dist}(\ell_{i^0},\ell_i)}+\sum_{i\in X}\frac{1}{\text{dist}(\ell_{i^0},\ell_i)}}\\
&=\frac{\frac{1-g_s}{g_s}}{\frac{1-g_s}{g_s}+1}\\
&=\frac{{1-g_s}}{1-g_s+g_s}={1-g_s}
\end{align*}

Unlike the other locality distances, the pseudo-distance is dependent on $\alpha_k$ and $s$, so it must be recalculated at every new location assignment step.
This definition for $\text{dist}^0(\ell_{i^0},\ell_{i^0})$ will be important in Section~\ref{sec:allocs}.

%
\subsubsection{Allocation Considerations}\label{sec:allocs}
Another complication that is still present with locality-based selection is that each probability is still roughly proportional to the inverse distance from the agent's current location.
There is no consideration for the size of each locality $|L_i|$, so locations in smaller localities are actually favoured because they are generally competing in a smaller pool (once their locality has been chosen).
This seems counter-intuitive, and we should expect a heavier traffic flow towards more bigger localities rather than away from them.
Whilst this is mitigated by a larger proportion of agents having their \texttt{home} in built up areas, it can still cause a significant and unusual build up of traffic in areas that should not be so busy.
This can be exacerbated if there are a lot of $s$-agents with a $g_s$ that encourages them to travel; imagine a one pub town that suddenly and inexplicably has an influx of hundreds of tourists!
Whilst this is a plausible scenario, it should not exist by default, and we would prefer to be able to control this.
\

We can add this control by introducing an additional weighting to each locality that reflects its overall size.
Not only should we account for the number of locations in the locality, but also have some measure of their popularity and expected usage.
Each location $\ell$ is assigned an allocation number $a_\ell$.
Conceptually, $a_\ell$ represents the expected number of agents that will visit  $\ell$ during the given scenario, but its main purpose here is to weight the probabilities for each locality.
We now give the probability that the next activity $\alpha_k$ occurs in locality $L_i$ as:
\begin{align}
  \text{Pr}(\alpha_k \text{ occurs in } L_i)\propto \frac{1}{\text{dist}(\ell_{i^0},\ell_i)}\sum_{\ell\in L_i\cap M_{s,\alpha_k} } {a_\ell} \label{prob}
\end{align}

 Localities with higher overall allocations for the relevant activity are given a higher weighting.
Importantly, the current locality is equally affected by this metric, with a `distance' being prescribed as in \eqref{pseud} (note that $g_s$ is now not a proper probability, but a factor that determines the pseudo-distance) so an agent's decision to stay within their current locality will also consider the options on offer there.
This method of selecting destinations is a version of the gravity model of transportation \citep{AndersonGravityModel2011}.
It allows the dispersion of agents throughout the day to be more tightly controlled, and for hotspots to be set according to the prevailing conditions.
For instance, if on a given day there is an event at one particular beach, then the allocation for that beach should be raised to encourage those agents with beach activities to go there.

\subsubsection{Home Selection}
We have discussed how agents will select a new location based on their current location, but have not yet established where they are initialised.
For most agents, the location they begin at will heavily influence all further trip locations.
Each plan has been explicitly constructed to ensure that an agent begins and ends their day with the same $\alpha_1$ activity, so unlike with other activities, we guarantee that an $s$-agent's \texttt{home} location remains the same every time it is assigned (note that as defined in Section~\ref{Dist}, \texttt{home} and $\alpha_1$ are synonymous).
As it will always be the first activity in an $s$-agent's ordered daily plan, $\alpha_1$ is also the only activity that needs to be assigned without the context of a prior $\ell^0$ location.
Instead, we randomly choose $L_i$ from the set of eligible localities  $$\{L_i|L_i\cap M_{s,\alpha_1}\neq\varnothing\}.$$
and then consequently we can choose $\ell\in L_i\cap M_{s,\alpha_1}$ as the location for $\alpha_1$.
\

The allocation numbers $a_\ell$ for $\ell \in M_{s,\alpha_1}$ play a special role in the \texttt{home} selection process.
In this case, the conceptual understanding of $a_\ell$ as the expected number of people attending the location during the scenario does impose a hard cap on the number of agents that can reside at $\ell$.
Whenever an agent (from \texttt{any} subgroup) is assigned $\ell$ as their \texttt{home} location we deduct $1$ from $a_\ell$, globally affecting the likelihood of that location being assigned to another agent.
If $a_\ell=0$, then $\ell$ is removed from $\mathcal{L}$.
This ensures that the population is initially distributed in a manner that matches the region.
It is therefore especially important that all specified \texttt{home} locations $$\bigcup_{s\in S} M_{s,\alpha_1}$$
and the associated allocation numbers reflect the true distribution and density of people in each area of interest, because this will dictate where traffic is heaviest for all other activities.
Algorithm~\ref{alg:locations} describes the location section of the algorithm.
The function \texttt{choose} randomly samples from the specified set.
\begin{algorithm}
\caption{Choosing activity locations for all agents}
\label{alg:locations}
Form \texttt{plan} for each \texttt{agent}$\in \{1,2,...,P\}$ via Algorithm \ref{alg:time-select}\\
\For{$s \in S$}{
   \For{$\texttt{agent}  \in\{1,2,...,P_s\}$}{
   $\texttt{home\_locality}=$\texttt{choose} $\{L_i|L_i\cap M_{s,\alpha_1}\neq\varnothing\}$ \\
   $\texttt{home\_location}=$\texttt{choose} $\{\ell\in \texttt{home\_locality}\cap M_{s,\alpha_1}|a_\ell>0\}$ \\
   $a_\ell=a_\ell-1$\\
   \If{$a_\ell==0$}
   {
     $\mathcal{L}=\mathcal{L}\setminus\{\ell\}$\\
   }
   $\ell_0=\texttt{home\_location}$\\
    \For{$\texttt{activity} \in \texttt{plan}_{\texttt{agent}}\setminus\{\alpha_1\} $}{
$\texttt{locality}=$\texttt{choose} $ L_i\in R $ with weighted prob. from \eqref{prob}\\
$\texttt{activity\_location}=$\texttt{choose} $\ell\in L_i\cap M_{s,\texttt{activity}}$ \\
}
}
}
\end{algorithm}

%% file: chapter-theory/subsec-behaviours.tex
\subsection{Adding Bushfire Behaviours}

\begin{figure}[ht]
  \begin{center}
\input{figs/fig-integration}
\caption{A diagram representing the BDI-ABM integration \citep{SinghIntegratingBDIAgents2016}.}\label{fig:brainbod}
\end{center}
\end{figure}
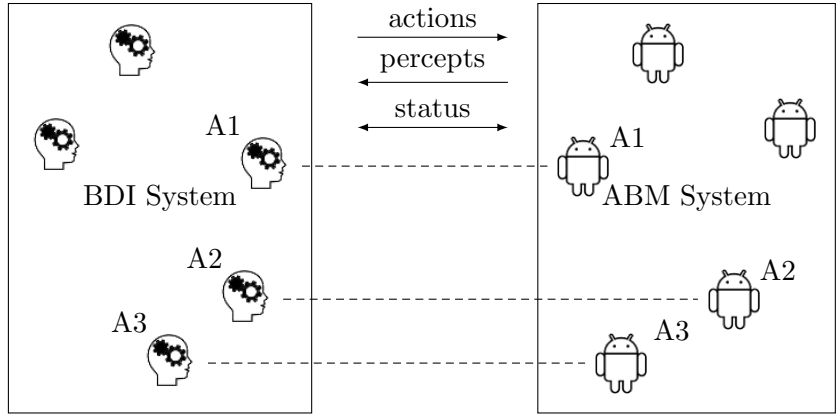

The above algorithm creates a \texttt{population.xml} that is ready for MATSim to use on a regular day, but it will not contain any of the information required to describe how the agents will act in a bushfire.
We now describe a behaviour model that allows an agent to disrupt these usual daily plans by taking the exceptional circumstances into account.
This model is designed to be implemented in the BDI-ABM integration framework described in \citep{SinghIntegratingBDIAgents2016}, where agents running in the BDI engine Jill serve as cognitive counterparts to the active ABM agents in MATSim (see Figure~\ref{fig:brainbod}).
The behaviour model thus follows Jill's architecture, where goal-plan hierarchies filter abstract belief states down to actions for the ABM agent execute.
The design of the hierarchies is based around the same broad input subgroups that govern the population generation algorithm, and accordingly we initialise behaviours as appended attributes in the generated MATSim \texttt{population.xml}.
These BDI attributes-- together with the associated characteristics that emerge from the population generation inputs-- form a behaviour profile that describes how an agent is going to act in a given bushfire scenario.\footnote{As we move into discussing the bushfire behaviour model, our language will shift towards a more descriptive tone. From here on, the focus is less on formally defining concepts, and more on explaining the methods and rationale behind our attempt to capture human behaviour in the model.}

\subsubsection{Threshold Model} \label{Thresh}
Since the 2009 Black Saturday bushfires in Victoria, most Australian research  regarding bushfire behaviour has concerned if and when people will evacuate in the face of the threat. This decision process is central to the behavioural model.
We need to accommodate the temporal aspect of incoming information in an unfolding situation, where multiple warning sources can have a cumulative effect on a person's actions.
The scheme should also ensure that there is variation in behaviour even within subgroups; some people will intrinsically be more inclined to leave early, whereas others might never evacuate even in the face of severe danger.
\

For this reason, we employ a threshold-based model, where each agent has an intrinsic tolerance value which determines how risk averse they are, and therefore how likely they are to respond to the new information about the bushfire threat.
Incoming information, or percepts, are broken up into two categories.
\texttt{Environmental} alerts pertain to information gleaned directly from the surroundings.
They will include embers, smoke, flames, the movement of neighbouring agents, and the actions of emergency services in the vicinity.
\texttt{Transmitted} alerts include those that are issued by the emergency authorities before and during a bushfire emergency and other messages received via the media or other communications networks.
Let $E$ and $\mathcal{T}$ denote the sets of all alerts $e$ and $\tau$ for each category.
Both sets are totally ordered
\begin{align*}
  e_1\leq e_2 \leq ...\leq e_q\leq...\leq e_Q ,\\
  \tau_1\leq \tau_2 \leq ... \leq \tau_m\leq...\leq \tau_M
\end{align*}

i.e. we denote the severity or value of each percept relative to the others in its category.
Whenever an agent receives a percept, it will update its `barometer' for the corresponding alert type.
The two categories differ here in that the \texttt{Environmental} barometer only updates monotonically when the new alert is of greater severity than the previous alert, whereas the \texttt{Transmitted} barometer can also be decreased based on a reduced level of threat.
This reflects the nature of the different alert types; if you see smoke, it doesn't reduce the threat of the flames you had seen earlier, but if you receive a less severe warning from an authority it is likely to decrease the concern caused by an earlier message.
We now have a way of capturing the different types of threat percepts that might trigger an evacuation, defined a method of updating the severity level of the threat, and separated the percepts into groups that broadly maintain a distinction between primary and secondary information.
\\

We now must combine these two barometers into one value that reflects a person's overall threat level. If we consider
$$
E\times \mathcal{T}=\{ \{0,0\},\{e_1,0\},\{0,\tau_1\},\{e_1,\tau_1\},...,\{e_Q,\tau_M\}\}
$$

representing the set of all combinations of alerts that an agent may have registered as their barometer at some time (with $0$ representing no threat of that type being received).
There is naturally a partial ordering on the ordered pairs for each fixed value of $e_q$ or $t_m$ (stemming from the ordering in the other category set).
However, there is no defined order of, say, $\{e_1,t_3\}$ and $\{e_2,t_2\}$.
We therefore need to define a mapping

$$r: E\times \mathcal{T} \longrightarrow (0,1)$$

which maintains all partial orders and also defines an order between each undefined element pair. Note that the value set that $r$ maps to is arbitrary, and serves only to quantify the ordering into a ranking score.
We may then have the case that, say, $r(\{e_1,t_3\})=r(\{e_2,t_2\})$ i.e. an agent can arrive at a ranking score through various different combinations of alerts. The ranking $r$ of the current barometer levels is then compared against an agent's intrinsic tolerance score \texttt{tol}, and if $r>\tt tol$ then the agent will change their daily plan.

\subsubsection{Possible Bushfire Behaviours}
We now show how changes of plan are decided upon by an agentusing the Belief-Desire-Intention model of cognition.
We design and implement behaviours based on what is understood to occur in a bushfire situation.
Fundamentally, we have constrained variation of behaviour to three end goals:

\begin{itemize}
  \item \texttt{Go\_home\_now}: The agent will take a route towards their \texttt{home} location. It is important that every agent has a \texttt{home} location, as specified in \ref{Dist}.
  \item \texttt{Go\_to\_dependant\_now}: The agent will take a route towards the location of a dependant. Here a dependant may represent a child, elderly relative, farm animal or some other responsibility that an agent may have that will require them to detour before going home or before leaving.
  \item \texttt{Leave\_now}: The agent will decide on an evacuation destination and begin to move there.

\end{itemize}

These goals are  triggered as the result of the reasoning process that accounts for current location, whether the agent is also planning on going home, and if there are dependants to attend to.
We do this using \textit{two} threshold levels for each agent. The first, \texttt{tol=INIT} represents the level at which an agent decides to stop their daily plan, but has not yet committed to seeking shelter, and triggers the high-level \texttt{Initial\_Response} goal.
The second, \texttt{tol=ACT}, defines the level at which an agent will decide to leave, and will trigger the \texttt{Act\_Now} goal. Both high-level goal structures are described as goal-plan tree diagrams in Figure~\ref{fig:initresp} and Figure~\ref{fig:actnow}.

\begin{figure}[!t]
  \centering
\input{figs/init-tree.tex}
\caption{Initial Response goal-plan tree (goals are in green, plans are in yellow).}\label{fig:initresp}
\end{figure}
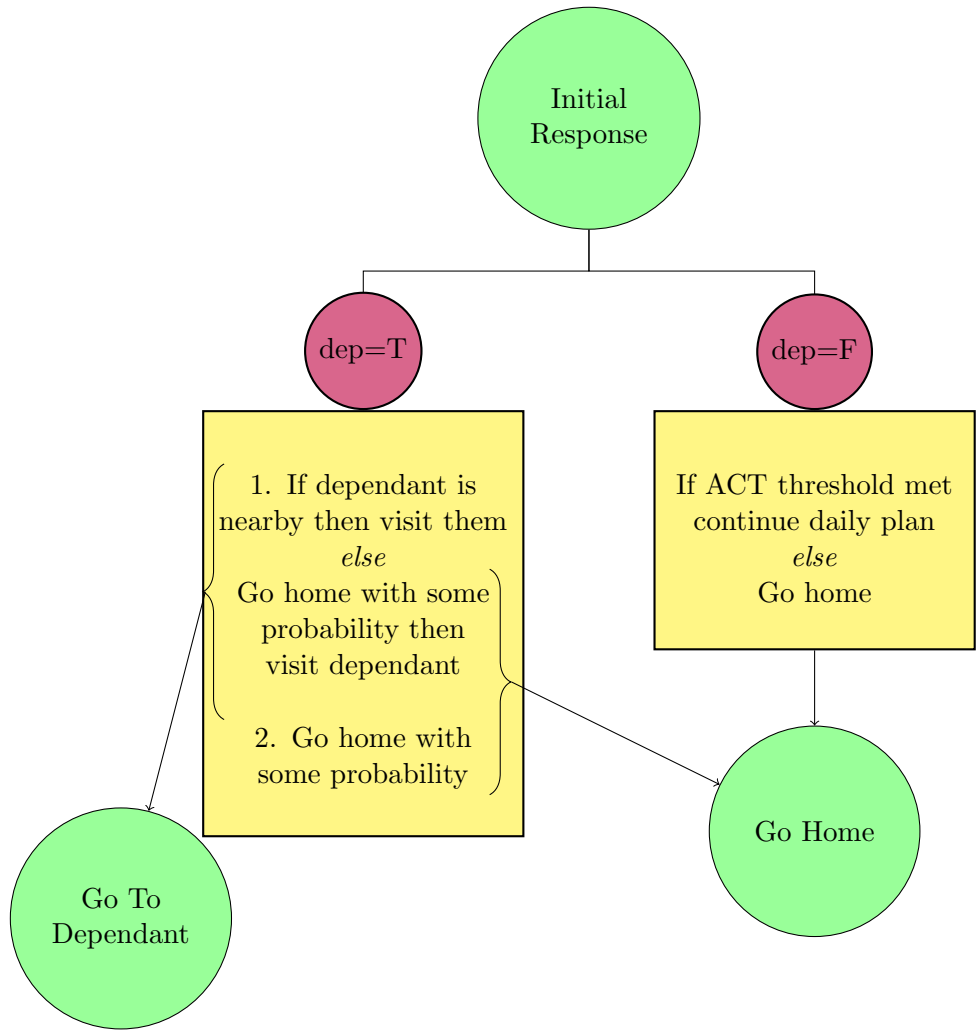

\begin{figure}[!htb]
  \centering
\input{figs/act-tree.tex}
\caption{Final Response goal-plan tree (goals are in green, plans are in yellow).}\label{fig:actnow}
\end{figure}
\

Both  $\texttt{INIT},\texttt{ACT}\in [0,1]$, with $0$ and $1$ indicating that the threshold is always or never broken respectively.
It is clear too that the two decisions happen in order, so $\texttt{INIT} \leq \texttt{ACT}$.
The distinction between the two thresholds is important as it allows a diverse range of responses. An agent may have its \texttt{INIT} threshold broken, returning to \texttt{home} and then never exceeding its \texttt{ACT} threshold; this would represent someone who has chosen to stay and defend their property.
Likewise, we can define agents who will never be allowed to do this by setting $\texttt{INIT}=\texttt{ACT}$ i.e. both thresholds will be triggered at the same time.
Note that if the agent has a dependant, then they are guaranteed to execute the \texttt{Go\_to\_dependant} goal even in the case that $\texttt{INIT}=\texttt{ACT}$; this is illustrated in the goal-plan tree.
\joelComment{Still need to add para about potential behaviour examples (slides)}

\subsubsection{Assigning Attributes}

Despite the simplicity of the BDI structure above, it can result in a range of outcomes for an agent depending on their threshold levels, whether they have a dependant, and on the probability that dictates whether the agent should go home.
This is further diversified when you consider the number of destination options implicit in the \texttt{Leave\_now} goal.
To tie this model to the population generation algorithm, we choose an allocation of these possibilities that matches the expected behaviour of the predefined subgroups in a bushfire.
\\

For a subgroup $s$, we define the following parameters:

\begin{itemize}
  \item $\texttt{prob\_of\_dependant}_s$: a probability determining how likely an $s$-agent is to have a dependant.
  \item $\texttt{stay}_s$: a Boolean variable determining whether an $s$-agent will be permitted to stay and defend i.e. whether the strict inequality \texttt{INIT}$<$\texttt{ACT} is possible.
  \item $\texttt{prob\_of\_go\_home}_s$: a probability determining how likely an $s$-agent is to return home before leaving.
  \item $[\texttt{threshold\_min}_s$,$\texttt{threshold\_max}_s]$: an interval within the chosen ranking range which determines where \texttt{INIT} and \texttt{ACT} should lie within.
\end{itemize}

We can then attribute to each agent their \texttt{INIT} and \texttt{ACT} thresholds, determine if they have a dependant and indicate whether they will decide to \texttt{Go\_home\_now}.\

Lastly, we must assign dependant, invacuation and evacuation locations.
The dependant location can either be chosen within some radius of \texttt{home}, or if there are particular areas likely to have dependants (e.g. schools, retirement homes), these can be specified as dependant locations and selected in the same manner as in Section~\ref{Locs}.
The invacuation and evacuation points will determine preferred destinations to evacuate to once \texttt{ACT} has been triggered.
An agent will first attempt to reach its preferred evacuation location, but if that is not possible, it will seek refuge within its current locality (with the invacuation point as preference if possible).
Clearly there are many potential situations that might prevent the agent accessing either location, but having these defined preferences allows behaviours to be further shaped on the subgroup level.
To determine the known invacuation and evacuation points, we also use a similar
method as in Section~\ref{Locs}.
We want invacuation points to be close to the agent's \texttt{home}, so we enforce a strict preference for invacuation locations within the locality of the agent's home.
Conversely, evacuation points should be far from the \texttt{home}, so we use probabilities proportional to distance rather than inverse distance.
In this case, a person is more interested in getting out of the area than evacuating to a specific location, so entire localities are weighted by their total capacity, and the destination point is the central node of the locality.
\

Figure \ref{fig:att-xml} provides an example of these attributes within a MATSim \texttt{population.xml} file.
We conclude this section by highlighting how the chosen subgroups $S$ have fundamentally shaped both the population generation and the assignment of BDI attributes.
Indeed, a particular agent belonging to a subgroup $s$ will have a listable set of potential behaviours that completely capture how it might act during a given bushfire scenario.
Within these behaviours there is enough variation to ensure that over a population of interacting agents the simulation will not be deterministic, and instead will approximate the complex and unpredictable nature of a real bushfire.

\begin{figure}[!ht]
  \centering
  \input{figs/attribute-xml-ex.tex}
  \caption{An agent plan with attributes appended.}\label{fig:att-xml}
\end{figure}

%% file: figs/fig-integration.tex
\begin{tikzpicture}
\tikzstyle{active}=[draw=red, fill=red!20, very thick, draw]

\node[above=0.25cm, xshift=2.75cm, label={[label distance=-0.3cm]120:A1}]
(bdi1) {\includegraphics[height=0.8cm]{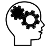}};

\node[below=0.45cm,xshift=2.5cm, label={[label distance=-0.3cm]120:A2}]
(bdi2) {\includegraphics[height=0.8cm]{figs/fig-braincogs.png}};

\node[below=1.3cm, xshift=1.5cm, label={[label distance=-0.3cm]120:A3}]
(bdi3) {\includegraphics[height=0.8cm]{figs/fig-braincogs.png}};

\node[above=0.5cm]
(bdi4) {\includegraphics[height=0.8cm]{figs/fig-braincogs.png}};

\node[above=1.75cm, xshift=1cm]
(bdi5){\includegraphics[height=0.8cm]{figs/fig-braincogs.png}};

\node[above=0.25cm,xshift=7cm, label={[label distance=-0.3cm]30:A1}]
(abm1) {\includegraphics[height=0.8cm]{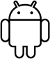}};

\node[below=0.45cm, xshift=9cm, label={[label distance=-0.3cm]30:A2}]
(abm2) {\includegraphics[height=0.8cm]{figs/fig-robot.png}};

\node[below=1.3cm, xshift=7.5cm, label={[label distance=-0.2cm]30:A3}]
(abm3) {\includegraphics[height=0.8cm]{figs/fig-robot.png}};

\node[above=1.75cm, xshift=8cm]
(abm4) {\includegraphics[height=0.8cm]{figs/fig-robot.png}};

\node[above=0.75cm, xshift=9.75cm]
(abm5) {\includegraphics[height=0.8cm]{figs/fig-robot.png}};
\draw[densely dashed](bdi1)--(abm1);
\draw[densely dashed](bdi2)--(abm2);
\draw[densely dashed](bdi3)--(abm3);
\begin{pgfonlayer}{background}
\node [fit=(bdi1) (bdi2) (bdi3) (bdi4) (bdi5)]
[rectangle, minimum height=4cm, minimum width=2.5cm, draw]{BDI System};
\node [fit=(abm1) (abm2) (abm3) (abm4) (abm5)]
[rectangle, minimum height=4cm, minimum width=2.5cm, draw]{ABM System};
\draw [->, >=latex,] (4,2.5) -- node[above]{actions} (6,2.5);
\draw [<-, >=latex,] (4,1.9) -- node[above]{percepts} (6,1.9);
\draw [<->, >=latex,] (4,1.3) -- node[above]{status} (6,1.3);
\end{pgfonlayer}
\end{tikzpicture}

%% file: figs/init-tree.tex
\begin{tikzpicture}[
    pc/.style={circle,thick,draw,fill=purple!60,rotate=0,
		anchor=south,minimum width=0.8cm},
    label distance=3mm,
    every label/.style={,scale=10,black!90},
    plan/.style={rectangle,thick,draw,fill=yellow!60,text width=4cm,
		text centered,anchor=north},
    goal/.style={circle,draw,fill=green!40,align=center,text width=2.5cm},
    precon/.style={circle,draw,fill=orange!40},
    edge from parent/.style={draw=black!100},
    edge from parent path={(\tikzparentnode.south) -- ++(0,-0.55cm)
			-| (\tikzchildnode.north)},
    level 1/.style={sibling distance=6cm,level distance=2.4cm,
			growth parent anchor=south,nodes=plan},
    ]

    \node (goal_init) [goal] {Initial Response}
        child {node (plan_depT) {\vspace{0.5cm}\\{1. If dependant is nearby then visit them\\ \emph{else}\\ Go home with some probability then visit dependant}\vspace{0.5cm}\\ 2. Go home with some probability\vspace{0.5cm}}}
	     	child {node (plan_depF) {\vspace{0.5cm}\\ If ACT threshold met continue daily plan \\ \emph{else}\\ Go home\vspace{0.5cm}}};
    \node (goal_home) [goal] [below=of plan_depF]  {Go Home} edge [<-] (plan_depF);
    \node (goal_dep) [goal] [below left=1pt and 1pt of plan_depT] {Go To Dependant} ;

   \node [pc]	at (plan_depT.north)	[label={}]	{dep=T};
   \node [pc]	at (plan_depF.north)	[label={}]	{dep=F};
  \path[->,draw] (-1.03,-7.5)-- (goal_home)[label={\}}];
  \path[->,draw] (-5.1,-6.3)-- (goal_dep)[label={\}}];
  \draw [decorate,decoration={brace,amplitude=8pt},xshift=-5.35cm,yshift=-8.5cm]
  (0.5,0.5) -- (0.5,3.9) ;
  \draw [decorate,decoration={brace,amplitude=8pt,mirror},xshift=-1.82cm,yshift=-9.5cm]
  (0.5,0.5) -- (0.5,3.5) ;
\end{tikzpicture}

%% file: figs/act-tree.tex
\begin{tikzpicture}[
    pc/.style={circle,thick,draw,fill=purple!60,rotate=0,
		anchor=south,minimum width=0.8cm},
    label distance=3mm,
    every label/.style={,scale=10,black!90},
    plan/.style={rectangle,thick,draw,fill=yellow!60,text width=4cm,
		text centered,anchor=north},
    goal/.style={circle,draw,fill=green!40,align=center,text width=2.5cm},
    precon/.style={circle,draw,fill=orange!40},
    edge from parent/.style={draw=black!100},
    edge from parent path={(\tikzparentnode.south) -- ++(0,-0.55cm)
			-| (\tikzchildnode.north)},
    level 1/.style={sibling distance=6cm,level distance=2.4cm,
			growth parent anchor=south,nodes=plan},
    ]

    \node (goal_init) [goal] {Act Now}
        child {node (plan_depT) {\vspace{0.5cm}\\ {1. Go home with some probability} \vspace{0.5cm}\\ {2. Leave now}\vspace{0.5cm}}}
	     	child {node (plan_depF) {\vspace{0.5cm}\\Leave now\vspace{0.5cm}}};
    \node (goal_home) [goal] [below=of plan_depF]  {\,\,\,Leave Now\,\,\,} edge [<-] (plan_depF);
    \node (goal_dep) [goal] [below left=1pt and 1pt of plan_depT] {\,\,\,Go Home\,\,\,} ;

   \node [pc]	at (plan_depT.north)	[label={}]	{dep=F};
   \node [pc]	at (plan_depF.north)	[label={}]	{dep=T};
  \path[->,draw] (-1.8,-6.2)-- (goal_home)[label={\}}];
  \path[->,draw] (-4.5,-4.8)-- (goal_dep)[label={\}}];
\end{tikzpicture}

%% file: figs/attribute-xml-ex.tex
\begin{lstlisting}[%
language=XML,
numbers=none]

<person id= "0" >
    <attributes>
      <attribute name="BDIAgentType" class="java.lang.String" >io.github.agentsoz.ees.agents.bushfire.Resident</attribute>
      <attribute name="HasxsAtLocation" class="java.lang.String" >766722.480977607,5754455.48885175</attribute>
      <attribute name="InitialResponseThreshold" class="java.lang.Double" >0.1</attribute>
      <attribute name="FinalResponseThreshold"   class="java.lang.Double" >0.1</attribute>
      <attribute name="WillGoHomeAfterVisitingDependants" class="java.lang.Boolean" >true</attribute>
      <attribute name="WillGoHomeBeforeLeaving" class="java.lang.Boolean" >false</attribute>
      <attribute name="EvacLocationPreference" class="java.lang.String">,791118.102569293,5753461.09876635</attribute>
      <attribute name="InvacLocationPreference" class="java.lang.String">,766728.618195617,5754449.11666823</attribute>
    </attributes>
    ...
  </person>

  \end{lstlisting}

%% file: chapter-applications/sec-applications.tex
\section{Applications}
\joelComment{-future work/discussion: can initial MATSim plans be iteratively improved prior to evacuation scenario?}
We will now apply the population generation model to the Surf Coast Shire region, an area that has been affected by a number of severe fire incidents (Black Saturday, Ash Wednesday) and contains a community that is regularly exposed to the bushfire threat.
We construct a population for a typical summer weekday in the region, analysing the algorithm's performance and examining the resultant simulation runs to verify that the outputted agent plans adequately reflect the input data.
The specific inputs for the scenario have been co-developed with domain experts from the Surf Coast Shire Council (SCSC) and the Department of Environment, Land, Water and Planning (DELWP).

\input{chapter-applications/subsec-plan-algorithm-inputs}
\input{chapter-applications/subsec-plan-algorithm-results}
\input{chapter-applications/subsec-attributes}
\input{chapter-applications/subsec-application}
\input{chapter-applications/subsec-discussion}

%% file: chapter-applications/subsec-plan-algorithm-inputs.tex
\subsection{Algorithm Inputs in Surf Coast Shire}
In this section we describe the inputs that the population generation algorithm takes and justify them within the context of the Surf Coast shire region.
The chosen inputs are designed to easily translate across a large number of potential scenarios, and for this reason the number of parameters is kept to a minimum while at the same time giving sufficient flexibility in the kinds of populations that can be generated.
Another motivation for this is to provide a user-friendly input process using data and expert opinion most likely to be available to emergency personnel.
\subsubsection{Subgroups in Surf Coast Shire}
The fire season in Victoria falls between November-April and coincides with the busiest tourism period on the Surf Coast, and this also brings a large transient and seasonal population of workers, semi-permanent residents and holiday home owners.
Small townships along the Great Ocean Road can have up to 6-7 times the resident population, with permanent residents far outnumbered by tourists and other short-term visitors.
This gives a basis upon which to split the population into subgroups, with visitors not only having very different reasons and priorities in forming their daily plans, but also a reduced knowledge of the area and routes between locations.

The five identified subgroups for the region are:

\begin{itemize}
\item \verb|Resident| ($R$): those who live in the region, have local knowledge of roads and places of congregation.
They are connected to the community, are likely to have a concern for others, pets, and property, and are most likely to defend property.
\item \verb|ResidentPartTime| ($RP$): those who own property in the region, but may only live there for several months of the year.
Are somewhat familiar with the area, but do not have a large community network. They are less likely to be prepared for a bushfire threat and less likely to defend property.
\item \verb|VisitorRegular| ($VR$): those who have visited the region on several occasions and may have a holiday home there.
Will know the area somewhat, but are unlikely to have pets or relatives.
Very unlikely to stay and defend property.
\item\verb|VisitorOvernight|($VO$): those who are unfamiliar with the area and living in short term accommodation.
They will not defend property but are likely to gather belongings and then follow instructions or leave the region.
\item\verb|VisitorDaytime| ($VD$): those who are mostly unfamiliar with local roads and places of congregation.
They will either ignore a fire alert or leave region.
\end{itemize}

Thus we set

\begin{align*}
S= \{R,RP,VR,VO,
VD\}
\end{align*}

and require as a first input the number of required agents for each $s\in S$:
\begin{align*}
  P_R&=10000\\
P_{RP}&=5000\\
P_{VR}&=5000\\
P_{VO}&=15000\\
P_{VD}&=15000\\
\end{align*}

giving a total population of $P=50000$ in the region.
\subsubsection{Activities in Surf Coast Shire}
We now consider the most likely or significant types of activities that occur during summer in the region, taking into account both where people might congregate and which trips are important in shaping a person's daily plan.
Whilst the actual nature and location of trips made by individuals of different subgroups may differ significantly, it is useful to maintain a broad description of activities here and add granularity via the location maps for each subgroup as this makes the activity distributions easier to construct and understand.

The five activities that broadly capture the types of trips that an agent (of any subgroup) will make are:

\begin{itemize}
\item $\alpha_1=$ \texttt{home}: as required by the algorithm; \texttt{home} locations will range from: residential addresses for \texttt{Resident}, \texttt{ResidentPartTime} and a percentage of \texttt{VisitorRegular}; to hotels, caravan parks and other short-term accommodation options for
\texttt{VisitorOvernight} and the remaining \texttt{VisitorRegular}; to specific `source' locations outside of the region for \texttt{VisitorDaytime}.
Duration is set to $2$ hours for all subgroups. Note that duration here
\item $\alpha_2=$ \texttt{work}: the work activity will only apply to \texttt{Resident} and \newline \texttt{ResidentPartTime}, and will occur at specified place-of-business locations.
Duration is set to $4$ hours for all subgroups, which allows for people having breaks and/or working multiple jobs.
This activity requires the most consideration when it is being distributed throughout the day, with peak \texttt{work} hours occurring during a typical 9am-5pm window.
\item  $\alpha_3=$ \texttt{shop}: similarly, these will be designated in places specified as shopping districts.
The focus will be on larger shop areas, where people may tend to gather at times in the day.
These locations are also more relevant from an emergency perspective as many of these shopping centres will be places of congregation for invacuation.
Duration is set to $2$ hours for all subgroups.
\item $\alpha_4=$ \texttt{beach}: this is a major activity during summer and will be particularly popular with all of the visitor subgroups.
Here the peak times in the distribution should be quite clear based on the forecast and could potentially consider tidal information on a per-day basis. \texttt{Resident} agents may have patterns for the \texttt{beach} activity that avoid peak tourist times.
Duration is set to $2$ hours for all subgroups.
\item $\alpha_5=$ \texttt{other}: will serve as a catch-all for other less significant activities that occur during the day.
It will allow for more diversity in plans between subgroups, and also ensure that there are not too many agents evacuating from the same spots in an emergency.
Should usually be maintained as a minor but constant presence at all time steps to allow for this diversity.
Duration is set to $2$ hours for all subgroups.
\end{itemize}
We therefore define the set of activities $$A=\{\alpha_1,\alpha_2,\alpha_3,\alpha_4,\alpha_5\}.$$

$A$ is mapped per subgroup to the set of locations $\mathcal{L}$, which are provided as a ESRI shapefile (\texttt{.shp}) by SCSC. Also included in this file are the locality and allocation number for each location.
Apart from the distinction in \texttt{home} locations described above, some other interesting differences between subgroup activity mappings include only mapping more popular beaches to the visitor groups, assigning schools to \texttt{other} for the \texttt{Resident} subgroup and assigning landmarks and national parks to \texttt{other} for \texttt{VisitorDaytime} and \texttt{VisitorOvernight}.

\subsubsection{Distributions in Surf Coast Shire}

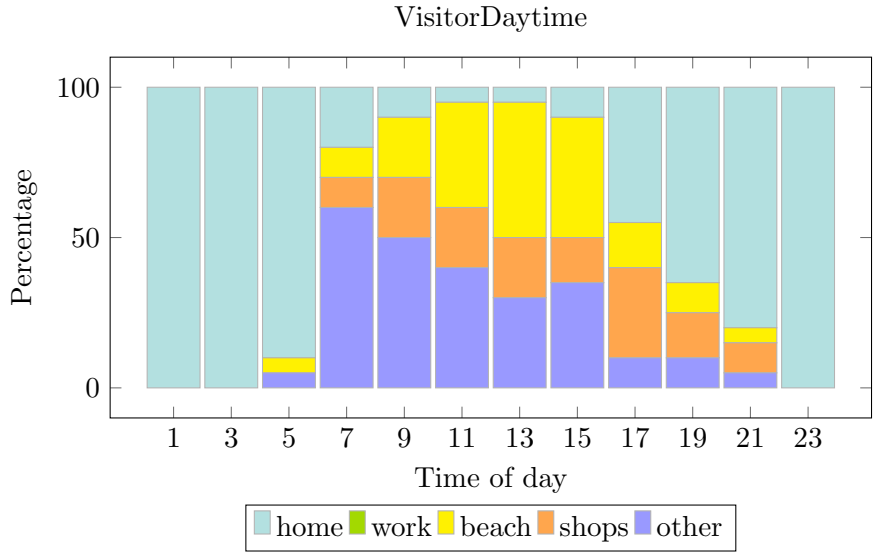
\begin{figure}[!ht]
  \begin{center}
\input{figs/visitor-daytime-distribution.tex}
\caption{Distribution table for \texttt{VisitorDaytime} subgroup on a typical summer day.}\label{fig:vd-actdist}
\end{center}
\end{figure}

We now take the set of activities $A$ and formulate distributions for each subgroup.
In keeping with our choices for durations, the time-step size is set to 2 hour blocks, meaning that there are 12 `bins' that need to be established for each subgroup.
Note that the durations listed for each activity do not necessarily reflect their \textit{typical} duration, but indicate the approximate time interval at which an agent will re-evaluate its activity.
It is quite possible that a \texttt{Resident} agent could spend the entire day performing the \texttt{home} activity if during activity selection \texttt{home} is chosen for every time-step.
Figures~\ref{fig:resident-actdist} and \ref{fig:vd-actdist} show some examples of activity distributions on a typical summer day for \texttt{Resident} and \texttt{VisitorDaytime}:

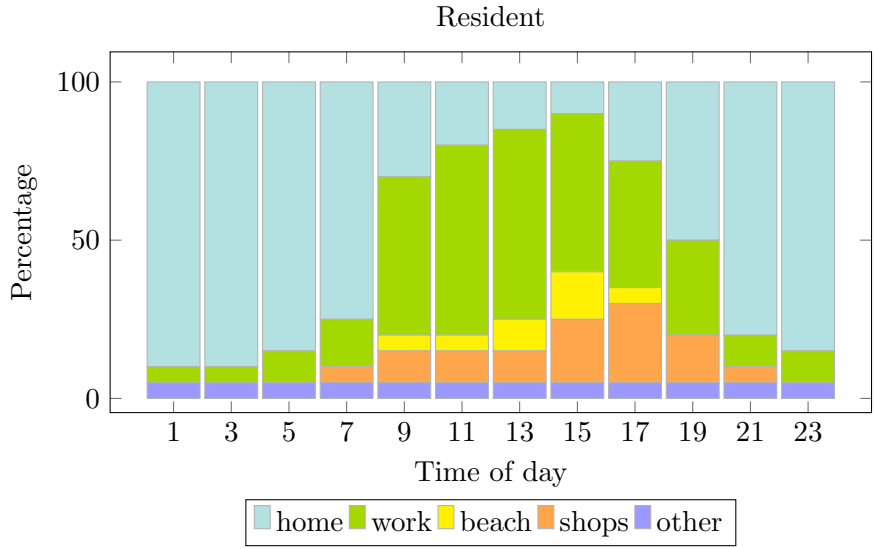
\begin{figure}[!ht]
  \begin{center}
\input{figs/resident-distribution.tex}
\caption{Activity distribution for \texttt{Resident} subgroup on a typical summer day. }\label{fig:resident-actdist}
\end{center}
\end{figure}

Figure~\ref{fig:resident-actdist} shows that for a \texttt{Resident}, the major working time is between 8 a.m. and 6 p.m. in keeping with a regular weekday.
In both cases, shopping activity increases in the afternoon and continues until dinnertime, when people (particularly visitors) are likely to be at restaurants and bars.
Note that these distributions have not been formally validated by SCSC, and rather represent a first attempt to capture the dynamics of typical weekday activities in the region.
These serve for demonstration purposes, but activity distributions will ultimately be derived from actual daily trip data.
\\

\subsubsection{Locations in Surf Coast Shire}
Locations are supplied by  SCSC and exist in a number of different forms.
At the most basic level, a small number of source nodes throughout the region serve to distribute the population, with a limited set of polygons describing areas of work, beaches etc.
This approach is crude but useful for quickly describing patterns of movement.
However, to create accurate day plans that vary across the population (and thus give realistic traffic movement) a greater level of detail is required.
Per-address data allows a population to be precisely mapped at an appropriate density to residential areas and tourist spots. It also allows the possibility of specific, niche trips (e.g. to a golf club, national park or fishing pier) that further diversify the population's movements or capture known patterns that may be of concern in an emergency.
\begin{figure}[ht]
  \begin{center}
\includegraphics[scale=0.3]{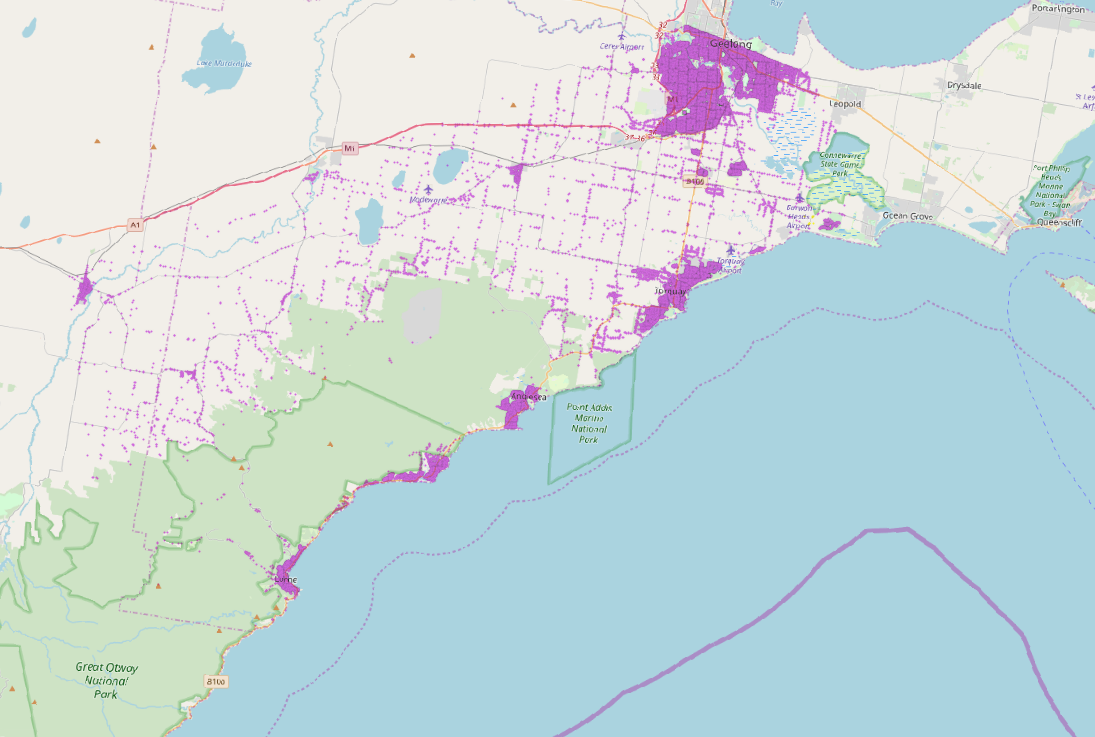}
\caption{Per-address location data on the Surf Coast Shire. The location points (pink) cluster around larger coastal communities, with hardly any addresses in the national park area (green).}\label{fig:address-data}
\end{center}
\end{figure}
Figure~\ref{fig:address-data} illustrates why realistic address data can have a significant impact on resultant traffic flow.
\

The trade-off is that the increased detail increases algorithm runtime.
The locality-based approach (see Section~\ref{sec:locale} mitigates this somewhat, and also allows us to frame our subgroups in terms of how likely they are to move through the region. The final input we need to form agent plans is a travel factor measure for each subgroup.
We would expect that visitors are more likely to move between localities than residents, and so we assign the following provisional probabilities:

\begin{align*}
g_{R}=0.2\\
g_{RP}=0.3\\
g_{VR}=0.4\\
g_{VO}=0.6\\
g_{VD}=0.8
\end{align*}


%% file: figs/resident-distribution.tex
\pgfplotstableread{
hour home work beach shops other
01 90 05 00 00 05
03 90 05 00 00 05
05 85 10 00 00 05
07 75 15 00 05 05
09 30 50 05 10 05
11 20 60 05 10 05
13 15 60 10 10 05
15 10 50 15 20 05
17 25 40 05 25 05
19 50 30 00 15 05
21 80 10 00 05 05
23 85 10 00 00 05

}\loadedtable

\begin{tikzpicture}
  \begin{axis}
    [
      scale only axis, 
      title=Resident,
      height=4.8cm,
      ylabel=Percentage,
      ylabel near ticks,
      xlabel=Time of day,
      width=0.8\textwidth,
      bar width = 20pt,
      ybar stacked,
      xtick=data,
      reverse legend,
      legend columns=-1,
      legend style={at={(0.5,-0.24)}, anchor=north}
    ]
    \addplot+[black!30,fill=blue!40] table[x=hour, y=other] {\loadedtable}; \addlegendentry{other};
    \addplot+[black!30,fill=orange!70] table[x=hour, y=shops] {\loadedtable}; \addlegendentry{shops};
    \addplot+[black!30,fill=yellow!100!black] table[x=hour, y=beach] {\loadedtable}; \addlegendentry{beach};
    \addplot+[black!30, fill=lime!85!black] table[x=hour, y=work] {\loadedtable}; \addlegendentry{work};
    \addplot+[black!30,fill=cyan!30] table[x=hour, y=home] {\loadedtable}; \addlegendentry{home};
  \end{axis}
\end{tikzpicture}

%% file: figs/visitor-daytime-distribution.tex
\pgfplotstableread{
hour home work beach shops other
01 100 00 00 00 00
03 100 00 00 00 00
05  90 00 05 00 05
07  20 00 10 10 60
09  10 00 20 20 50
11  05 00 35 20 40
13  05 00 45 20 30
15  10 00 40 15 35
17  45 00 15 30 10
19  65 00 10 15 10
21  80 00 05 10 05
23 100 00 00 00 00

}\loadedtable

\begin{tikzpicture}
  \begin{axis}
    [
      scale only axis, 
      title=VisitorDaytime,
      height=4.8cm,
      ylabel=Percentage,
      ylabel near ticks,
      xlabel=Time of day,
      width=0.8\textwidth,
      bar width = 20pt,
      ybar stacked,
      xtick=data,
      reverse legend,
      legend columns=-1,
      legend style={at={(0.5,-0.24)}, anchor=north}
    ]
    \addplot+[black!30,fill=blue!40] table[x=hour, y=other] {\loadedtable}; \addlegendentry{other};
    \addplot+[black!30,fill=orange!70] table[x=hour, y=shops] {\loadedtable}; \addlegendentry{shops};
    \addplot+[black!30,fill=yellow!100!black] table[x=hour, y=beach] {\loadedtable}; \addlegendentry{beach};
    \addplot+[black!30, fill=lime!85!black] table[x=hour, y=work] {\loadedtable}; \addlegendentry{work};
    \addplot+[black!30,fill=cyan!30] table[x=hour, y=home] {\loadedtable}; \addlegendentry{home};
  \end{axis}
\end{tikzpicture}

%% file: chapter-applications/subsec-plan-algorithm-results.tex
\subsection{Results of the Algorithm}
We now run the algorithm with the above inputs using the programming language R to generate a viable MATSim \texttt{population.xml} file. For testing we used a 2013 Macbook Air with 2 cores and 8GB of RAM.
With $50,000$ agents, the process takes $25.14$ minutes, but this duration is largely due to the fine granularity of the full locations \texttt{.shp} file ($\sim75,000$ address points).
If we reduce that down to $\sim 40,000$ addresses, which is still very fine-grain, we get a runtime of $9.30$ minutes.
It is therefore important to balance a realistic representation of locations in the region with restricting the location options down to match the number required by the given population (note that many visitor agents will be mapped to the same \texttt{home} nodes; in fact we may have 10,000-15,000 all originating from the same out-of-region source node, so we generally require less nodes than agents).
We proceed with the full locations data as the most general case.
\\

\subsubsection{Plan Times}
To verify that the generated plans reflect the required population dynamics, we compare the input distributions, which detail the expected number of agents performing each activity during each time bin, to the generated number of agents completing those activities in the \texttt{population.xml} file.

\begin{table}[!htbp]
\caption{Percentage variation from input distribution at each time step for \texttt{Resident} subgroups.}\label{table:Rdiff}
\centering
\begin{tabular}{c c c c c c c}
\texttt{Resident}\\ [1ex]
\hline\hline
Activity & 12am & 2am & 4am & 6am & 8am & 10am \\
 & -2am & -4am & -6am & -8am & -10am & -12pm\\ [0.5ex]
\hline
\texttt{home} & 0.03 &	0.16 & 0.13 &	0.36 &	0.19 &	0.09 \\
\texttt{work} & -0.41 &	-0.41 &	0.05 & -0.25 &	-0.12	& 0.04 \\
\texttt{beach}	& 0	& 0 &	0 &	0	& -0.08 &	-0.18	\\
\texttt{shops} &	0&	0&	0	&-0.3&	-0.13&	-0.15\\
\texttt{other} &	0.38&	0.25&	-0.18&	0.19&	0.14&	0.2\\ [1ex] 
\hline 
Activity & 12pm& 2pm & 4pm & 6pm & 8pm& 10pm \\
& -2pm & -4pm & -6pm & -8pm& -10pm& -12am\\ [0.5ex]
\hline
\texttt{home} &	-0.2 & -0.03 & 0.35 &	\textbf{-3.83} &	-0.57 &	-0.09\\
\texttt{work} & -0.39 &	-0.49 &	-0.2&	\textbf{5.04}&	0.19&	0.19\\
\texttt{beach}& 0.08 &	0.79 &	-0.2 &	\textbf{0} &	0	& 0\\
\texttt{shops}&	0.52&	-0.26&	0.12&	\textbf{-1.07}&	-0.01&	0\\
\texttt{other}&	-0.01&	-0.01&	-0.07&	\textbf{-0.14}&	0.39&	-0.1\\ [1ex]
\end{tabular}
\end{table}

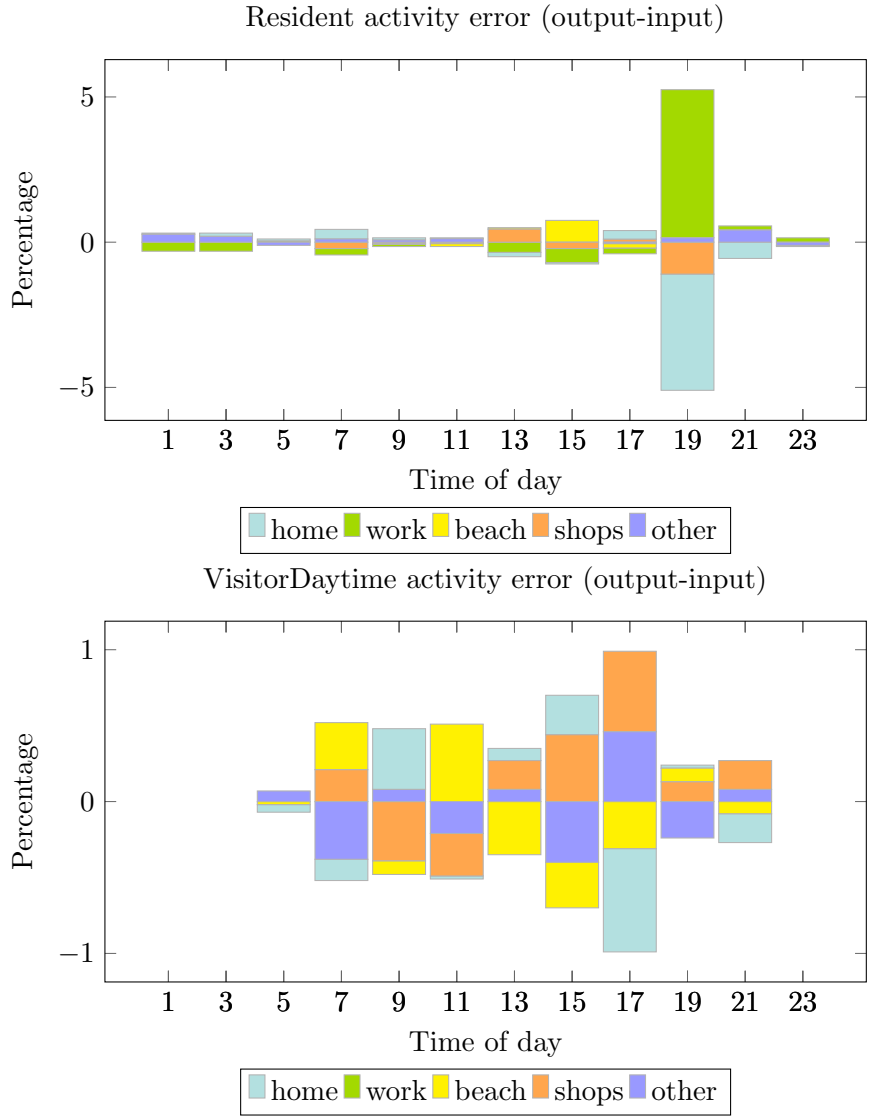
\begin{figure}[!htbp]
  \begin{center}
\input{figs/visdiffres.tex}
\input{figs/visdiffvisday.tex}
\caption{Percentage error in both the \texttt{Resident} and \texttt{VisitorDaytime} subgroup. The former has a very obvious error outlier. The latter has a maximum error $<1\%$, with the error spread evenly throughout the `busy' part of the day.   }\label{fig:visdiff}
\end{center}
\end{figure}

\begin{table}[!htbp]
  \centering
  \caption{Percentage variation from input distribution at each time step for  the \texttt{ResidentPartTime} subgroup.}\label{table:RPdiff}
  \begin{tabular}{c c c c c c c}
\texttt{ResidentPartTime}\\ [1ex]
\hline\hline
Activity & 12am & 2am & 4am & 6am & 8am & 10am \\
 & -2am & -4am & -6am & -8am & -10am & -12pm\\ [0.5ex]
\hline
\texttt{home} &-0.12&0&0.04&0.1&0.48&-0.5\\
\texttt{work} &0&0&-0.16&-0.08&-0.02&-0.2\\
\texttt{beach}&0&0&0&0.18&0.24&-0.24\\
\texttt{shops}&0&0&0&0.04&-0.3&-0.02\\
\texttt{other}&0.12&0&0.12&-0.24&-0.4&0.96 \\[1ex] 
\hline 
Activity & 12pm& 2pm & 4pm & 6pm & 8pm& 10pm \\
& -2pm & -4pm & -6pm & -8pm& -10pm& -12am\\ [0.5ex]
\hline
\texttt{home} &-0.2&-0.34&\textbf{-2.06}&-0.64&0.42&-0.92\\
\texttt{work} &0.24&0.3&\textbf{4.96}&0.02&0.02&0\\
\texttt{beach}&-1.2&0.06&\textbf{0.28}&0&0&0\\
\texttt{shops}&1.18&0.58&\textbf{-1.42}&0.56&-0.24&0.9\\
\texttt{other}&-0.02&-0.6&\textbf{-1.76}&0.06&-0.2&0.02\\ [1ex]
\end{tabular}

\label{table:RPdiff}
\end{table}
In Table~\ref{table:Rdiff} and Table~\ref{table:RPdiff}, we see for the two resident subgroups that the cumulative effect of \texttt{work} having a longer duration than other activities results in a $\sim 5\%$ discrepancy towards the end of the peak working period.
In the other subgroups (Table~\ref{table:VRdiff}-\ref{table:VDdiff}) where \texttt{work} is not a valid activity, we see that at all time-steps the generated plans match the input distributions to a maximum error of $\sim 1\%$. The difference in outcome for the two activity distributions for \texttt{Resident} and \texttt{VisitorDaytime}, first introduced in Figures ~\ref{fig:resident-actdist} and \ref{fig:vd-actdist}, is emphasised visually in Figure~\ref{fig:visdiff}.

\begin{table}[!htbp]
\caption{Percentage variation from input distribution at each time step for \texttt{VisitorRegular} subgroup.}\label{table:VRdiff}
\centering
\begin{tabular}{c c c c c c c}
\texttt{VisitorRegular}\\ [1ex]
\hline\hline
Activity & 12am & 2am & 4am & 6am & 8am & 10am \\
 & -2am & -4am & -6am & -8am & -10am & -12pm\\ [0.5ex]
\hline
\texttt{home} &0.22&0.16&0.26&-0.5&-0.76&-1.2\\
\texttt{beach}&0&0&0&0.38&0.26&0.46\\
\texttt{shops}&0&0&0&0&-0.16&0.1\\
\texttt{other}&-0.22&-0.16&-0.26&0.12&0.66&0.64\\[1ex] 
\hline 
Activity & 12pm& 2pm & 4pm & 6pm & 8pm& 10pm \\
& -2pm & -4pm & -6pm & -8pm& -10pm& -12am\\ [0.5ex]
\hline
\texttt{home} &-0.34&0.24&0.2&0.3&-0.5&0.74\\
\texttt{beach}&0.94&0.48&0.08&0.02&-0.32&0\\
\texttt{shops}&-0.6&-0.52&0.12&-0.2&0.6&-0.38\\
\texttt{other}&0&-0.2&-0.4&-0.12&0.22&-0.36\\ [1ex]
\hline\hline
\end{tabular}
\end{table}

\begin{table}[!htbp]
  \caption{Percentage variation from input distribution at each time step for  the \texttt{VisitorOvernight} subgroup.}\label{table:VOdiff}
  \centering
  \begin{tabular}{c c c c c c c}
\texttt{VisitorOvernight}\\ [1ex]
\hline\hline
Activity & 12am & 2am & 4am & 6am & 8am & 10am \\
 & -2am & -4am & -6am & -8am & -10am & -12pm\\ [0.5ex]
\hline
\texttt{home} &0.11&-0.04&-0.41&0.09&0.23&0\\
\texttt{beach}&0&0&0.37&0.11&0.13&-0.21\\
\texttt{shops}&-0.11&0.04&0&-0.34&-0.48&-0.01\\
\texttt{other}&0&0&0.03&0.14&0.12&0.21\\[1ex] 
\hline 
Activity & 12pm& 2pm & 4pm & 6pm & 8pm& 10pm \\
& -2pm & -4pm & -6pm & -8pm& -10pm& -12am\\ [0.5ex]
\hline
\texttt{home} &0&0.03&-0.03&0.03&-0.13&0\\
\texttt{beach}&-0.21&0.23&-0.25&0.31&0.05&0\\
\texttt{shops}&0.15&0&0.09&-0.09&0.19&0\\
\texttt{other}&0.07&-0.26&0.19&-0.25&-0.11&0\\ [1ex]
\hline\hline
\end{tabular}
\end{table}

\begin{table}[!htbp]
  \centering
  \caption{Percentage variation from input distribution at each time step for  the \texttt{VisitorDaytime} subgroup.}\label{table:VDdiff}
  \begin{tabular}{c c c c c c c}

\texttt{VisitorDaytime}\\ [1ex]
\hline\hline
Activity & 12am & 2am & 4am & 6am & 8am & 10am \\
 & -2am & -4am & -6am & -8am & -10am & -12pm\\ [0.5ex]
\hline
\texttt{home}&0&0&-0.05&-0.19&0.38&-0.04\\
\texttt{beach}&0&0&-0.03&0.22&-0.11&0.53\\
\texttt{shops}&0&0&0&0.32&-0.36&-0.27\\
\texttt{other}&0&0&0.08&-0.35&0.09&-0.21\\[1ex] 
\hline 
Activity & 12pm& 2pm & 4pm & 6pm & 8pm& 10pm \\
& -2pm & -4pm & -6pm & -8pm& -10pm& -12am\\ [0.5ex]
\hline
\texttt{home} &0.07&0.25&-0.63&0.03&-0.23&0\\
\texttt{beach}&-0.32&-0.26&-0.35&0.09&-0.08&0\\
\texttt{shops}&0.19&0.43&0.54&0.12&0.23&0\\
\texttt{other}&0.07&-0.41&0.45&-0.25&0.07&0\\ [1ex]
\end{tabular}

\end{table}

\newpage
\subsubsection{Plan Locations}

\begin{figure}[!htb]

\includegraphics[width=\linewidth]{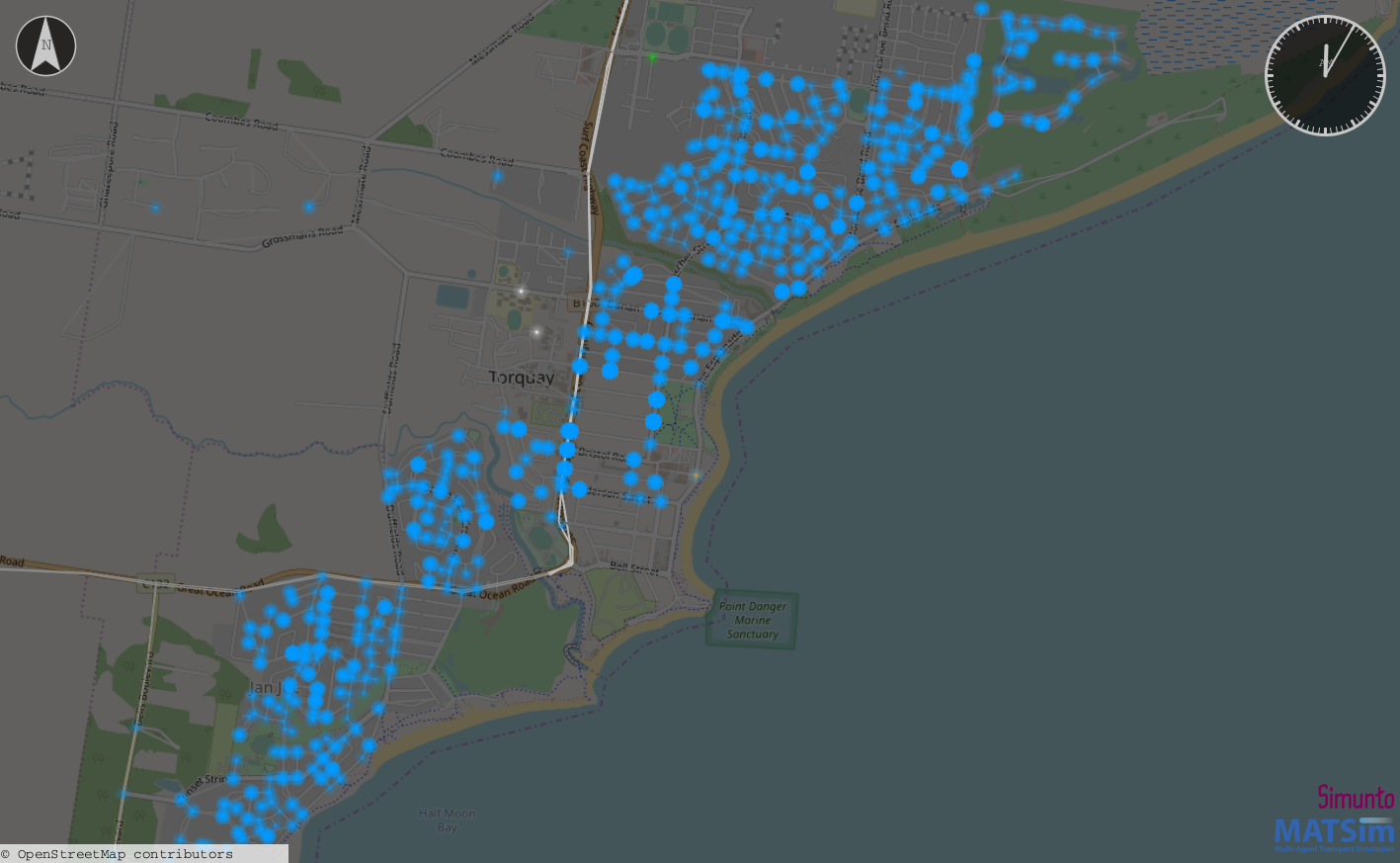}
\caption{Agent activity locations at midnight. The vast majority of the population are at home (blue), which is not surprising just after initialisation (recall that \textit{every} agent begins at home). We can see one or two agents have already started performing \texttt{other} activities (white) and one worker has already begun their shift (green).
We can note that the algorithm has successfully allocated a concentration of people to individual address points in residential areas, with CBD areas less populated during off-peak times.}
\label{fig:immediate}
\end{figure}

We must also analyse a) how successfully the algorithm distributes activities, and b) if the trips between activities are realistic enough to provide an appropriate traffic background.
To do this, we run the plan in MATSim and observe the output events data.
Figures~\ref{fig:immediate}-\ref{fig:combined} provide a snapshot run-through of how the morning plays out in this scenario for a particular community in the region (Torquay).
In these maps, location nodes are coloured if they are being used by an agent for an activity.
The colour mapping is
\begin{align*}
  \texttt{home}&\longrightarrow \text{blue}\\
  \texttt{work}&\longrightarrow \text{green}\\
  \texttt{beach}&\longrightarrow \text{yellow}\\
  \texttt{shops}&\longrightarrow \text{orange}\\
  \texttt{other}&\longrightarrow \text{white}
\end{align*}

\begin{figure}[!htb]
\includegraphics[width=\linewidth]{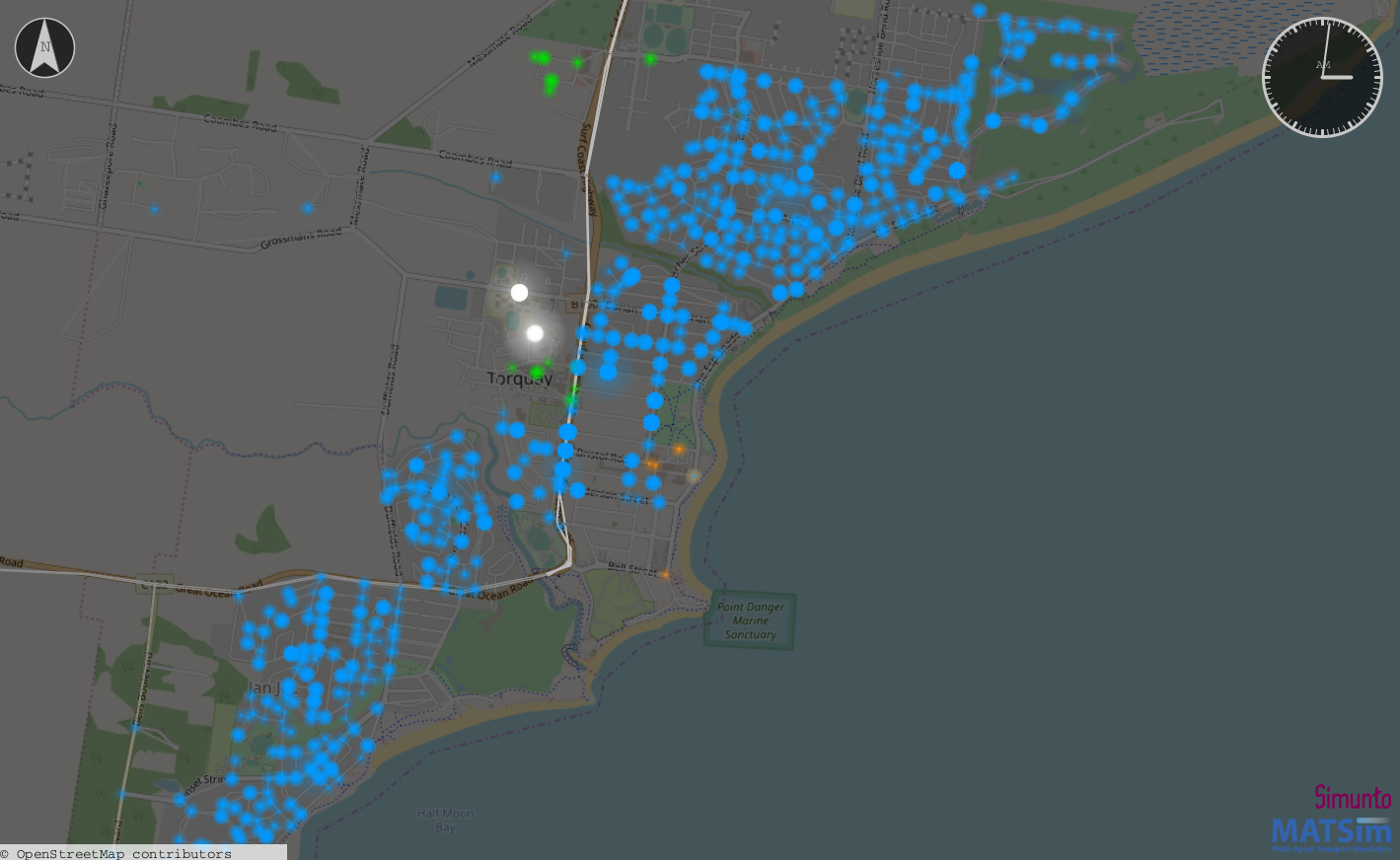}
\caption{Activity locations at 3am. More \texttt{other} activities have begun, and a few more agents are now at \texttt{work}. We also see the introduction of a few early morning \texttt{shops} activities (orange).
In this visualisation, glowing or bright points indicate that activity is starting at that location, with continuing activities maintaining their colour until they are terminated.}
\label{fig:proximal}

\end{figure}

\begin{figure}[!htb]
\includegraphics[width=\linewidth]{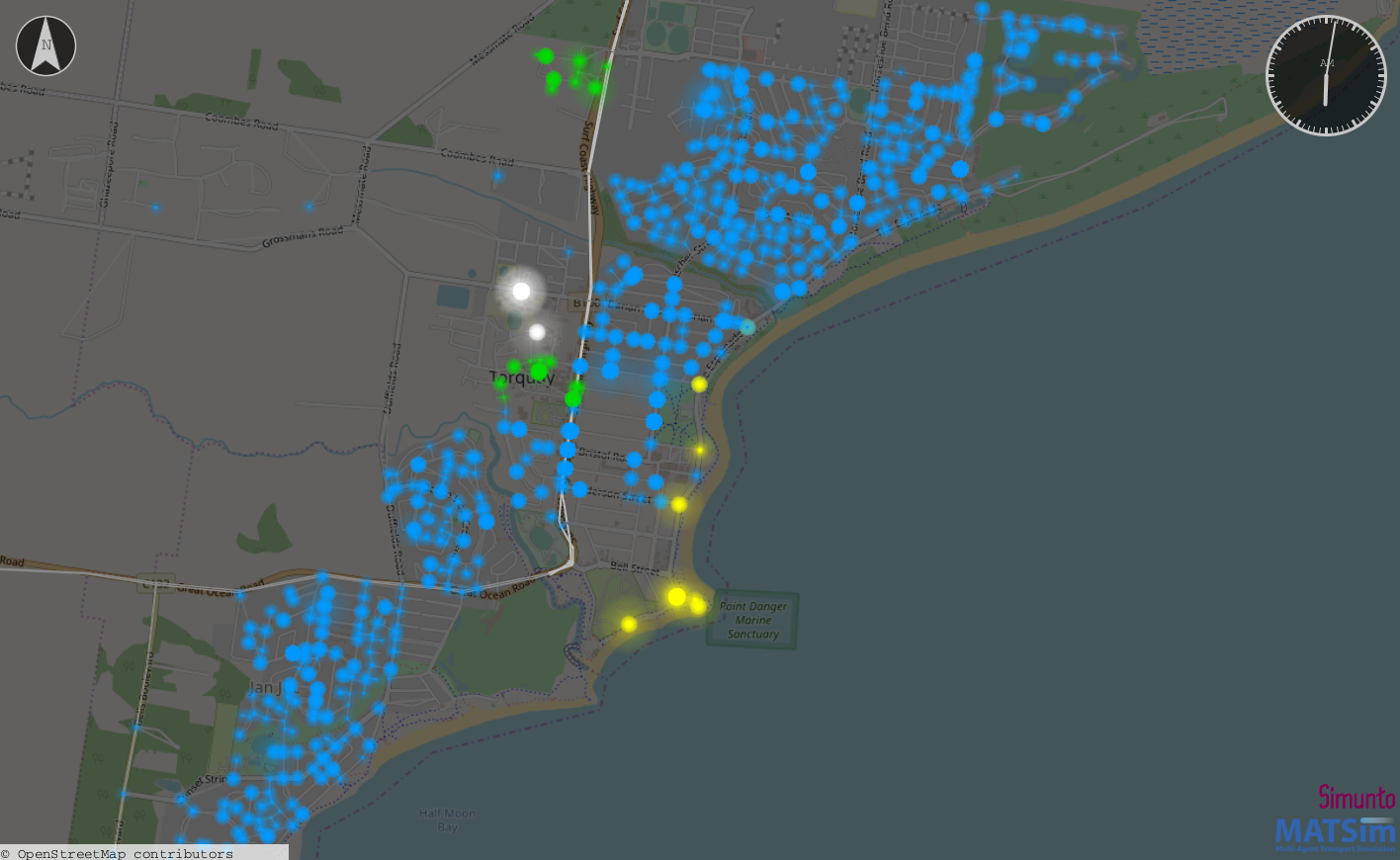}
\caption{Activity locations at 6am. There are a few more people at \texttt{work}, but the most significant change is that dawn has brought out a few \texttt{beach}goers (yellow).
Notice how both \texttt{work} and \texttt{beach} activities are concentrated within one or two areas. The input location data used here tends to focus on a few busy or popular areas as centres for particular locations.}
\label{fig:distal}
\end{figure}

\begin{figure}[!htb]
\includegraphics[width=\linewidth]{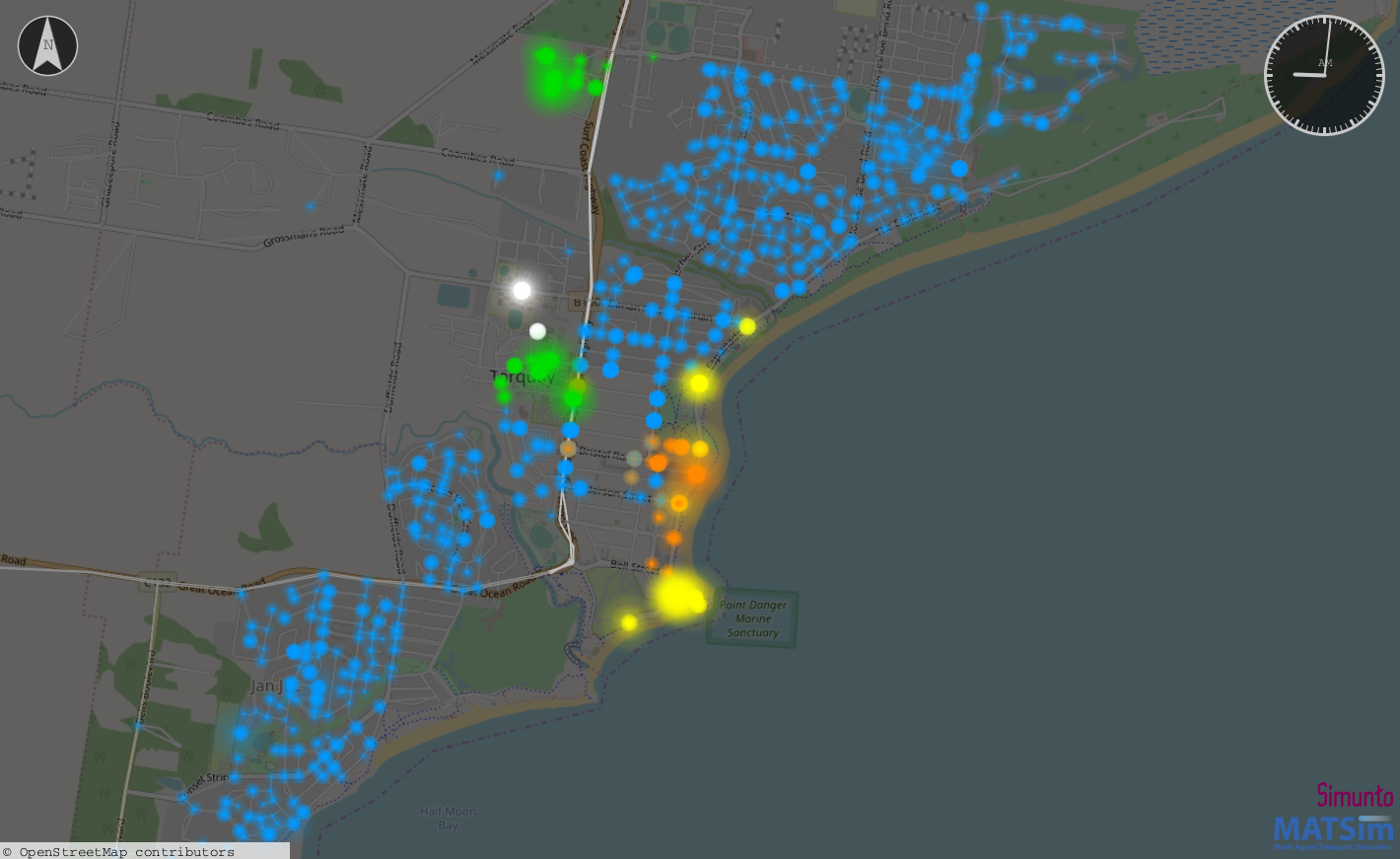}
\caption{At 9am the activity locations have reached a point where all 5 activities are well represented in the snapshot. We note that there is now a significant number of agents performing \texttt{work}, coinciding with the beginning of the peak \texttt{work} times for a typical summer weekday.
Beach activitiy is also peaking here, with many activities occurring along the coastal strip.}
\label{fig:combined}

\end{figure}

\begin{figure}[!htb]
  \includegraphics[width=\linewidth]{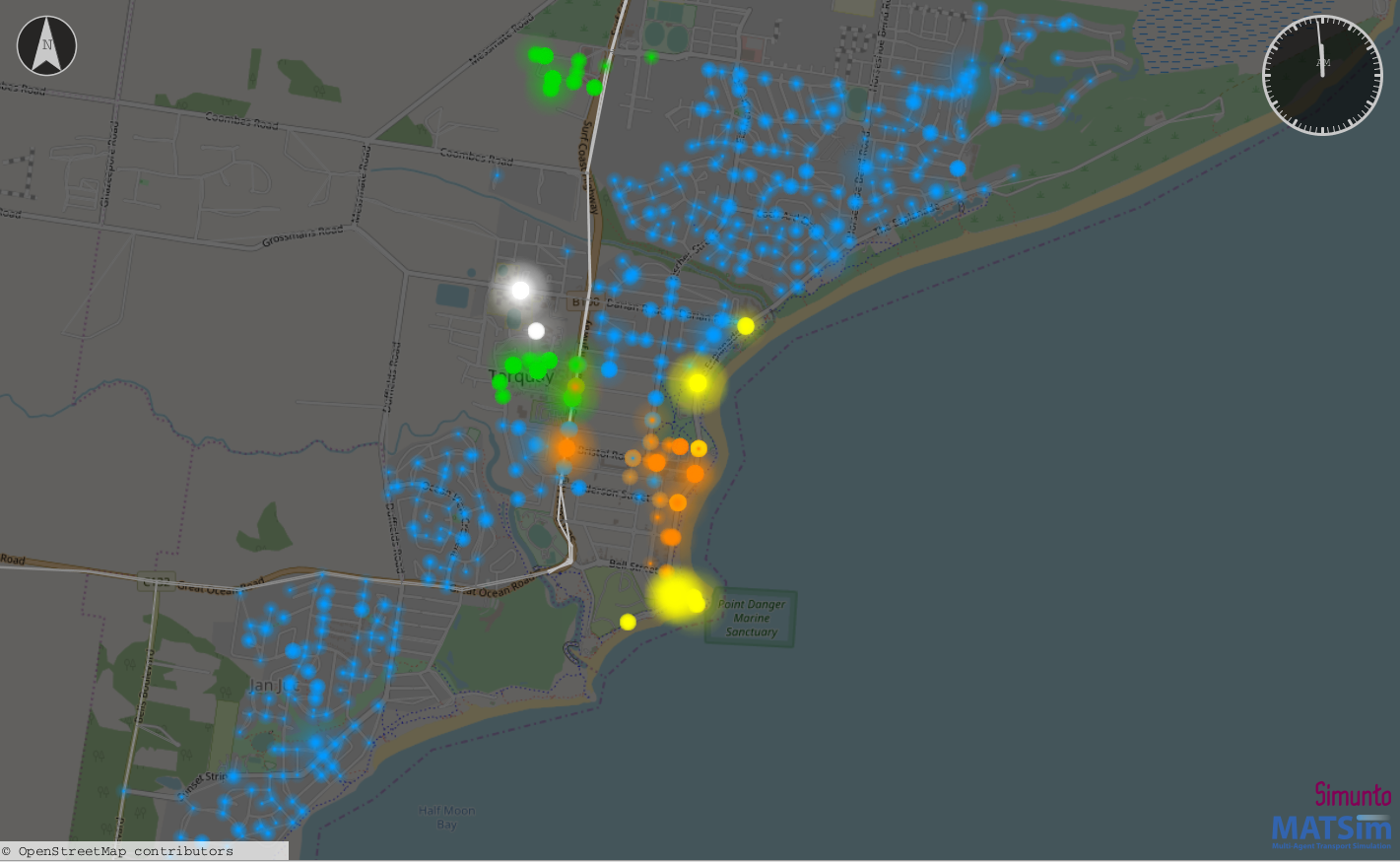}
  \caption{By midday we have less people beginning \texttt{work}, but there is still a steady set of working agents. Trips to \texttt{shops} have increased significantly, and are almost the most popular activity around lunch time. Many of those people who began work during the first peak work starting time would now be finished, and may have decided to go to the shops after work.
  The percentage of people at home has now very noticeably dropped from earlier in the morning, indicating that many agents are out and travelling to/undertaking other activities. }
  \label{fig:combined}

\end{figure}

\begin{figure}[!htb]
  \includegraphics[width=\linewidth]{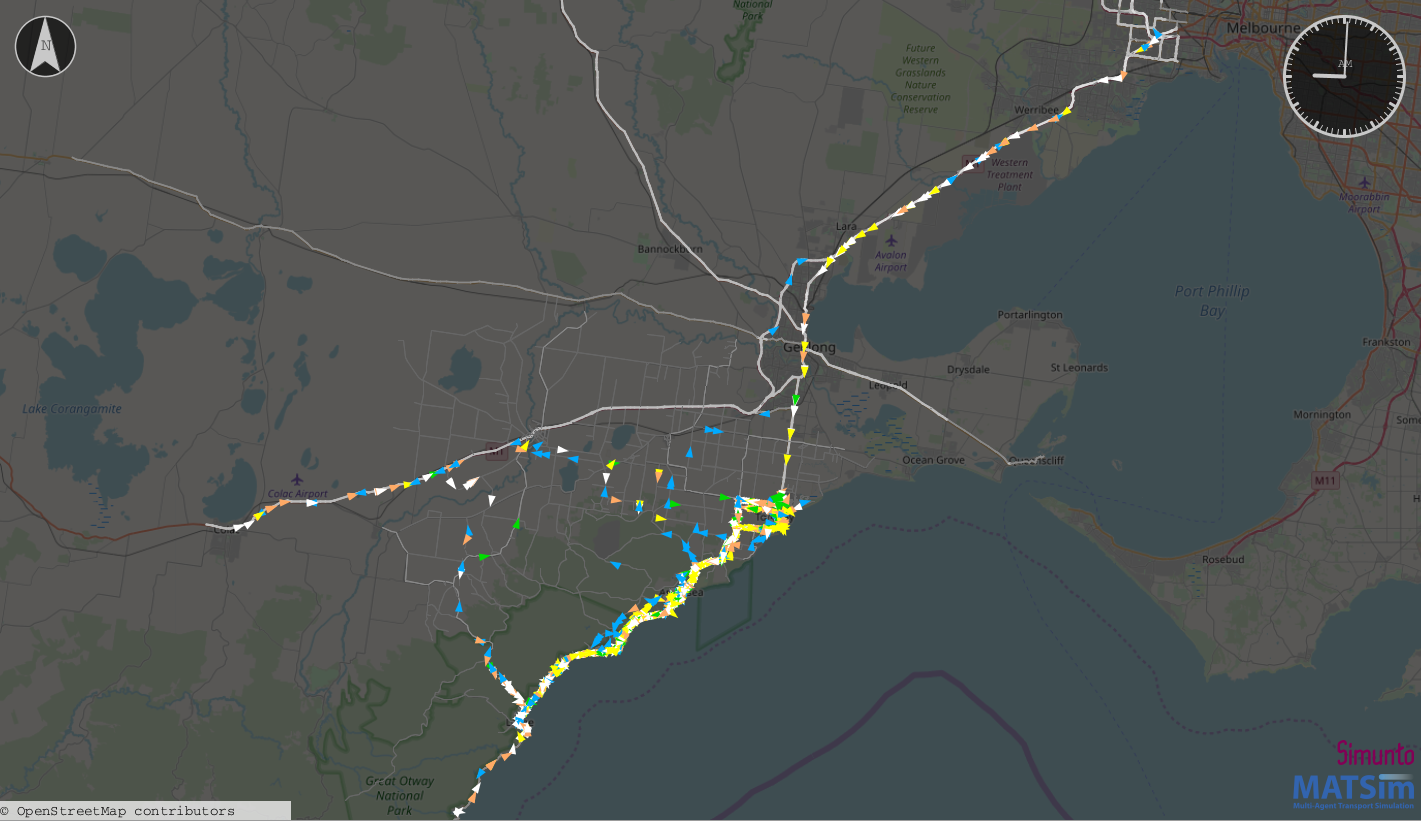}
  \caption{Here we have an overview of the entire region at 9am. We can see that a significant flow of traffic has already begun to come in from the East (from Melbourne), with smaller contributions also coming from the west.
  As we would expect, the most congested road in the region is along the coast (the Great Ocean Rd).
  Most of the work traffic seems to be focussed in the area we observed already in Torquay.}
  \label{fig:frogger}
\end{figure}

\begin{figure}[!htb]
  \includegraphics[width=\linewidth]{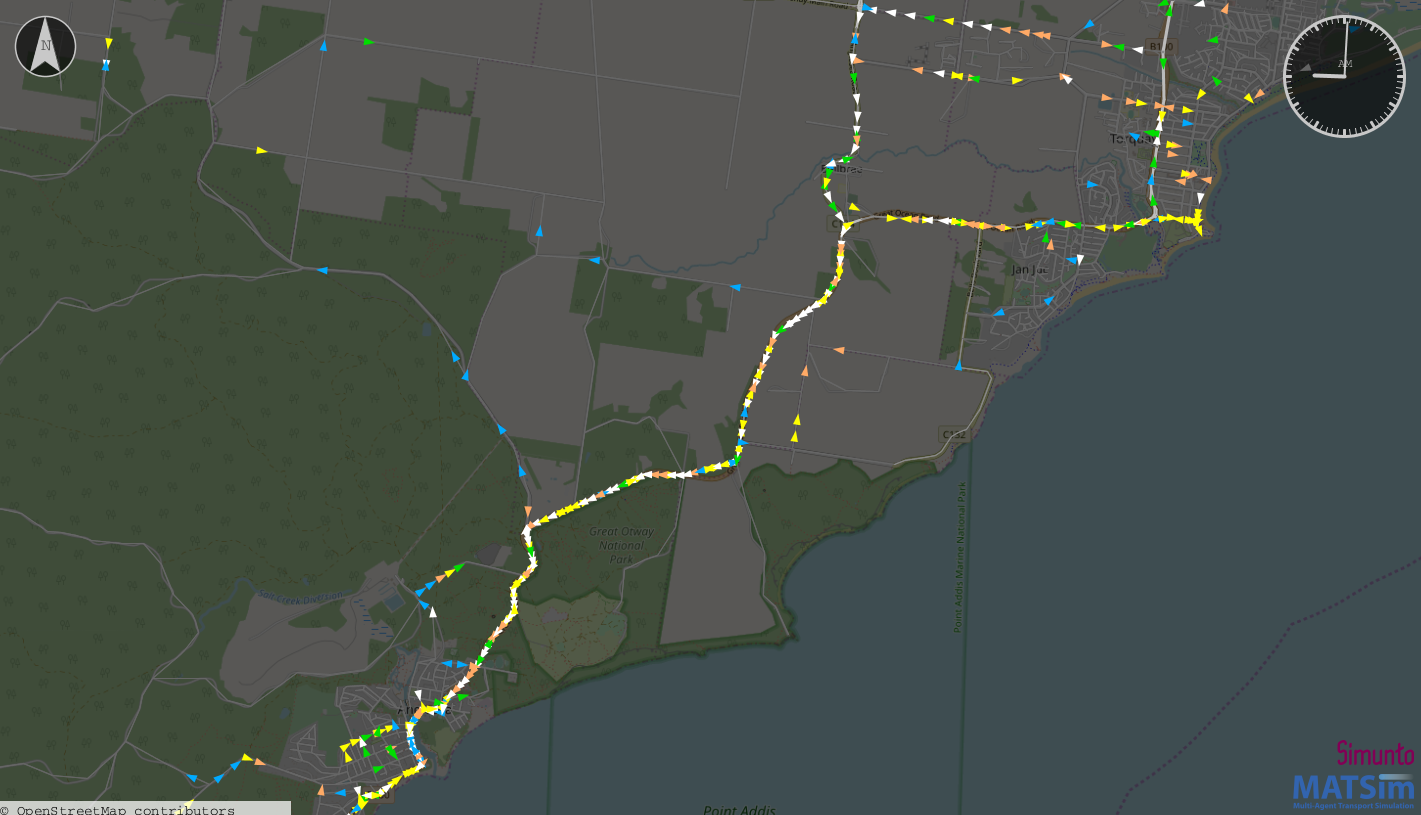}
  \caption{Zooming in on commuters between Anglesea and Torquay, we can see in greater detail how congested the coastal routes can get. Congestion within Anglesea seems comparatively light, and it seems the majority of travel is \textit{through} Anglesea rather than to or from it.}
  \label{fig:bogger}
\end{figure}

To verify that the background traffic is working as planned, we view snapshots of traffic flow in Figures ~\ref{fig:frogger} and ~\ref{fig:bogger}.
The colour mapping here is the same as previously, but now the triangles represent moving vehicles on the road that are travelling to a location to complete the activity corresponding to their colour.

%% file: figs/visdiffres.tex
\pgfplotstableread{
hour home work beach shops other
0.9999     0 -0.3125 0 0 0
1          0.04 0 00 00 0.2725
2.9999     0 -0.3125 0 0 0
03         0.1125 0 0 0 0.2
4.9999     0 0 0 0 -0.11
5          0.07 0.04 0 0 0
6.9999     0 -0.22 0 -0.22 0
7          0.33 0 0 0 0.11
8.9999     0 -0.075 -0.03 -0.045 0
9          0.075 0 0 0 0.075
10.9999    0 0 -0.1 -0.05 0
11         0.03 0 0 0 0.12
12.9999    -0.15 -0.35 0 0 0
13         0 0 0.05 0.45 0
14.9999    -0.05 -0.48 0 -0.22 0
15         0 0 0.75 0 0
16.9999    0 -0.2 -0.15 0 -0.05
17         0.3 0 0 0.1 0
18.9999    -4 0 0 -1.1 0
19         0 5.1 0 0 0.15
20.9999    -0.56 0 0 0 0
21         0 0.14 0 0 0.42
22.9999    -0.04 0 0 0 -0.11
23         0 0.15 0 0 0

}\loadedtable

\begin{tikzpicture}
  \begin{axis}
    [
      scale only axis, 
      title=Resident activity error (output-input),
      height=4.8cm,
      ylabel=Percentage,
      ylabel near ticks,
      xlabel=Time of day,
      width=0.8\textwidth,
      bar width = 20pt,
      ybar stacked,
      xtick=data,
      reverse legend,
      legend columns=-1,
      legend style={at={(0.5,-0.24)}, anchor=north}
    ]
    \addplot+[black!30,fill=blue!40] table[x=hour, y=other] {\loadedtable}; \addlegendentry{other};
    \addplot+[black!30,fill=orange!70] table[x=hour, y=shops] {\loadedtable}; \addlegendentry{shops};
    \addplot+[black!30,fill=yellow!100!black] table[x=hour, y=beach] {\loadedtable}; \addlegendentry{beach};
    \addplot+[black!30, fill=lime!85!black] table[x=hour, y=work] {\loadedtable}; \addlegendentry{work};
    \addplot+[black!30,fill=cyan!30] table[x=hour, y=home] {\loadedtable}; \addlegendentry{home};
  \end{axis}
\end{tikzpicture}

%% file: figs/visdiffvisday.tex
\pgfplotstableread{
hour home work beach shops other
0.9999     0 0 0 0 0
1          0 0 0 0 0
2.9999     0 0 0 0 0
03         0 0 0 0 0
4.9999     -0.05 0 -0.02 0 0
5          0 0 0 0 0.07
6.9999     -0.14 0 0 0 -0.38
7          0 0 .31 .21 0
8.9999     0 0 -0.09 -0.39 0
9          0.4 0 0 0 0.08
10.9999    -0.02 0 0 -0.28 -0.21
11         0 0 0.51 0 0
12.9999    0 0 -0.35 0 0
13         0.08 0 0 0.19 0.08
14.9999    0 0 -0.3 0 -0.4
15         0.26 0 0 0.44 0
16.9999    -0.68 0 -0.31 0 0
17         0 0 0 0.53 0.46
18.9999    0 0 0 0 -0.24
19         0.02 0 0.09 0.13 0
20.9999    -0.19 0 -0.08 0 0
21         0 0 0 0.19 0.08
22.9999    0 0 0 0 0
23         0 0 0 0 0

}\loadedtable

\begin{tikzpicture}
  \begin{axis}
    [
      scale only axis, 
      title=VisitorDaytime activity error (output-input),
      height=4.8cm,
      ylabel=Percentage,
      ylabel near ticks,
      xlabel=Time of day,
      width=0.8\textwidth,
      bar width = 20pt,
      ybar stacked,
      xtick=data,
      reverse legend,
      legend columns=-1,
      legend style={at={(0.5,-0.24)}, anchor=north}
    ]
    \addplot+[black!30,fill=blue!40] table[x=hour, y=other] {\loadedtable}; \addlegendentry{other};
    \addplot+[black!30,fill=orange!70] table[x=hour, y=shops] {\loadedtable}; \addlegendentry{shops};
    \addplot+[black!30,fill=yellow!100!black] table[x=hour, y=beach] {\loadedtable}; \addlegendentry{beach};
    \addplot+[black!30, fill=lime!85!black] table[x=hour, y=work] {\loadedtable}; \addlegendentry{work};
    \addplot+[black!30,fill=cyan!30] table[x=hour, y=home] {\loadedtable}; \addlegendentry{home};
  \end{axis}
\end{tikzpicture}

%% file: chapter-applications/subsec-attributes.tex
\clearpage
\subsection{Adding BDI Attributes}
We now append the attributes that will determine how each agent will act once their plans change due to a bushfire threat in the SCS region.
This involves first establishing sets $E$ and $\mathcal{T}$ of \texttt{Environmental} and \texttt{Transmitted} alerts, defining a ranking $r$ on $E\times\mathcal{T}$, and determining the parameters for each subgroup that will assign attributes.

\subsubsection{Alerts in the Surf Coast Shire}

There are any number of events that might trigger a person to evacuate during a bushfire, but here we focus only on two environmental cues, smoke and fire, and three core message types that may be issued by emergency services: `Advice', `Watch and Act', and `Evacuate Now'.
These warnings can be ordered in terms of severity, and we assign both types of alert values in $[0,1]$ as follows:
\begin{table}[!htbp]
\caption{Barometer scores for each alert.}
\centering
\begin{tabular}{c c c}
\hline\hline
\textsc{Environmental} &\textsc{Transmitted} & Value  \\ [0.5ex]
\hline
-&     Advice & 0.1 \\
-&Watch and Act& 0.2\\
Smoke & Evacuate Now &0.3 \\
Fire  & -& 0.4\\
[1ex]
\hline
\end{tabular}
\label{table:alerts} 
\end{table}

We then define the total barometer score  to be

$$
r(e,\tau)=e+\tau
$$

e.g. an agent who has both seen smoke and received a `Watch and Act' message will have a total barometer score of $0.5$.

There are a number of assumptions that we make with these messages.
Firstly, it is assumed that all messages are able to be sent to specific areas of the region.
We then also assume that any agent who is in that area when the message is sent will see the message.
The same holds for environmental alerts.
We assume that people only see and respond to the visual cues when they are within a 5km of the fire front (for smoke) or 1km (for fire).
Smoke is particularly hard to quantify in this sense, because in a bushfire it is very likely that people will be able to see smoke from further than 5km away.
But seeing smoke from afar is also not as likely to invoke a response.
We may conceptualise the smoke alert here as the point when smoke begins to interfere with people's line of sight.
Indeed, it may be useful in future work to have an additional, lesser, smoke alert which represents this smoke awareness.

\subsubsection{Attribute Parameters}
The only other inputs that the BDI model requires are the per-subgroup parameters which determine their attributes.
These are:
\begin{itemize}
  \item $\texttt{prob\_of\_dependant}_s$: the probability that an $s-agent$ has a dependant.
  \item $\texttt{stay}_s$: \texttt{true} if the strict inequality \texttt{INIT}$<$\texttt{ACT} is possible i.e. determines whether an $s$-agent can stay and defend.
  \item $\texttt{prob\_of\_go\_home}_s$: the probability that an $s$-agent will return home before leaving.
  \item $[\texttt{threshold\_min}_s$,$\texttt{threshold\_max}_s]$: the interval from which the two threshold score \texttt{INIT} and \texttt{ACT} are drawn from.
\end{itemize}

The values that we use for these variables here are given in Table~\ref{table:attributes}.

\begin{table}[!htbp]
\caption{Attribute parameters for each subgroup.} \label{table:attributes} 
\centering
\begin{tabular}{|p{3cm}|p{2cm}|p{2cm}|p{2cm}|}
\hline\hline
Subgroup & \texttt{prob\_of\_dependant} &  \texttt{prob\_of\_go\_home}& \texttt{stay}\\ [0.5ex]
\hline
\texttt{Resident}&  0.3  & 0.5&1\\
\texttt{ResidentPartTime}& 0.05&0.4&1\\
\texttt{VisitorRegular}& 0& 0.4&1\\
\texttt{VisitorOvernight}&0&0.8&0\\
\texttt{VisitorDaytime}&0 & 0 &0\\
\hline\hline
Subgroup & \texttt{threshold\_min}&\texttt{threshold\_max}&\\ [0.5ex]
\hline
\texttt{Resident}& 0.1&0.8&\\
\texttt{ResidentPartTime}& 0.1 & 0.6& \\
\texttt{VisitorRegular}& 0.1& 0.4&\\
\texttt{VisitorOvernight}&0.2&0.4 & \\
\texttt{VisitorDaytime}& 0.3&0.7 &\\[1ex]
\hline
\end{tabular}

\end{table}
\begin{figure}[!htbp]
  \centering
  \caption{An agent plan with attributes appended.}\label{fig:plan-xml}
  \input{figs/plan-attribute-xml.tex}
\end{figure}

With these inputs, we can now add the BDI attributes.
Using a separate R script, we assign Boolean values to each agents' plan that reflect whether it has a dependant and whether it will go home (note that this decision depends on whether the agent has a dependant).
We also determine a \texttt{INIT} and \texttt{ACT} tolerance and choose a location for the dependant (if applicable), as well as evacuation and invacuation preference locations.
These last two are determined by a separate \texttt{Refuges.shp} file provided by SCSC.
The resulting attributes appear at the start of the agent's MATSim plan, as shown in Figure~\ref{fig:plan-xml}.

Note that the blank \texttt{HasdependantsAtLocation} field occurs whenever the agent does not have a dependant.
The \texttt{InitialResponseThreshold} and \texttt{FinalResponseThreshold} fields represent \texttt{INIT} and \texttt{ACT} respectively.
\texttt{WillGoHomeAfterVisitingDependants} and \texttt{WillGoHomeBeforeLeaving} encompass two variations of the same behaviour, and at most one of them will be \texttt{true}-- this is determined via the \texttt{prob\_of\_go\_home} parameter.

%% file: figs/plan-attribute-xml.tex
\begin{lstlisting}[%
language=XML,
numbers=none]

<person id= "11154" >
  <attributes>
    <attribute name="BDIAgentType" class="java.lang.String" >io.github.agentsoz.ees.agents.bushfire.ResidentPartTime</attribute>
    <attribute name="HasDependantsAtLocation" class="java.lang.String" ></attribute>
    <attribute name="InitialResponseThreshold" class="java.lang.Double" >0.5</attribute>
    <attribute name="FinalResponseThreshold"   class="java.lang.Double" >0.6</attribute>
    <attribute name="WillGoHomeAfterVisitingDependants" class="java.lang.Boolean" >false</attribute>
    <attribute name="WillGoHomeBeforeLeaving" class="java.lang.Boolean" >false</attribute>
    <attribute name="EvacLocationPreference" class="java.lang.String">Lorne 34,759271.09371883,5729851.2967029</attribute>
    <attribute name="InvacLocationPreference" class="java.lang.String">,794020.893984802,5771845.08224784</attribute>
  </attributes>
  <plan selected="yes">
    <activity type="home" x="794363.976349215" y="5771367.18274888" end_time="08:04:00" />
    <leg mode="car" />
    <activity type="shops" x="790524.047786298" y="5752219.83828972" end_time="10:44:00" />
    <leg mode="car" />
    <activity type="other" x="789816.421674447" y="5753213.33592403" end_time="12:49:00" />
    <leg mode="car" />
    <activity type="work" x="789778.144754022" y="5752608.5932469" end_time="16:01:00" />
    <leg mode="car" />
    <activity type="shops" x="771117.511429056" y="5738680.57730346" end_time="19:06:00" />
    <leg mode="car" />
    <activity type="home" x="794363.976349215" y="5771367.18274888" />
  </plan>
</person>
  \end{lstlisting}

%% file: chapter-applications/subsec-application.tex
\subsection{SCS Application}

\begin{figure}[!htbp]
  \centering
  \includegraphics[scale=0.2]{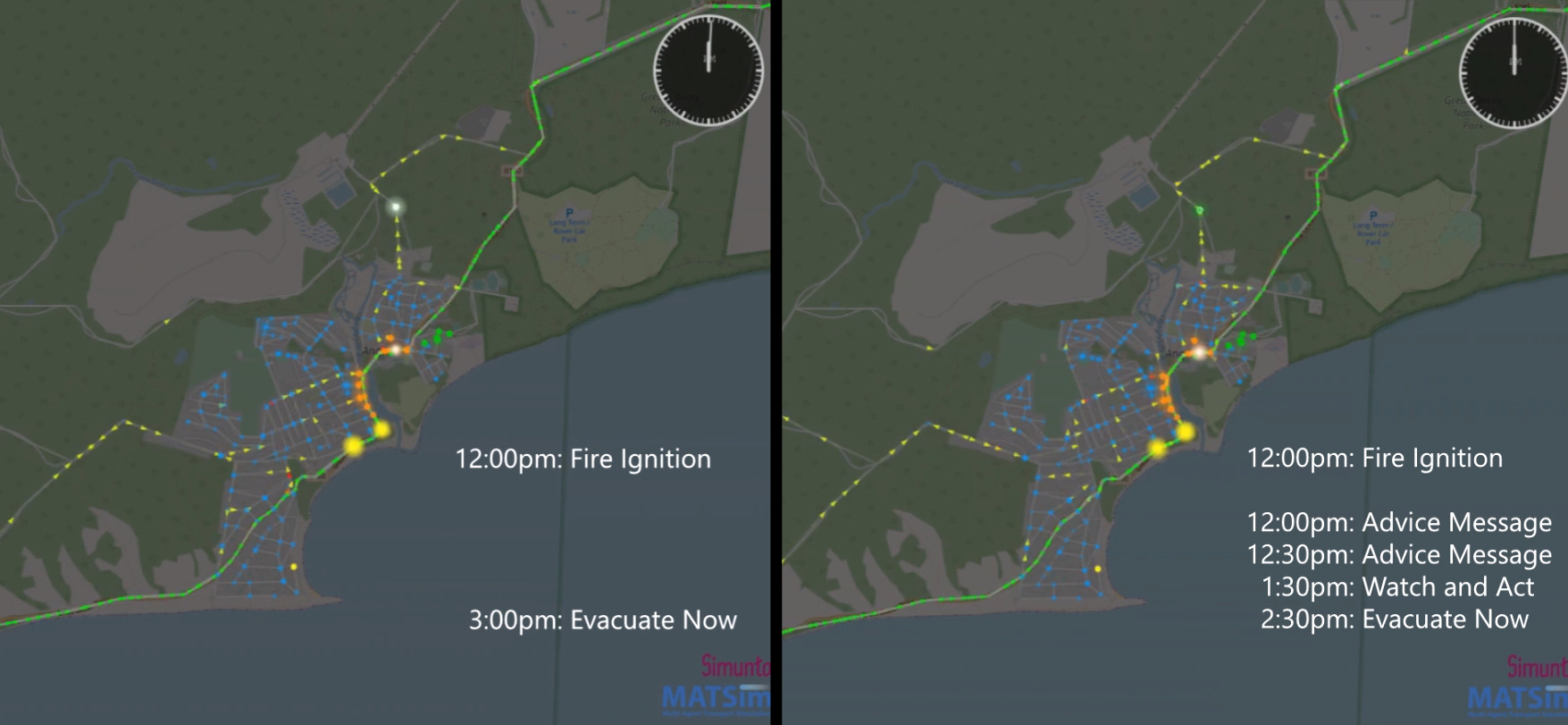}
  \caption{At 12pm when the fire ignites, the state is exactly the same in each scenario. We observe the familiar flow of traffic through Anglesea, and that there are still many people at \texttt{home}. As was the case in Torquay, midday sees a high level of \texttt{shops} and \texttt{beach} activity around the coastal strip. }\label{fig:comp-1200}
 \end{figure}

\begin{figure}[!htbp]
  \centering
    \includegraphics[scale=0.2]{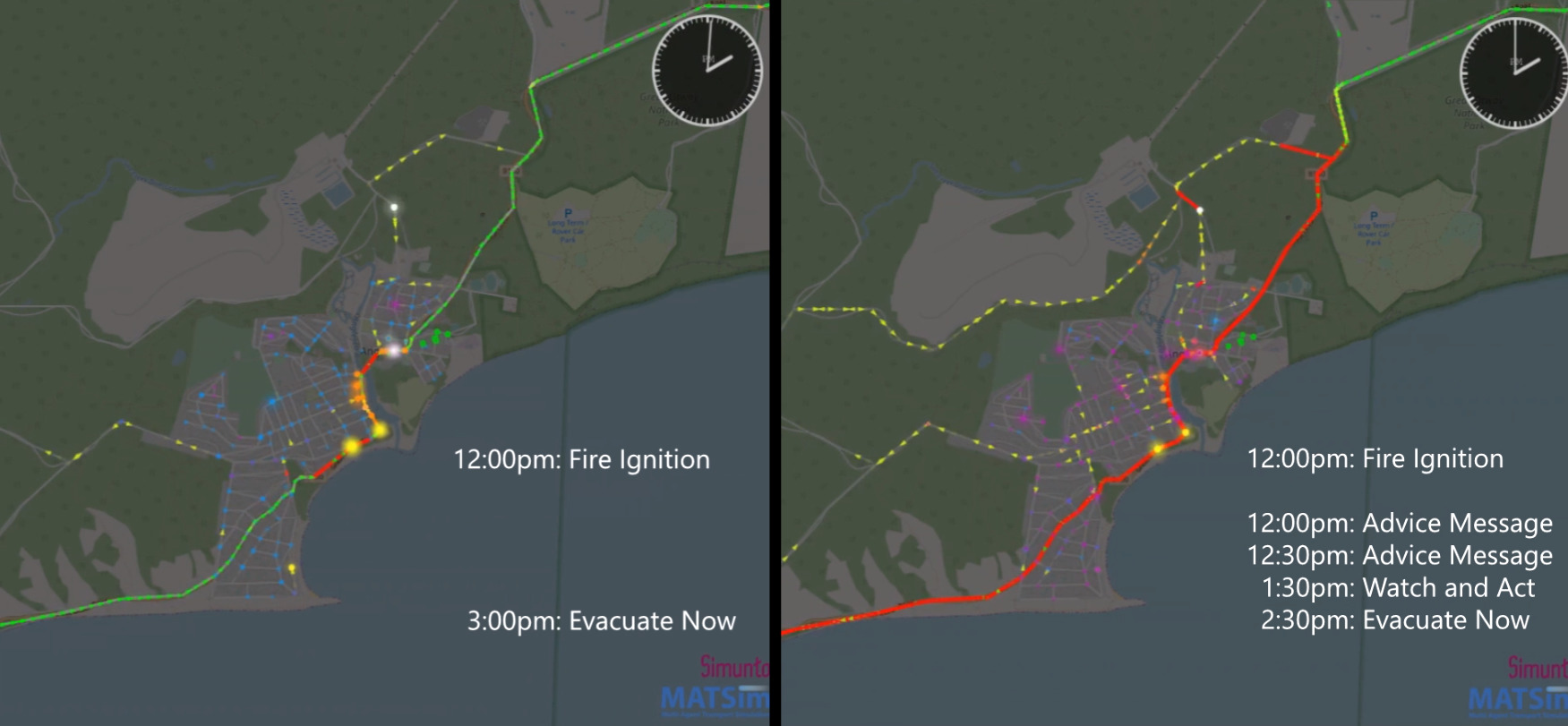}
  \caption{By 2pm, we see the two scenarios have diverged significantly. On the left, traffic has increased in the middle of Anglesea proper, but the traffic flow in and out has not changed.
  The purple activity location nodes indicate that some have decided to take action, but a lot of people are still at \texttt{home} oblivious to or unconcerned by the threat.
  On the right side, we see that congestion increases as people attempt to evacuate from the danger zone.
  Three messages have already been issued, with the severity raised at 1.30pm. There are hardly any people left in their homes, except those that have returned there to prepare and/or defend.
   }\label{fig:comp-1400}
\end{figure}

\begin{figure}[!htbp]
  \centering
    \includegraphics[scale=0.2]{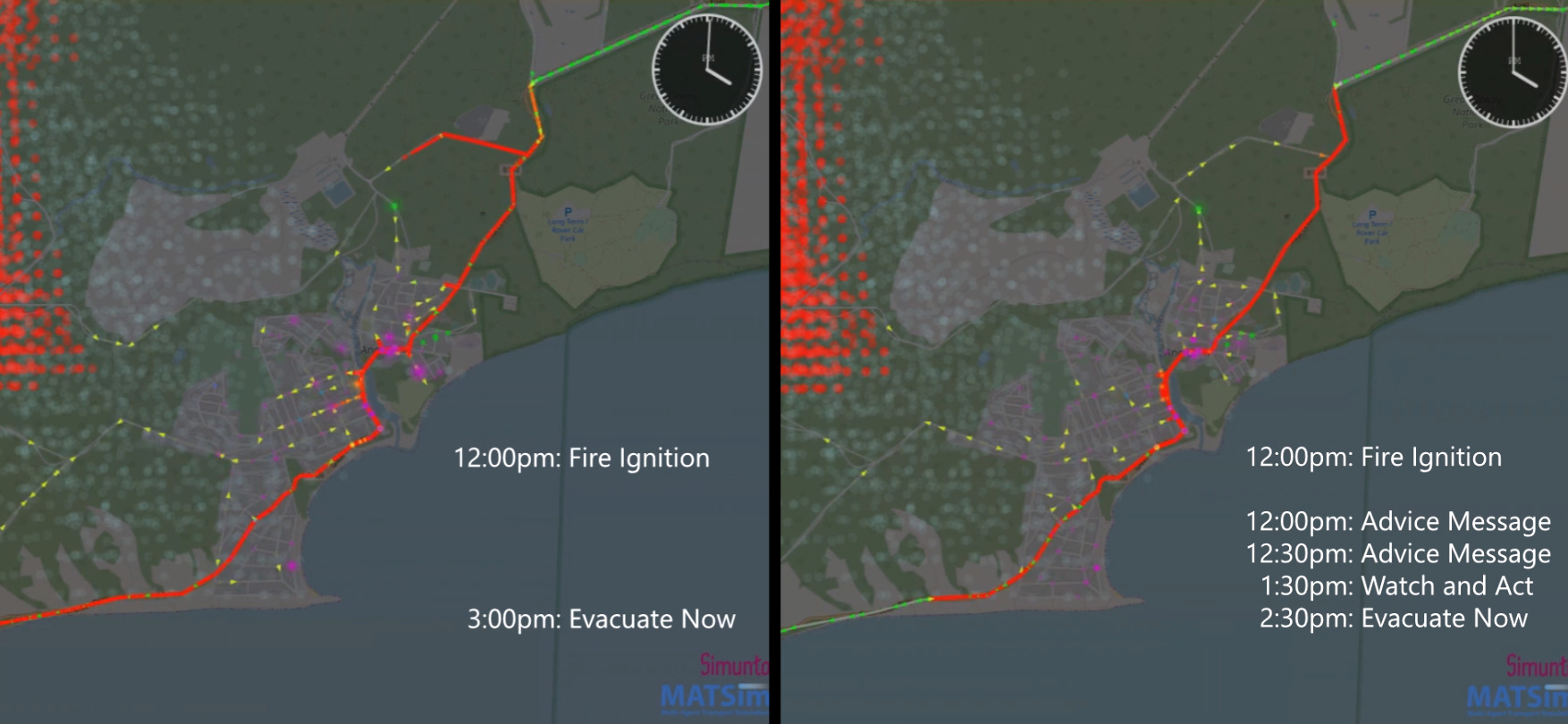}
  \caption{By 4pm, we begin to see the fire encroach upon the outskirts of Anglesea. The smoke front will now have alerted most people that they need to act even if the messages did not. The community on the left has by now received an `Evacuate Now' message, and we see that they are roughly at the level of congestion stage that the other community was two hours ago.
  In the scenario on the right, the congestion has eased somewhat by now towards the south of Anglesea, and pretty much everyone, bar a few defenders, has left or is trying to leave.}\label{fig:comp-1600}
\end{figure}

\begin{figure}[!htbp]
  \centering
    \includegraphics[scale=0.2]{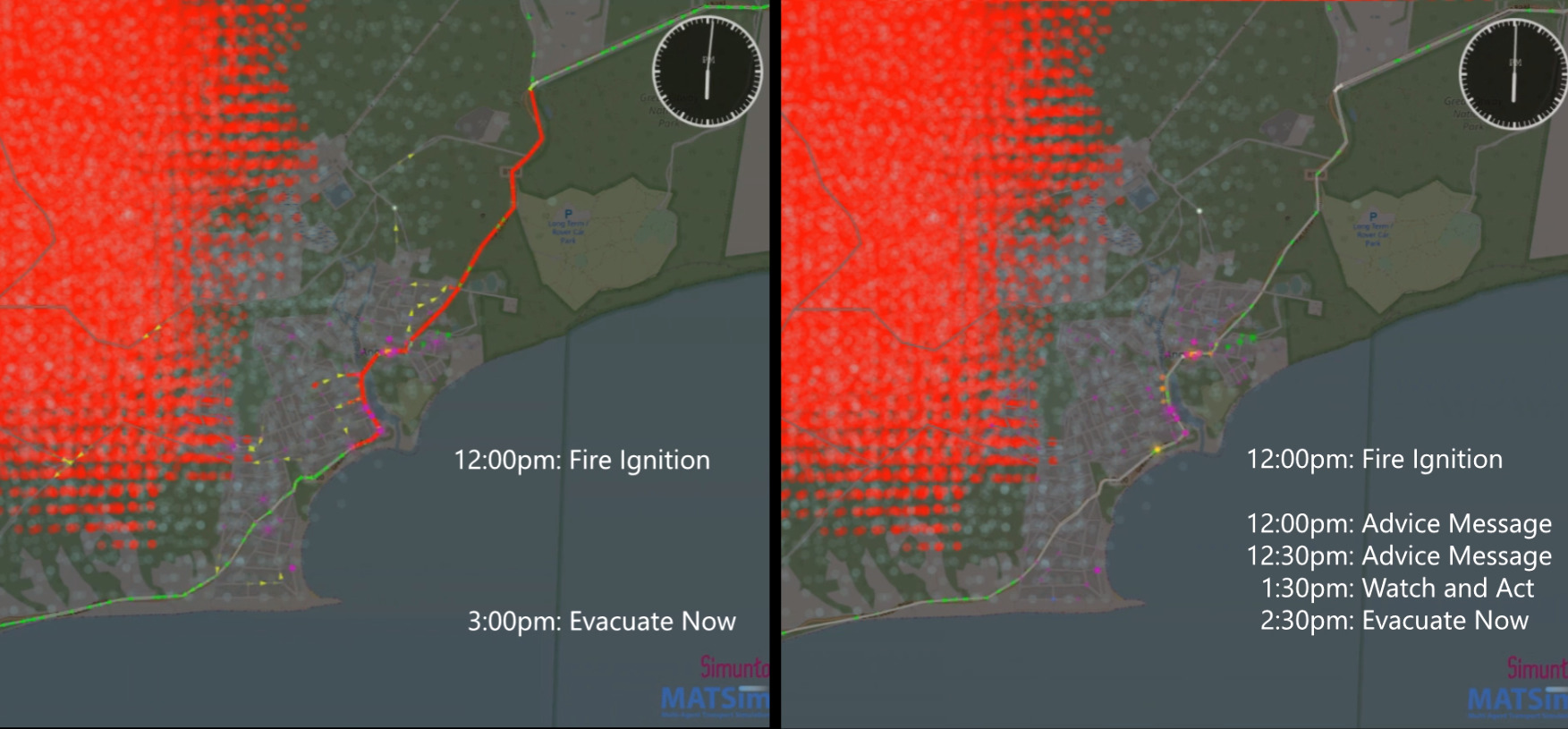}
  \caption{The fire is now engulfing the edge of the town. By 6pm, everyone in the community on the right who wanted to escape has done so, and there is no congestion on the roads leaving Anglesea. On the left, however, the situation is far more urgent, with lots of people still stuck on the road as the fire approaches. }\label{fig:comp-1800}
\end{figure}

To fully demonstrate the usefulness of the algorithm, we apply the plans with attributes in a bushfire evacuation simulation.
Using the EES simulation model, this will show both the importance of having the population realistically situated and engaged in daily plans, and the ability for BDI attributes to directly impact an agent's response to a bushfire threat.
In the following scenario, the response of a community in Anglesea is tested under two different scenarios.
Each scenario has the same generated plan and the same bushfire threat that ignites at midday and eventually encroaches into parts of the town.
In one test, the agents are left to respond to the fire on their own (with an eventual `Evacuate Now' warning at 3pm).
In the other, a sequence of messages of increasing severity are delivered to the population, directing them to act earlier.
We will proceed again with a series of snapshots.
This time, vehicles are coloured on a spectrum based on their speed in the network, where green represents free-speed and red represents heavy congestion.
The previous activity location node colours are also present in the background.
We proceed via Figures ~\ref{fig:comp-1200}-\ref{fig:comp-1800}.

\clearpage

Whether or not the stark contrast between outcomes in the  previous EES example totally reflects the effectiveness that emergency warnings can have, we did see the capability of agents to take in information from various sources and react to a threat accordingly.
Further, we saw variation in behaviour, even within each scenario, where a fair number of local agents decided to stay and defend, at least until it became clear that the danger was too great.
The movement of people prior to the evacuation situation was also important, as it meant that the main arterial roads out of Anglesea were already busy.
This affected the ability of people to evacuate, particularly on the left scenario.
Thus this application provides some useful insights into what benefits our population generation model with appended BDI attributes can provide.

%
%
%
%
%

%% file: chapter-applications/subsec-discussion.tex
\clearpage
\subsection{Discussion}
The error tables (Tables~\ref{table:Rdiff} and \ref{table:VDdiff}) reveal that variation in activity durations can affect the ability of the algorithm to reproduce the input distributions.
Whilst the algorithm takes into account the durations of each activity and adjusts the probability matrix accordingly, it still relies on the input distributions giving a `feasible path' for agents to complete activities according to their durations.
If the distribution of people working drops off too dramatically from one time step to the next, the algorithm may not be able to shift enough agents away from the \texttt{work} activity.
This is because the distributions are essentially decoupled from activity durations.
One solution is to smooth the changes between time-steps in the distribution.
However, if the distribution data does accurately reflect the whereabouts of the population, then modifying them essentially results in a discrepancy anyway, so this process should be minimised.
The other alternative is to make small adjustments to the durations of each activity, in conjunction with smaller time-steps (i.e. more detailed activity distributions).
This requires a more complicated set of distribution inputs, and these may be difficult or impossible to accurately obtain.
We therefore must also weigh the need for the output to match the distributions exactly against the accuracy of the distributions themselves and the appropriate duration of each activity.
If there was no level of duration control, then total correspondence with the distributions would be attained (agents are always where they should be) but with the drawback that there will be more movement between activities (because each agent can change activity at every time-step).
This may result in background traffic that does not realistically represent the region.
To strike an appropriate balance between these needs, a maximum error of $\sim5\%$ was decided upon, which we observe in the tables.
\

One potential improvement to the population generation process would be to apply MATsim's iterative re-planning capabilities to the generated \texttt{population.xml} file.
The agent plans will then converge to a traffic equilibrium, ensuring that the daily plans we apply prior to the bushfire threat are collectively feasible and optimised at the individual level.
However, this process could serve to compound the error that we observe above, as the iterative process would likely diverge away from the idealised expected version of events presented by the input distributions.
Similarly, attempting to calibrate the model outputs to actual traffic counts would add to the tension that exists between matching the static inputs to a dynamic output process that is designed to replicate an even more complex reality.
This calibration approach is planned for future work, and will likely see the algorithm modified to factor in these traffic counts as part of the generation process.
\

The alert ranking system provides a way of ordering alerts that is easy to understand, but is somewhat arbitrary in the way it orders the possible combinations of environmental and transmitted alerts.
It results in some equivalences that may not exist, e.g.

$$ (\text{Smoke}, \text{Advice})\equiv (\text{Fire},0)$$

A more thorough ranking system would explicitly order every possible combination of environmental and transmitted alert.
With the number of alerts we consider during this demonstration this would be a simple task, but in practice there will be other alerts of both types that may or may not be used in a given scenario.
This current system may be viewed as a useful temporary measure to allow quick interchanging or introduction of new alerts without necessarily needing to place every new combination into some order.
Once the full set of alerts is established, a more concrete ranking system can be applied.
Some other potential environmental alerts are observing embers, observing neighbours leaving and aural cues.
There are also a number possible of transmitted messages that are possible, often as variations of the three included here.

\

%% file: sec-conclusion.tex
\section{Conclusion}
Bushfires threaten communities every year, and due to their sometimes rapid onset and unpredictable nature it is imperative that at-risk regions prepare themselves to respond.
Planning that accounts for nuances in a specific population and region can help improve resilience and preparedness of both the authorities and the community in the face of a threat.
In this research we have presented a method for capturing these dynamics and measuring their significance through bushfire evacuation simulation.

Our contribution is an algorithm that takes a set of activity profiles and generates agent daily plans in the form of a MATSim \texttt{population.xml} file.
 The algorithm was designed to generate synthetic populations for regions that are not easily approximated by incremental sampling techniques due to an actual population that has a significant transient element.
 Further, we have implemented this algorithm as a tool to generate populations for emergency evacuation simulations. This requires a set of flexible inputs that can be modified to suit a particular scenario and that are be easily understandable to domain experts.

 We have also utilised the BDI cognitive framework to construct a bushfire behaviour model that allows our generated populations to reason and respond to a bushfire threat. Using threshold parameters, we have introduced a method to vary these responses across our populations in a manner that mimics the way that humans process new information. Input attributes of this bushfire behaviour model have been integrated into the population generation algorithm to provide a single tool that can create synthetic populations for a wide range of different bushfire scenarios.

 Lastly, we have applied this tool to Surf Coast Shire in Victoria and tested it in a number of scenarios. The algorithm has contributed to the wider project by bringing a dynamic element to the notion of evacuation messaging, and the behaviour model will allow emergency personnel to test different approaches to bushfire response when the population has a varying level of engagement.
 Ultimately it is hoped that the algorithm and associated model presented here will improve understanding of the risks associated with bushfire evacuation.

\subsection{Future Work}
This research is part of a larger project looking at improving the evacuation planning process for at-risk communities.
This thesis presents work to date in the wider context of this project, with the ultimate aim being to develop a model that gives a realistic representation of any population, both in the form of their daily plans and their eventual bushfire responses.
Most of the input data used in Section 4 has only been informally validated and does not necessarily reflect the actual make-up of the population in the Surf Coast Shire.
In this sense, much future work will be devoted to ensuring that the model is validated and calibrated.
For the algorithm, this will mean using traffic counts and travel times to match observed data with the movements of the synthetic population.
For the behavioural model, we plan to continue to work with behavioural experts, especially in the area of behaviour profiles and archetypes \citep{StrahanSelfevacuationarchetypesAustralian2018}.
On particular area of interest is the introduction of social networks  as another means for agents to become aware of the threat level. This could be implemented as a third alert type in the threshold model, and allow a bigger set of alert combinations to diversify the kinds of possible responses further.

In terms of application, the  project  aims to encompass the whole of Victoria.
The required inputs therefore must be easily obtainable and general enough to apply to each region, whilst still maintaining the ability to distinguish communities by their demographic make-up.
This broader approach will also necessitate that the model allows for interaction between regions, and how evacuation movements in one place can have a flow-on effect in another.

%% file: ms.bbl
\begin{thebibliography}{}

\bibitem [\protect \citeauthoryear {%
Adam%
, Danet%
, Thangarajah%
\BCBL {}\ \BBA {} Dugdale%
}{%
Adam%
\ \protect \BOthers {.}}{%
{\protect \APACyear {2016}}%
}]{%
AdamBdimodellingsimulation2016}
\APACinsertmetastar {%
AdamBdimodellingsimulation2016}%
\begin{APACrefauthors}%
Adam, C.%
, Danet, G.%
, Thangarajah, J.%
\BCBL {}\ \BBA {} Dugdale, J.%
\end{APACrefauthors}%
\unskip\
\newblock
\APACrefYearMonthDay{2016}{}{}.
\newblock
{\BBOQ}\APACrefatitle {{{BDI}} Modelling and Simulation of Human Behaviours in
  Bushfires} {{{BDI}} modelling and simulation of human behaviours in
  bushfires}.{\BBCQ}
\newblock
\BIn{} \APACrefbtitle {Information {{Systems}} for {{Crisis Response}} and
  {{Management}} in {{Mediterranean Countries}}.} {Information {{Systems}} for
  {{Crisis Response}} and {{Management}} in {{Mediterranean Countries}}.}
\newblock
\APACaddressPublisher{Madrid}{{Springer Germany}}.
\PrintBackRefs{\CurrentBib}

\bibitem [\protect \citeauthoryear {%
Adam%
\ \BBA {} Gaudou%
}{%
Adam%
\ \BBA {} Gaudou%
}{%
{\protect \APACyear {2017}}%
}]{%
AdamModellingHumanBehaviours2017}
\APACinsertmetastar {%
AdamModellingHumanBehaviours2017}%
\begin{APACrefauthors}%
Adam, C.%
\BCBT {}\ \BBA {} Gaudou, B.%
\end{APACrefauthors}%
\unskip\
\newblock
\APACrefYearMonthDay{2017}{}{}.
\newblock
{\BBOQ}\APACrefatitle {Modelling {{Human Behaviours}} in {{Disasters}} from
  {{Interviews}}: {{Application}} to {{Melbourne Bushfires}}} {Modelling
  {{Human Behaviours}} in {{Disasters}} from {{Interviews}}: {{Application}} to
  {{Melbourne Bushfires}}}.{\BBCQ}
\newblock
\APACjournalVolNumPages{Journal of Artificial Societies and Social
  Simulation}{20}{3}{}.
\newblock
\begin{APACrefDOI} \doi{10.18564/jasss.3395} \end{APACrefDOI}
\PrintBackRefs{\CurrentBib}

\bibitem [\protect \citeauthoryear {%
J\BPBI E.~Anderson%
}{%
J\BPBI E.~Anderson%
}{%
{\protect \APACyear {2011}}%
}]{%
AndersonGravityModel2011}
\APACinsertmetastar {%
AndersonGravityModel2011}%
\begin{APACrefauthors}%
Anderson, J\BPBI E.%
\end{APACrefauthors}%
\unskip\
\newblock
\APACrefYearMonthDay{2011}{}{}.
\newblock
{\BBOQ}\APACrefatitle {The {{Gravity Model}}} {The {{Gravity Model}}}.{\BBCQ}
\newblock
\APACjournalVolNumPages{Annual Review of Economics}{3}{1}{133-160}.
\newblock
\begin{APACrefDOI} \doi{10.1146/annurev-economics-111809-125114}
  \end{APACrefDOI}
\PrintBackRefs{\CurrentBib}

\bibitem [\protect \citeauthoryear {%
J\BPBI R.~Anderson%
}{%
J\BPBI R.~Anderson%
}{%
{\protect \APACyear {1996}}%
}]{%
AndersonACTsimpletheory1996}
\APACinsertmetastar {%
AndersonACTsimpletheory1996}%
\begin{APACrefauthors}%
Anderson, J\BPBI R.%
\end{APACrefauthors}%
\unskip\
\newblock
\APACrefYearMonthDay{1996}{}{}.
\newblock
{\BBOQ}\APACrefatitle {{{ACT}}: {{A}} Simple Theory of Complex Cognition}
  {{{ACT}}: {{A}} simple theory of complex cognition}.{\BBCQ}
\newblock
\APACjournalVolNumPages{American Psychologist}{51}{4}{355-365}.
\newblock
\begin{APACrefDOI} \doi{10.1037/0003-066X.51.4.355} \end{APACrefDOI}
\PrintBackRefs{\CurrentBib}

\bibitem [\protect \citeauthoryear {%
Antonovics%
, Iwasa%
\BCBL {}\ \BBA {} Hassell%
}{%
Antonovics%
\ \protect \BOthers {.}}{%
{\protect \APACyear {1995}}%
}]{%
AntonovicsGeneralizedModelParasitoid1995}
\APACinsertmetastar {%
AntonovicsGeneralizedModelParasitoid1995}%
\begin{APACrefauthors}%
Antonovics, J.%
, Iwasa, Y.%
\BCBL {}\ \BBA {} Hassell, M\BPBI P.%
\end{APACrefauthors}%
\unskip\
\newblock
\APACrefYearMonthDay{1995}{{\APACmonth{05}}}{}.
\newblock
{\BBOQ}\APACrefatitle {A {{Generalized Model}} of {{Parasitoid}}, {{Venereal}},
  and {{Vector}}-{{Based Transmission Processes}}} {A {{Generalized Model}} of
  {{Parasitoid}}, {{Venereal}}, and {{Vector}}-{{Based Transmission
  Processes}}}.{\BBCQ}
\newblock
\APACjournalVolNumPages{The American Naturalist}{145}{5}{661-675}.
\newblock
\begin{APACrefDOI} \doi{10.1086/285761} \end{APACrefDOI}
\PrintBackRefs{\CurrentBib}

\bibitem [\protect \citeauthoryear {%
Auchincloss%
\ \BBA {} Garcia%
}{%
Auchincloss%
\ \BBA {} Garcia%
}{%
{\protect \APACyear {2015}}%
}]{%
AuchinclossBriefintroductoryguide2015}
\APACinsertmetastar {%
AuchinclossBriefintroductoryguide2015}%
\begin{APACrefauthors}%
Auchincloss, A\BPBI H.%
\BCBT {}\ \BBA {} Garcia, L\BPBI M\BPBI T.%
\end{APACrefauthors}%
\unskip\
\newblock
\APACrefYearMonthDay{2015}{{\APACmonth{11}}}{}.
\newblock
{\BBOQ}\APACrefatitle {Brief Introductory Guide to Agent-Based Modeling and an
  Illustration from Urban Health Research} {Brief introductory guide to
  agent-based modeling and an illustration from urban health research}.{\BBCQ}
\newblock
\APACjournalVolNumPages{Cadernos de saude publica}{31}{Suppl 1}{65-78}.
\newblock
\begin{APACrefDOI} \doi{10.1590/0102-311X00051615} \end{APACrefDOI}
\PrintBackRefs{\CurrentBib}

\bibitem [\protect \citeauthoryear {%
Axelrod%
}{%
Axelrod%
}{%
{\protect \APACyear {1997}}%
}]{%
AxelrodAdvancingArtSimulation1997}
\APACinsertmetastar {%
AxelrodAdvancingArtSimulation1997}%
\begin{APACrefauthors}%
Axelrod, R.%
\end{APACrefauthors}%
\unskip\
\newblock
\APACrefYearMonthDay{1997}{}{}.
\newblock
{\BBOQ}\APACrefatitle {Advancing the {{Art}} of {{Simulation}} in the {{Social
  Sciences}}} {Advancing the {{Art}} of {{Simulation}} in the {{Social
  Sciences}}}.{\BBCQ}
\newblock
\BIn{} R.~Conte, R.~Hegselmann\BCBL {}\ \BBA {} P.~Terna\ (\BEDS),
  \APACrefbtitle {Simulating {{Social Phenomena}}} {Simulating {{Social
  Phenomena}}}\ (\BPG~21-40).
\newblock
\APACaddressPublisher{}{{Springer Berlin Heidelberg}}.
\PrintBackRefs{\CurrentBib}

\bibitem [\protect \citeauthoryear {%
Ballas%
\ \protect \BOthers {.}}{%
Ballas%
\ \protect \BOthers {.}}{%
{\protect \APACyear {2005}}%
}]{%
BallasSimBritainspatialmicrosimulation2005}
\APACinsertmetastar {%
BallasSimBritainspatialmicrosimulation2005}%
\begin{APACrefauthors}%
Ballas, D.%
, Clarke, G.%
, Dorling, D.%
, Eyre, H.%
, Thomas, B.%
\BCBL {}\ \BBA {} Rossiter, D.%
\end{APACrefauthors}%
\unskip\
\newblock
\APACrefYearMonthDay{2005}{{\APACmonth{01}}}{}.
\newblock
{\BBOQ}\APACrefatitle {{{SimBritain}}: A Spatial Microsimulation Approach to
  Population Dynamics} {{{SimBritain}}: A spatial microsimulation approach to
  population dynamics}.{\BBCQ}
\newblock
\APACjournalVolNumPages{Population, Space and Place}{11}{1}{13-34}.
\newblock
\begin{APACrefDOI} \doi{10.1002/psp.351} \end{APACrefDOI}
\PrintBackRefs{\CurrentBib}

\bibitem [\protect \citeauthoryear {%
Barrett%
, Ran%
\BCBL {}\ \BBA {} Pillai%
}{%
Barrett%
\ \protect \BOthers {.}}{%
{\protect \APACyear {2000}}%
}]{%
BarrettDevelopingDynamicTraffic2000}
\APACinsertmetastar {%
BarrettDevelopingDynamicTraffic2000}%
\begin{APACrefauthors}%
Barrett, B.%
, Ran, B.%
\BCBL {}\ \BBA {} Pillai, R.%
\end{APACrefauthors}%
\unskip\
\newblock
\APACrefYearMonthDay{2000}{{\APACmonth{01}}}{}.
\newblock
{\BBOQ}\APACrefatitle {Developing a {{Dynamic Traffic Management Modeling
  Framework}} for {{Hurricane Evacuation}}} {Developing a {{Dynamic Traffic
  Management Modeling Framework}} for {{Hurricane Evacuation}}}.{\BBCQ}
\newblock
\APACjournalVolNumPages{Transportation Research Record: Journal of the
  Transportation Research Board}{1733}{}{115-121}.
\newblock
\begin{APACrefDOI} \doi{10.3141/1733-15} \end{APACrefDOI}
\PrintBackRefs{\CurrentBib}

\bibitem [\protect \citeauthoryear {%
Bazzan%
, Wahle%
\BCBL {}\ \BBA {} Kl\"ugl%
}{%
Bazzan%
\ \protect \BOthers {.}}{%
{\protect \APACyear {1999}}%
}]{%
BazzanAgentsTrafficModelling1999}
\APACinsertmetastar {%
BazzanAgentsTrafficModelling1999}%
\begin{APACrefauthors}%
Bazzan, A\BPBI L\BPBI C.%
, Wahle, J.%
\BCBL {}\ \BBA {} Kl\"ugl, F.%
\end{APACrefauthors}%
\unskip\
\newblock
\APACrefYear{1999}.
\newblock
\APACrefbtitle {Agents in {{Traffic Modelling}} - from {{Reactive}} to {{Social
  Behaviour}}} {Agents in {{Traffic Modelling}} - from {{Reactive}} to {{Social
  Behaviour}}}.
\PrintBackRefs{\CurrentBib}

\bibitem [\protect \citeauthoryear {%
Bellifemine%
, Poggi%
\BCBL {}\ \BBA {} Rimassa%
}{%
Bellifemine%
\ \protect \BOthers {.}}{%
{\protect \APACyear {2001}}%
}]{%
BellifemineDevelopingmultiagentsystems2001}
\APACinsertmetastar {%
BellifemineDevelopingmultiagentsystems2001}%
\begin{APACrefauthors}%
Bellifemine, F.%
, Poggi, A.%
\BCBL {}\ \BBA {} Rimassa, G.%
\end{APACrefauthors}%
\unskip\
\newblock
\APACrefYearMonthDay{2001}{}{}.
\newblock
{\BBOQ}\APACrefatitle {Developing Multi-Agent Systems with a {{FIPA}}-Compliant
  Agent Framework} {Developing multi-agent systems with a {{FIPA}}-compliant
  agent framework}.{\BBCQ}
\newblock
\APACjournalVolNumPages{Software: Practice and Experience}{31}{2}{103-128}.
\newblock
\begin{APACrefDOI}
  \doi{10.1002/1097-024X(200102)31:2<103::AID-SPE358>3.0.CO;2-O}
  \end{APACrefDOI}
\PrintBackRefs{\CurrentBib}

\bibitem [\protect \citeauthoryear {%
Beloglazov%
, Almashor%
, Abebe%
, Richter%
\BCBL {}\ \BBA {} Steer%
}{%
Beloglazov%
\ \protect \BOthers {.}}{%
{\protect \APACyear {2016}}%
}]{%
BeloglazovSimulationwildfireevacuation2016}
\APACinsertmetastar {%
BeloglazovSimulationwildfireevacuation2016}%
\begin{APACrefauthors}%
Beloglazov, A.%
, Almashor, M.%
, Abebe, E.%
, Richter, J.%
\BCBL {}\ \BBA {} Steer, K\BPBI C\BPBI B.%
\end{APACrefauthors}%
\unskip\
\newblock
\APACrefYearMonthDay{2016}{{\APACmonth{01}}}{}.
\newblock
{\BBOQ}\APACrefatitle {Simulation of Wildfire Evacuation with Dynamic Factors
  and Model Composition} {Simulation of wildfire evacuation with dynamic
  factors and model composition}.{\BBCQ}
\newblock
\APACjournalVolNumPages{Simulation Modelling Practice and
  Theory}{60}{}{144-159}.
\newblock
\begin{APACrefDOI} \doi{10.1016/j.simpat.2015.10.002} \end{APACrefDOI}
\PrintBackRefs{\CurrentBib}

\bibitem [\protect \citeauthoryear {%
Birkin%
\ \BBA {} Clarke%
}{%
Birkin%
\ \BBA {} Clarke%
}{%
{\protect \APACyear {1988}}%
}]{%
BirkinSynthesisSyntheticSpatial1988}
\APACinsertmetastar {%
BirkinSynthesisSyntheticSpatial1988}%
\begin{APACrefauthors}%
Birkin, M.%
\BCBT {}\ \BBA {} Clarke, M.%
\end{APACrefauthors}%
\unskip\
\newblock
\APACrefYearMonthDay{1988}{{\APACmonth{12}}}{}.
\newblock
{\BBOQ}\APACrefatitle {Synthesis\textemdash{{A Synthetic Spatial Information
  System}} for {{Urban}} and {{Regional Analysis}}: {{Methods}} and
  {{Examples}}} {Synthesis\textemdash{{A Synthetic Spatial Information System}}
  for {{Urban}} and {{Regional Analysis}}: {{Methods}} and
  {{Examples}}}.{\BBCQ}
\newblock
\APACjournalVolNumPages{Environment and Planning A}{20}{12}{1645-1671}.
\newblock
\begin{APACrefDOI} \doi{10.1068/a201645} \end{APACrefDOI}
\PrintBackRefs{\CurrentBib}

\bibitem [\protect \citeauthoryear {%
Bonabeau%
}{%
Bonabeau%
}{%
{\protect \APACyear {2002}}%
}]{%
BonabeauAgentbasedmodelingMethods2002}
\APACinsertmetastar {%
BonabeauAgentbasedmodelingMethods2002}%
\begin{APACrefauthors}%
Bonabeau, E.%
\end{APACrefauthors}%
\unskip\
\newblock
\APACrefYearMonthDay{2002}{{\APACmonth{05}}}{}.
\newblock
{\BBOQ}\APACrefatitle {Agent-Based Modeling: {{Methods}} and Techniques for
  Simulating Human Systems} {Agent-based modeling: {{Methods}} and techniques
  for simulating human systems}.{\BBCQ}
\newblock
\APACjournalVolNumPages{Proceedings of the National Academy of Sciences of the
  United States of America}{99}{Suppl 3}{7280-7287}.
\newblock
\begin{APACrefDOI} \doi{10.1073/pnas.082080899} \end{APACrefDOI}
\PrintBackRefs{\CurrentBib}

\bibitem [\protect \citeauthoryear {%
Bordini%
, H\"ubner%
\BCBL {}\ \BBA {} Vieira%
}{%
Bordini%
\ \protect \BOthers {.}}{%
{\protect \APACyear {2005}}%
}]{%
BordiniJasonGoldenFleece2005}
\APACinsertmetastar {%
BordiniJasonGoldenFleece2005}%
\begin{APACrefauthors}%
Bordini, R\BPBI H.%
, H\"ubner, J\BPBI F.%
\BCBL {}\ \BBA {} Vieira, R.%
\end{APACrefauthors}%
\unskip\
\newblock
\APACrefYearMonthDay{2005}{}{}.
\newblock
{\BBOQ}\APACrefatitle {Jason and the {{Golden Fleece}} of {{Agent}}-{{Oriented
  Programming}}} {Jason and the {{Golden Fleece}} of {{Agent}}-{{Oriented
  Programming}}}.{\BBCQ}
\newblock
\BIn{} R\BPBI H.~Bordini, M.~Dastani, J.~Dix\BCBL {}\ \BBA {}
  A.~El~Fallah~Seghrouchni\ (\BEDS), \APACrefbtitle {Multi-{{Agent
  Programming}}: {{Languages}}, {{Platforms}} and {{Applications}}}
  {Multi-{{Agent Programming}}: {{Languages}}, {{Platforms}} and
  {{Applications}}}\ (\BPG~3-37).
\newblock
\APACaddressPublisher{Boston, MA}{{Springer US}}.
\newblock
\begin{APACrefDOI} \doi{10.1007/0-387-26350-0-1} \end{APACrefDOI}
\PrintBackRefs{\CurrentBib}

\bibitem [\protect \citeauthoryear {%
Bratman%
, Israel%
\BCBL {}\ \BBA {} Pollack%
}{%
Bratman%
\ \protect \BOthers {.}}{%
{\protect \APACyear {1988}}%
}]{%
BratmanPlansresourceboundedpractical1988}
\APACinsertmetastar {%
BratmanPlansresourceboundedpractical1988}%
\begin{APACrefauthors}%
Bratman, M\BPBI E.%
, Israel, D\BPBI J.%
\BCBL {}\ \BBA {} Pollack, M\BPBI E.%
\end{APACrefauthors}%
\unskip\
\newblock
\APACrefYearMonthDay{1988}{{\APACmonth{09}}}{}.
\newblock
{\BBOQ}\APACrefatitle {Plans and Resource-Bounded Practical Reasoning} {Plans
  and resource-bounded practical reasoning}.{\BBCQ}
\newblock
\APACjournalVolNumPages{Computational Intelligence}{4}{3}{349-355}.
\newblock
\begin{APACrefDOI} \doi{10.1111/j.1467-8640.1988.tb00284.x} \end{APACrefDOI}
\PrintBackRefs{\CurrentBib}

\bibitem [\protect \citeauthoryear {%
Busetta%
, Ronnquist%
, Hodgson%
\BCBL {}\ \BBA {} Lucas%
}{%
Busetta%
\ \protect \BOthers {.}}{%
{\protect \APACyear {1999}}%
}]{%
BusettaJACKIntelligentAgents1999}
\APACinsertmetastar {%
BusettaJACKIntelligentAgents1999}%
\begin{APACrefauthors}%
Busetta, P.%
, Ronnquist, R.%
, Hodgson, A.%
\BCBL {}\ \BBA {} Lucas, A.%
\end{APACrefauthors}%
\unskip\
\newblock
\APACrefYear{1999}.
\newblock
\APACrefbtitle {{{JACK Intelligent Agents}} - {{Components}} for {{Intelligent
  Agents}} in {{Java}}} {{{JACK Intelligent Agents}} - {{Components}} for
  {{Intelligent Agents}} in {{Java}}}.
\PrintBackRefs{\CurrentBib}

\bibitem [\protect \citeauthoryear {%
Chan%
, Son%
\BCBL {}\ \BBA {} Macal%
}{%
Chan%
\ \protect \BOthers {.}}{%
{\protect \APACyear {2010}}%
}]{%
ChanAgentbasedsimulationtutorial2010}
\APACinsertmetastar {%
ChanAgentbasedsimulationtutorial2010}%
\begin{APACrefauthors}%
Chan, W\BPBI K\BPBI V.%
, Son, Y\BPBI J.%
\BCBL {}\ \BBA {} Macal, C\BPBI M.%
\end{APACrefauthors}%
\unskip\
\newblock
\APACrefYearMonthDay{2010}{{\APACmonth{12}}}{}.
\newblock
{\BBOQ}\APACrefatitle {Agent-Based Simulation Tutorial - Simulation of Emergent
  Behavior and Differences between Agent-Based Simulation and Discrete-Event
  Simulation} {Agent-based simulation tutorial - simulation of emergent
  behavior and differences between agent-based simulation and discrete-event
  simulation}.{\BBCQ}
\newblock
\BIn{} \APACrefbtitle {Proceedings of the 2010 {{Winter Simulation
  Conference}}} {Proceedings of the 2010 {{Winter Simulation Conference}}}\
  (\BPG~135-150).
\newblock
\begin{APACrefDOI} \doi{10.1109/WSC.2010.5679168} \end{APACrefDOI}
\PrintBackRefs{\CurrentBib}

\bibitem [\protect \citeauthoryear {%
Chen%
\ \BBA {} Zhan%
}{%
Chen%
\ \BBA {} Zhan%
}{%
{\protect \APACyear {2008}}%
}]{%
ChenAgentbasedmodellingsimulation2008}
\APACinsertmetastar {%
ChenAgentbasedmodellingsimulation2008}%
\begin{APACrefauthors}%
Chen, X.%
\BCBT {}\ \BBA {} Zhan, F\BPBI B.%
\end{APACrefauthors}%
\unskip\
\newblock
\APACrefYearMonthDay{2008}{{\APACmonth{01}}}{}.
\newblock
{\BBOQ}\APACrefatitle {Agent-Based Modelling and Simulation of Urban
  Evacuation: Relative Effectiveness of Simultaneous and Staged Evacuation
  Strategies} {Agent-based modelling and simulation of urban evacuation:
  Relative effectiveness of simultaneous and staged evacuation
  strategies}.{\BBCQ}
\newblock
\APACjournalVolNumPages{Journal of the Operational Research
  Society}{59}{1}{25-33}.
\newblock
\begin{APACrefDOI} \doi{10.1057/palgrave.jors.2602321} \end{APACrefDOI}
\PrintBackRefs{\CurrentBib}

\bibitem [\protect \citeauthoryear {%
Christakis%
}{%
Christakis%
}{%
{\protect \APACyear {2007}}%
}]{%
ChristakisSpreadObesityLarge2007}
\APACinsertmetastar {%
ChristakisSpreadObesityLarge2007}%
\begin{APACrefauthors}%
Christakis, N\BPBI A.%
\end{APACrefauthors}%
\unskip\
\newblock
\APACrefYearMonthDay{2007}{}{}.
\newblock
{\BBOQ}\APACrefatitle {The {{Spread}} of {{Obesity}} in a {{Large Social
  Network}} over 32 {{Years}}} {The {{Spread}} of {{Obesity}} in a {{Large
  Social Network}} over 32 {{Years}}}.{\BBCQ}
\newblock
\APACjournalVolNumPages{The New England Journal of Medicine}{}{}{370-379}.
\newblock
\begin{APACrefDOI} \doi{10.1056/NEJMsa066082} \end{APACrefDOI}
\PrintBackRefs{\CurrentBib}

\bibitem [\protect \citeauthoryear {%
\{Country Fire Authority\}%
\ \BBA {} \{Sweeney Research\}%
}{%
\{Country Fire Authority\}%
\ \BBA {} \{Sweeney Research\}%
}{%
{\protect \APACyear {2009}}%
}]{%
CountryFireAuthorityQualitativeReportCFA2009}
\APACinsertmetastar {%
CountryFireAuthorityQualitativeReportCFA2009}%
\begin{APACrefauthors}%
\{Country Fire Authority\}%
\BCBT {}\ \BBA {} \{Sweeney Research\}.%
\end{APACrefauthors}%
\unskip\
\newblock
\APACrefYearMonthDay{2009}{{\APACmonth{09}}}{}.
\newblock
\APACrefbtitle {A {{Qualitative Report}} on {{CFA Community Engagement}}} {A
  {{Qualitative Report}} on {{CFA Community Engagement}}}\
  \APACbVolEdTR{}{\BTR{}}.
\newblock
\APACaddressInstitution{}{{Country Fire Authority}}.
\PrintBackRefs{\CurrentBib}

\bibitem [\protect \citeauthoryear {%
Crooks%
\ \BBA {} Heppenstall%
}{%
Crooks%
\ \BBA {} Heppenstall%
}{%
{\protect \APACyear {2012}}%
}]{%
CrooksIntroductionAgentBasedModelling2012}
\APACinsertmetastar {%
CrooksIntroductionAgentBasedModelling2012}%
\begin{APACrefauthors}%
Crooks, A\BPBI T.%
\BCBT {}\ \BBA {} Heppenstall, A\BPBI J.%
\end{APACrefauthors}%
\unskip\
\newblock
\APACrefYearMonthDay{2012}{}{}.
\newblock
{\BBOQ}\APACrefatitle {Introduction to {{Agent}}-{{Based Modelling}}}
  {Introduction to {{Agent}}-{{Based Modelling}}}.{\BBCQ}
\newblock
\BIn{} \APACrefbtitle {Agent-{{Based Models}} of {{Geographical Systems}}}
  {Agent-{{Based Models}} of {{Geographical Systems}}}\ (\BPG~85-105).
\newblock
\APACaddressPublisher{}{{Springer, Dordrecht}}.
\PrintBackRefs{\CurrentBib}

\bibitem [\protect \citeauthoryear {%
Dash%
\ \BBA {} Gladwin%
}{%
Dash%
\ \BBA {} Gladwin%
}{%
{\protect \APACyear {2007}}%
}]{%
DashEvacuationDecisionMaking2007}
\APACinsertmetastar {%
DashEvacuationDecisionMaking2007}%
\begin{APACrefauthors}%
Dash, N.%
\BCBT {}\ \BBA {} Gladwin, H.%
\end{APACrefauthors}%
\unskip\
\newblock
\APACrefYearMonthDay{2007}{{\APACmonth{08}}}{}.
\newblock
{\BBOQ}\APACrefatitle {Evacuation {{Decision Making}} and {{Behavioral
  Responses}}: {{Individual}} and {{Household}}} {Evacuation {{Decision
  Making}} and {{Behavioral Responses}}: {{Individual}} and
  {{Household}}}.{\BBCQ}
\newblock
\APACjournalVolNumPages{Natural Hazards Review}{8}{3}{69-77}.
\newblock
\begin{APACrefDOI} \doi{10.1061/(ASCE)1527-6988(2007)8:3(69)} \end{APACrefDOI}
\PrintBackRefs{\CurrentBib}

\bibitem [\protect \citeauthoryear {%
Davis%
, Trapman%
, Leirs%
, Begon%
\BCBL {}\ \BBA {} Heesterbeek%
}{%
Davis%
\ \protect \BOthers {.}}{%
{\protect \APACyear {2008}}%
}]{%
Davisabundancethresholdplague2008}
\APACinsertmetastar {%
Davisabundancethresholdplague2008}%
\begin{APACrefauthors}%
Davis, S.%
, Trapman, P.%
, Leirs, H.%
, Begon, M.%
\BCBL {}\ \BBA {} Heesterbeek, J\BPBI a\BPBI P.%
\end{APACrefauthors}%
\unskip\
\newblock
\APACrefYearMonthDay{2008}{{\APACmonth{07}}}{}.
\newblock
{\BBOQ}\APACrefatitle {The Abundance Threshold for Plague as a Critical
  Percolation Phenomenon} {The abundance threshold for plague as a critical
  percolation phenomenon}.{\BBCQ}
\newblock
\APACjournalVolNumPages{Nature}{454}{7204}{634-637}.
\newblock
\begin{APACrefDOI} \doi{10.1038/nature07053} \end{APACrefDOI}
\PrintBackRefs{\CurrentBib}

\bibitem [\protect \citeauthoryear {%
Dawson%
, Peppe%
\BCBL {}\ \BBA {} Wang%
}{%
Dawson%
\ \protect \BOthers {.}}{%
{\protect \APACyear {2011}}%
}]{%
Dawsonagentbasedmodelriskbased2011}
\APACinsertmetastar {%
Dawsonagentbasedmodelriskbased2011}%
\begin{APACrefauthors}%
Dawson, R\BPBI J.%
, Peppe, R.%
\BCBL {}\ \BBA {} Wang, M.%
\end{APACrefauthors}%
\unskip\
\newblock
\APACrefYearMonthDay{2011}{{\APACmonth{10}}}{}.
\newblock
{\BBOQ}\APACrefatitle {An Agent-Based Model for Risk-Based Flood Incident
  Management} {An agent-based model for risk-based flood incident
  management}.{\BBCQ}
\newblock
\APACjournalVolNumPages{Natural Hazards}{59}{1}{167-189}.
\newblock
\begin{APACrefDOI} \doi{10.1007/s11069-011-9745-4} \end{APACrefDOI}
\PrintBackRefs{\CurrentBib}

\bibitem [\protect \citeauthoryear {%
{de Boer}%
, Hindriks%
, {van der Hoek}%
\BCBL {}\ \BBA {} Meyer%
}{%
{de Boer}%
\ \protect \BOthers {.}}{%
{\protect \APACyear {2002}}%
}]{%
deBoerAgentProgrammingDeclarative2002}
\APACinsertmetastar {%
deBoerAgentProgrammingDeclarative2002}%
\begin{APACrefauthors}%
{de Boer}, F\BPBI S.%
, Hindriks, K\BPBI V.%
, {van der Hoek}, W.%
\BCBL {}\ \BBA {} Meyer, J\BHBI J\BPBI C.%
\end{APACrefauthors}%
\unskip\
\newblock
\APACrefYearMonthDay{2002}{{\APACmonth{07}}}{}.
\newblock
{\BBOQ}\APACrefatitle {Agent {{Programming}} with {{Declarative Goals}}} {Agent
  {{Programming}} with {{Declarative Goals}}}.{\BBCQ}
\newblock
\APACjournalVolNumPages{arXiv:cs/0207008}{}{}{}.
\PrintBackRefs{\CurrentBib}

\bibitem [\protect \citeauthoryear {%
D'Inverno%
, Luck%
, Georgeff%
, Kinny%
\BCBL {}\ \BBA {} Wooldridge%
}{%
D'Inverno%
\ \protect \BOthers {.}}{%
{\protect \APACyear {2004}}%
}]{%
DInvernodMARSArchitectureSpecification2004}
\APACinsertmetastar {%
DInvernodMARSArchitectureSpecification2004}%
\begin{APACrefauthors}%
D'Inverno, M.%
, Luck, M.%
, Georgeff, M.%
, Kinny, D.%
\BCBL {}\ \BBA {} Wooldridge, M.%
\end{APACrefauthors}%
\unskip\
\newblock
\APACrefYearMonthDay{2004}{{\APACmonth{07}}}{}.
\newblock
{\BBOQ}\APACrefatitle {The {{dMARS Architecture}}: {{A Specification}} of the
  {{Distributed Multi}}-{{Agent Reasoning System}}} {The {{dMARS
  Architecture}}: {{A Specification}} of the {{Distributed Multi}}-{{Agent
  Reasoning System}}}.{\BBCQ}
\newblock
\APACjournalVolNumPages{Autonomous Agents and Multi-Agent
  Systems}{9}{1-2}{5-53}.
\newblock
\begin{APACrefDOI} \doi{10.1023/B:AGNT.0000019688.11109.19} \end{APACrefDOI}
\PrintBackRefs{\CurrentBib}

\bibitem [\protect \citeauthoryear {%
Epstein%
}{%
Epstein%
}{%
{\protect \APACyear {2009}}%
}]{%
EpsteinModellingcontainpandemics2009}
\APACinsertmetastar {%
EpsteinModellingcontainpandemics2009}%
\begin{APACrefauthors}%
Epstein, J\BPBI M.%
\end{APACrefauthors}%
\unskip\
\newblock
\APACrefYearMonthDay{2009}{{\APACmonth{08}}}{}.
\newblock
{\BBOQ}\APACrefatitle {Modelling to Contain Pandemics} {Modelling to contain
  pandemics}.{\BBCQ}
\newblock
\APACjournalVolNumPages{Nature}{460}{7256}{687-687}.
\newblock
\begin{APACrefDOI} \doi{10.1038/460687a} \end{APACrefDOI}
\PrintBackRefs{\CurrentBib}

\bibitem [\protect \citeauthoryear {%
Farmer%
\ \BBA {} Foley%
}{%
Farmer%
\ \BBA {} Foley%
}{%
{\protect \APACyear {2009}}%
}]{%
Farmereconomyneedsagentbased2009}
\APACinsertmetastar {%
Farmereconomyneedsagentbased2009}%
\begin{APACrefauthors}%
Farmer, J\BPBI D.%
\BCBT {}\ \BBA {} Foley, D.%
\end{APACrefauthors}%
\unskip\
\newblock
\APACrefYearMonthDay{2009}{{\APACmonth{08}}}{}.
\newblock
{\BBOQ}\APACrefatitle {The Economy Needs Agent-Based Modelling} {The economy
  needs agent-based modelling}.{\BBCQ}
\newblock
\APACjournalVolNumPages{Nature}{460}{7256}{685-686}.
\newblock
\begin{APACrefDOI} \doi{10.1038/460685a} \end{APACrefDOI}
\PrintBackRefs{\CurrentBib}

\bibitem [\protect \citeauthoryear {%
Funk%
, Salath\'e%
\BCBL {}\ \BBA {} Jansen%
}{%
Funk%
\ \protect \BOthers {.}}{%
{\protect \APACyear {2010}}%
}]{%
FunkModellinginfluencehuman2010}
\APACinsertmetastar {%
FunkModellinginfluencehuman2010}%
\begin{APACrefauthors}%
Funk, S.%
, Salath\'e, M.%
\BCBL {}\ \BBA {} Jansen, V\BPBI A\BPBI A.%
\end{APACrefauthors}%
\unskip\
\newblock
\APACrefYearMonthDay{2010}{{\APACmonth{05}}}{}.
\newblock
{\BBOQ}\APACrefatitle {Modelling the Influence of Human Behaviour on the Spread
  of Infectious Diseases: A Review} {Modelling the influence of human behaviour
  on the spread of infectious diseases: A review}.{\BBCQ}
\newblock
\APACjournalVolNumPages{Journal of The Royal Society
  Interface}{}{}{rsif20100142}.
\newblock
\begin{APACrefDOI} \doi{10.1098/rsif.2010.0142} \end{APACrefDOI}
\PrintBackRefs{\CurrentBib}

\bibitem [\protect \citeauthoryear {%
Galea%
, Riddle%
\BCBL {}\ \BBA {} Kaplan%
}{%
Galea%
\ \protect \BOthers {.}}{%
{\protect \APACyear {2010}}%
}]{%
GaleaCausalthinkingcomplex2010}
\APACinsertmetastar {%
GaleaCausalthinkingcomplex2010}%
\begin{APACrefauthors}%
Galea, S.%
, Riddle, M.%
\BCBL {}\ \BBA {} Kaplan, G\BPBI A.%
\end{APACrefauthors}%
\unskip\
\newblock
\APACrefYearMonthDay{2010}{{\APACmonth{02}}}{}.
\newblock
{\BBOQ}\APACrefatitle {Causal Thinking and Complex System Approaches in
  Epidemiology} {Causal thinking and complex system approaches in
  epidemiology}.{\BBCQ}
\newblock
\APACjournalVolNumPages{International Journal of Epidemiology}{39}{1}{97-106}.
\newblock
\begin{APACrefDOI} \doi{10.1093/ije/dyp296} \end{APACrefDOI}
\PrintBackRefs{\CurrentBib}

\bibitem [\protect \citeauthoryear {%
Gilbert%
}{%
Gilbert%
}{%
{\protect \APACyear {2006}}%
}]{%
GilbertWhendoessocial2006}
\APACinsertmetastar {%
GilbertWhendoessocial2006}%
\begin{APACrefauthors}%
Gilbert, N.%
\end{APACrefauthors}%
\unskip\
\newblock
\APACrefYearMonthDay{2006}{}{}.
\newblock
{\BBOQ}\APACrefatitle {When Does Social Simulation Need Cognitive Models?}
  {When does social simulation need cognitive models?}{\BBCQ}
\newblock
\BIn{} R.~Sun\ (\BED), \APACrefbtitle {Cognition and {{Multi}}-{{Agent
  Interaction}}: {{From Cognitive Modeling}} to {{Social Simulation}}}
  {Cognition and {{Multi}}-{{Agent Interaction}}: {{From Cognitive Modeling}}
  to {{Social Simulation}}}\ (\BPG~428-432).
\newblock
\APACaddressPublisher{Cambridge}{{Cambridge University Press}}.
\PrintBackRefs{\CurrentBib}

\bibitem [\protect \citeauthoryear {%
Gilbert%
\ \BBA {} Troitzsch%
}{%
Gilbert%
\ \BBA {} Troitzsch%
}{%
{\protect \APACyear {2011}}%
}]{%
GilbertSimulationSocialScientist2011}
\APACinsertmetastar {%
GilbertSimulationSocialScientist2011}%
\begin{APACrefauthors}%
Gilbert, N.%
\BCBT {}\ \BBA {} Troitzsch, K.%
\end{APACrefauthors}%
\unskip\
\newblock
\APACrefYear{2011}.
\newblock
\APACrefbtitle {Simulation for the {{Social Scientist}}} {Simulation for the
  {{Social Scientist}}}\ (\PrintOrdinal{2nd}\ \BEd).
\newblock
\APACaddressPublisher{UK}{{Open University Press}}.
\PrintBackRefs{\CurrentBib}

\bibitem [\protect \citeauthoryear {%
Gode%
\ \BBA {} Sunder%
}{%
Gode%
\ \BBA {} Sunder%
}{%
{\protect \APACyear {1993}}%
}]{%
GodeAllocativeEfficiencyMarkets1993}
\APACinsertmetastar {%
GodeAllocativeEfficiencyMarkets1993}%
\begin{APACrefauthors}%
Gode, D\BPBI K.%
\BCBT {}\ \BBA {} Sunder, S.%
\end{APACrefauthors}%
\unskip\
\newblock
\APACrefYearMonthDay{1993}{{\APACmonth{02}}}{}.
\newblock
{\BBOQ}\APACrefatitle {Allocative {{Efficiency}} of {{Markets}} with
  {{Zero}}-{{Intelligence Traders}}: {{Market}} as a {{Partial Substitute}} for
  {{Individual Rationality}}} {Allocative {{Efficiency}} of {{Markets}} with
  {{Zero}}-{{Intelligence Traders}}: {{Market}} as a {{Partial Substitute}} for
  {{Individual Rationality}}}.{\BBCQ}
\newblock
\APACjournalVolNumPages{Journal of Political Economy}{101}{1}{119-137}.
\newblock
\begin{APACrefDOI} \doi{10.1086/261868} \end{APACrefDOI}
\PrintBackRefs{\CurrentBib}

\bibitem [\protect \citeauthoryear {%
Grimm%
\ \BBA {} Railsback%
}{%
Grimm%
\ \BBA {} Railsback%
}{%
{\protect \APACyear {2005}}%
}]{%
GrimmIndividualbasedModellingEcology2005}
\APACinsertmetastar {%
GrimmIndividualbasedModellingEcology2005}%
\begin{APACrefauthors}%
Grimm, V.%
\BCBT {}\ \BBA {} Railsback, S.%
\end{APACrefauthors}%
\unskip\
\newblock
\APACrefYear{2005}.
\newblock
\APACrefbtitle {Individual-Based {{Modelling}} and {{Ecology}}}
  {Individual-based {{Modelling}} and {{Ecology}}}.
\newblock
\APACaddressPublisher{USA}{{Princeton University Press}}.
\PrintBackRefs{\CurrentBib}

\bibitem [\protect \citeauthoryear {%
Harland%
, Heppenstall%
, Smith%
\BCBL {}\ \BBA {} Birkin%
}{%
Harland%
\ \protect \BOthers {.}}{%
{\protect \APACyear {2012}}%
}]{%
HarlandCreatingrealisticsynthetic2012}
\APACinsertmetastar {%
HarlandCreatingrealisticsynthetic2012}%
\begin{APACrefauthors}%
Harland, K.%
, Heppenstall, A.%
, Smith, D.%
\BCBL {}\ \BBA {} Birkin, M\BPBI H.%
\end{APACrefauthors}%
\unskip\
\newblock
\APACrefYearMonthDay{2012}{}{}.
\newblock
{\BBOQ}\APACrefatitle {Creating Realistic Synthetic Populations at Varying
  Spatial Scales: {{A}} Comparative Critique of Population Synthesis
  Techniques} {Creating realistic synthetic populations at varying spatial
  scales: {{A}} comparative critique of population synthesis
  techniques}.{\BBCQ}
\newblock
\APACjournalVolNumPages{Journal of Artificial Societies and Social
  Simulation}{15}{}{}.
\PrintBackRefs{\CurrentBib}

\bibitem [\protect \citeauthoryear {%
Haynes%
, Handmer%
, McAneney%
, Tibbits%
\BCBL {}\ \BBA {} Coates%
}{%
Haynes%
\ \protect \BOthers {.}}{%
{\protect \APACyear {2010}}%
}]{%
HaynesAustralianbushfirefatalities2010}
\APACinsertmetastar {%
HaynesAustralianbushfirefatalities2010}%
\begin{APACrefauthors}%
Haynes, K.%
, Handmer, J.%
, McAneney, J.%
, Tibbits, A.%
\BCBL {}\ \BBA {} Coates, L.%
\end{APACrefauthors}%
\unskip\
\newblock
\APACrefYearMonthDay{2010}{{\APACmonth{05}}}{}.
\newblock
{\BBOQ}\APACrefatitle {Australian Bushfire Fatalities 1900\textendash{}2008:
  Exploring Trends in Relation to the `{{Prepare}}, Stay and Defend or Leave
  Early' Policy} {Australian bushfire fatalities 1900\textendash{}2008:
  Exploring trends in relation to the `{{Prepare}}, stay and defend or leave
  early' policy}.{\BBCQ}
\newblock
\APACjournalVolNumPages{Environmental Science \& Policy}{13}{3}{185-194}.
\newblock
\begin{APACrefDOI} \doi{10.1016/j.envsci.2010.03.002} \end{APACrefDOI}
\PrintBackRefs{\CurrentBib}

\bibitem [\protect \citeauthoryear {%
Horni%
, Nagel%
\BCBL {}\ \BBA {} Axhausen%
}{%
Horni%
\ \protect \BOthers {.}}{%
{\protect \APACyear {2016}}%
}]{%
HorniIntroducingMATSim2016}
\APACinsertmetastar {%
HorniIntroducingMATSim2016}%
\begin{APACrefauthors}%
Horni, A.%
, Nagel, K.%
\BCBL {}\ \BBA {} Axhausen, K.%
\end{APACrefauthors}%
\unskip\
\newblock
\APACrefYearMonthDay{2016}{{\APACmonth{08}}}{}.
\newblock
{\BBOQ}\APACrefatitle {Introducing {{MATSim}}} {Introducing {{MATSim}}}.{\BBCQ}
\newblock
\BIn{} A.~Horni, K.~Nagel\BCBL {}\ \BBA {} K.~Axhausen\ (\BEDS), \APACrefbtitle
  {The {{Multi}}-{{Agent Transport Simulation MATSim}}} {The {{Multi}}-{{Agent
  Transport Simulation MATSim}}}\ (\BPG~3-8).
\newblock
\APACaddressPublisher{London}{{Ubiquity Press}}.
\PrintBackRefs{\CurrentBib}

\bibitem [\protect \citeauthoryear {%
Ingrand%
, Georgeff%
\BCBL {}\ \BBA {} Rao%
}{%
Ingrand%
\ \protect \BOthers {.}}{%
{\protect \APACyear {1992}}%
}]{%
Ingrandarchitecturerealtimereasoning1992}
\APACinsertmetastar {%
Ingrandarchitecturerealtimereasoning1992}%
\begin{APACrefauthors}%
Ingrand, F\BPBI F.%
, Georgeff, M\BPBI P.%
\BCBL {}\ \BBA {} Rao, A\BPBI S.%
\end{APACrefauthors}%
\unskip\
\newblock
\APACrefYearMonthDay{1992}{{\APACmonth{12}}}{}.
\newblock
{\BBOQ}\APACrefatitle {An Architecture for Real-Time Reasoning and System
  Control} {An architecture for real-time reasoning and system control}.{\BBCQ}
\newblock
\APACjournalVolNumPages{IEEE Expert}{7}{6}{34-44}.
\newblock
\begin{APACrefDOI} \doi{10.1109/64.180407} \end{APACrefDOI}
\PrintBackRefs{\CurrentBib}

\bibitem [\protect \citeauthoryear {%
Jain%
, Ronald%
\BCBL {}\ \BBA {} Winter%
}{%
Jain%
\ \protect \BOthers {.}}{%
{\protect \APACyear {2015}}%
}]{%
JainCreatingSyntheticPopulation2015}
\APACinsertmetastar {%
JainCreatingSyntheticPopulation2015}%
\begin{APACrefauthors}%
Jain, S.%
, Ronald, N.%
\BCBL {}\ \BBA {} Winter, S.%
\end{APACrefauthors}%
\unskip\
\newblock
\APACrefYearMonthDay{2015}{{\APACmonth{12}}}{}.
\newblock
{\BBOQ}\APACrefatitle {Creating a {{Synthetic Population}}: {{A Comparison}} of
  {{Tools}}} {Creating a {{Synthetic Population}}: {{A Comparison}} of
  {{Tools}}}.{\BBCQ}
\newblock
\BIn{} \APACrefbtitle {Conference of {{Transportation Research Group}} of
  {{India}}} {Conference of {{Transportation Research Group}} of {{India}}}\
  (\BPG~13).
\newblock
\APACaddressPublisher{India}{}.
\PrintBackRefs{\CurrentBib}

\bibitem [\protect \citeauthoryear {%
Jarvis%
, Jarvis%
\BCBL {}\ \BBA {} R\"onnquist%
}{%
Jarvis%
\ \protect \BOthers {.}}{%
{\protect \APACyear {2008}}%
}]{%
JarvisHolonicExecutionBDI2008}
\APACinsertmetastar {%
JarvisHolonicExecutionBDI2008}%
\begin{APACrefauthors}%
Jarvis, J.%
, Jarvis, D.%
\BCBL {}\ \BBA {} R\"onnquist, R.%
\end{APACrefauthors}%
\unskip\
\newblock
\APACrefYear{2008}.
\newblock
\APACrefbtitle {Holonic {{Execution}}: {{A BDI Approach}}} {Holonic
  {{Execution}}: {{A BDI Approach}}}.
\newblock
\APACaddressPublisher{}{{Springer Science \& Business Media}}.
\PrintBackRefs{\CurrentBib}

\bibitem [\protect \citeauthoryear {%
Johnson%
, Johnson%
\BCBL {}\ \BBA {} Sutherland%
}{%
Johnson%
\ \protect \BOthers {.}}{%
{\protect \APACyear {2011}}%
}]{%
JohnsonStayGoHuman2011}
\APACinsertmetastar {%
JohnsonStayGoHuman2011}%
\begin{APACrefauthors}%
Johnson, P\BPBI F.%
, Johnson, C\BPBI E.%
\BCBL {}\ \BBA {} Sutherland, C.%
\end{APACrefauthors}%
\unskip\
\newblock
\APACrefYearMonthDay{2011}{}{}.
\newblock
{\BBOQ}\APACrefatitle {Stay or {{Go}}? {{Human Behavior}} in {{Major
  Evacuations}}} {Stay or {{Go}}? {{Human Behavior}} in {{Major
  Evacuations}}}.{\BBCQ}
\newblock
\BIn{} \APACrefbtitle {Pedestrian and {{Evacuation Dynamics}}} {Pedestrian and
  {{Evacuation Dynamics}}}\ (\BPG~675-684).
\newblock
\APACaddressPublisher{}{{Springer, Boston, MA}}.
\PrintBackRefs{\CurrentBib}

\bibitem [\protect \citeauthoryear {%
Jung%
}{%
Jung%
}{%
{\protect \APACyear {1969}}%
}]{%
JUNGCollectedWorksJung1969}
\APACinsertmetastar {%
JUNGCollectedWorksJung1969}%
\begin{APACrefauthors}%
Jung, C\BPBI G.%
\end{APACrefauthors}%
\unskip\
\newblock
\APACrefYear{1969}.
\newblock
\APACrefbtitle {Collected {{Works}} of {{C}}.{{G}}. {{Jung}}, {{Volume}} 9
  ({{Part}} 1): {{Archetypes}} and the {{Collective Unconscious}}} {Collected
  {{Works}} of {{C}}.{{G}}. {{Jung}}, {{Volume}} 9 ({{Part}} 1): {{Archetypes}}
  and the {{Collective Unconscious}}}\ (G.~Adler\ \BBA {} R\BPBI F\BPBI
  C.~Hull, \BEDS{}).
\newblock
\APACaddressPublisher{}{{Princeton University Press}}.
\PrintBackRefs{\CurrentBib}

\bibitem [\protect \citeauthoryear {%
Kennedy%
}{%
Kennedy%
}{%
{\protect \APACyear {2012}}%
}]{%
KennedyModellingHumanBehaviour2012}
\APACinsertmetastar {%
KennedyModellingHumanBehaviour2012}%
\begin{APACrefauthors}%
Kennedy, W\BPBI G.%
\end{APACrefauthors}%
\unskip\
\newblock
\APACrefYearMonthDay{2012}{}{}.
\newblock
{\BBOQ}\APACrefatitle {Modelling {{Human Behaviour}} in {{Agent}}-{{Based
  Models}}} {Modelling {{Human Behaviour}} in {{Agent}}-{{Based
  Models}}}.{\BBCQ}
\newblock
\BIn{} \APACrefbtitle {Agent-{{Based Models}} of {{Geographical Systems}}}
  {Agent-{{Based Models}} of {{Geographical Systems}}}\ (\BPG~167-179).
\newblock
\APACaddressPublisher{}{{Springer, Dordrecht}}.
\PrintBackRefs{\CurrentBib}

\bibitem [\protect \citeauthoryear {%
Kermack%
\ \BBA {} McKendrick%
}{%
Kermack%
\ \BBA {} McKendrick%
}{%
{\protect \APACyear {1927}}%
}]{%
Kermackcontributionmathematicaltheory1927}
\APACinsertmetastar {%
Kermackcontributionmathematicaltheory1927}%
\begin{APACrefauthors}%
Kermack, W\BPBI O.%
\BCBT {}\ \BBA {} McKendrick, A\BPBI G.%
\end{APACrefauthors}%
\unskip\
\newblock
\APACrefYearMonthDay{1927}{{\APACmonth{08}}}{}.
\newblock
{\BBOQ}\APACrefatitle {A {{Contribution}} to the {{Mathematical Theory}} of
  {{Epidemics}}} {A {{Contribution}} to the {{Mathematical Theory}} of
  {{Epidemics}}}.{\BBCQ}
\newblock
\APACjournalVolNumPages{Proc. R. Soc. Lond. A}{115}{772}{700-721}.
\newblock
\begin{APACrefDOI} \doi{10.1098/rspa.1927.0118} \end{APACrefDOI}
\PrintBackRefs{\CurrentBib}

\bibitem [\protect \citeauthoryear {%
Knoop%
}{%
Knoop%
}{%
{\protect \APACyear {2009}}%
}]{%
KnoopRoadIncidentsNetwork2009}
\APACinsertmetastar {%
KnoopRoadIncidentsNetwork2009}%
\begin{APACrefauthors}%
Knoop, V\BPBI L.%
\end{APACrefauthors}%
\unskip\
\newblock
\APACrefYearMonthDay{2009}{}{}.
\newblock
{\BBOQ}\APACrefatitle {Road {{Incidents}} and {{Network Dynamics}}: {{Effects}}
  on Driving Behaviour and Traffic Congestion} {Road {{Incidents}} and
  {{Network Dynamics}}: {{Effects}} on driving behaviour and traffic
  congestion}.{\BBCQ}
\newblock

\PrintBackRefs{\CurrentBib}

\bibitem [\protect \citeauthoryear {%
Kulash%
}{%
Kulash%
}{%
{\protect \APACyear {1990}}%
}]{%
KulashTraditionalNeighbourhoodDevelopment1990}
\APACinsertmetastar {%
KulashTraditionalNeighbourhoodDevelopment1990}%
\begin{APACrefauthors}%
Kulash, W.%
\end{APACrefauthors}%
\unskip\
\newblock
\APACrefYearMonthDay{1990}{{\APACmonth{10}}}{}.
\newblock
\APACrefbtitle {Traditional {{Neighbourhood Development}}: {{Will}} the
  {{Traffic Work}}?} {Traditional {{Neighbourhood Development}}: {{Will}} the
  {{Traffic Work}}?}
\newblock
\APACaddressPublisher{Bellevue WA}{}.
\PrintBackRefs{\CurrentBib}

\bibitem [\protect \citeauthoryear {%
Laird%
}{%
Laird%
}{%
{\protect \APACyear {2007}}%
}]{%
LairdExtendingSoarCognitive2007}
\APACinsertmetastar {%
LairdExtendingSoarCognitive2007}%
\begin{APACrefauthors}%
Laird, J\BPBI E.%
\end{APACrefauthors}%
\unskip\
\newblock
\APACrefYearMonthDay{2007}{{\APACmonth{07}}}{}.
\newblock
\APACrefbtitle {Extending the {{Soar Cognitive Architecture}}:} {Extending the
  {{Soar Cognitive Architecture}}:}\ \APACbVolEdTR{}{\BTR{}}.
\newblock
\APACaddressInstitution{Fort Belvoir, VA}{{Defense Technical Information
  Center}}.
\newblock
\begin{APACrefDOI} \doi{10.21236/ADA473738} \end{APACrefDOI}
\PrintBackRefs{\CurrentBib}

\bibitem [\protect \citeauthoryear {%
Lee%
, Son%
\BCBL {}\ \BBA {} Jin%
}{%
Lee%
\ \protect \BOthers {.}}{%
{\protect \APACyear {2010}}%
}]{%
LeeIntegratedHumanDecision2010}
\APACinsertmetastar {%
LeeIntegratedHumanDecision2010}%
\begin{APACrefauthors}%
Lee, S.%
, Son, Y\BHBI J.%
\BCBL {}\ \BBA {} Jin, J.%
\end{APACrefauthors}%
\unskip\
\newblock
\APACrefYearMonthDay{2010}{{\APACmonth{11}}}{}.
\newblock
{\BBOQ}\APACrefatitle {An {{Integrated Human Decision Making Model}} for
  {{Evacuation Scenarios Under}} a {{BDI Framework}}} {An {{Integrated Human
  Decision Making Model}} for {{Evacuation Scenarios Under}} a {{BDI
  Framework}}}.{\BBCQ}
\newblock
\APACjournalVolNumPages{ACM Trans. Model. Comput. Simul.}{20}{4}{23:1--23:24}.
\newblock
\begin{APACrefDOI} \doi{10.1145/1842722.1842728} \end{APACrefDOI}
\PrintBackRefs{\CurrentBib}

\bibitem [\protect \citeauthoryear {%
Lindell%
\ \BBA {} Prater%
}{%
Lindell%
\ \BBA {} Prater%
}{%
{\protect \APACyear {2007}}%
}]{%
LindellCriticalBehavioralAssumptions2007}
\APACinsertmetastar {%
LindellCriticalBehavioralAssumptions2007}%
\begin{APACrefauthors}%
Lindell, M\BPBI K.%
\BCBT {}\ \BBA {} Prater, C\BPBI S.%
\end{APACrefauthors}%
\unskip\
\newblock
\APACrefYearMonthDay{2007}{{\APACmonth{03}}}{}.
\newblock
{\BBOQ}\APACrefatitle {Critical {{Behavioral Assumptions}} in {{Evacuation Time
  Estimate Analysis}} for {{Private Vehicles}}: {{Examples}} from {{Hurricane
  Research}} and {{Planning}}} {Critical {{Behavioral Assumptions}} in
  {{Evacuation Time Estimate Analysis}} for {{Private Vehicles}}: {{Examples}}
  from {{Hurricane Research}} and {{Planning}}}.{\BBCQ}
\newblock
\APACjournalVolNumPages{Journal of Urban Planning and
  Development}{133}{1}{18-29}.
\newblock
\begin{APACrefDOI} \doi{10.1061/(ASCE)0733-9488(2007)133:1(18)}
  \end{APACrefDOI}
\PrintBackRefs{\CurrentBib}

\bibitem [\protect \citeauthoryear {%
{Lloyd-Smith}%
, Getz%
\BCBL {}\ \BBA {} Westerhoff%
}{%
{Lloyd-Smith}%
\ \protect \BOthers {.}}{%
{\protect \APACyear {2004}}%
}]{%
Lloyd-SmithFrequencydependentincidence2004}
\APACinsertmetastar {%
Lloyd-SmithFrequencydependentincidence2004}%
\begin{APACrefauthors}%
{Lloyd-Smith}, J\BPBI O.%
, Getz, W\BPBI M.%
\BCBL {}\ \BBA {} Westerhoff, H\BPBI V.%
\end{APACrefauthors}%
\unskip\
\newblock
\APACrefYearMonthDay{2004}{{\APACmonth{03}}}{}.
\newblock
{\BBOQ}\APACrefatitle {Frequency\textendash{}Dependent Incidence in Models of
  Sexually Transmitted Diseases: Portrayal of Pair\textendash{}Based
  Transmission and Effects of Illness on Contact Behaviour}
  {Frequency\textendash{}dependent incidence in models of sexually transmitted
  diseases: Portrayal of pair\textendash{}based transmission and effects of
  illness on contact behaviour}.{\BBCQ}
\newblock
\APACjournalVolNumPages{Proceedings of the Royal Society of London B:
  Biological Sciences}{271}{1539}{625-634}.
\newblock
\begin{APACrefDOI} \doi{10.1098/rspb.2003.2632} \end{APACrefDOI}
\PrintBackRefs{\CurrentBib}

\bibitem [\protect \citeauthoryear {%
Macal%
\ \BBA {} North%
}{%
Macal%
\ \BBA {} North%
}{%
{\protect \APACyear {2005}}%
}]{%
MacalTutorialagentbasedmodeling2005}
\APACinsertmetastar {%
MacalTutorialagentbasedmodeling2005}%
\begin{APACrefauthors}%
Macal, C\BPBI M.%
\BCBT {}\ \BBA {} North, M\BPBI J.%
\end{APACrefauthors}%
\unskip\
\newblock
\APACrefYearMonthDay{2005}{{\APACmonth{12}}}{}.
\newblock
{\BBOQ}\APACrefatitle {Tutorial on Agent-Based Modeling and Simulation}
  {Tutorial on agent-based modeling and simulation}.{\BBCQ}
\newblock
\BIn{} \APACrefbtitle {Proceedings of the {{Winter Simulation Conference}},
  2005.} {Proceedings of the {{Winter Simulation Conference}}, 2005.}\ (\BPG~14
  pp.-).
\newblock
\begin{APACrefDOI} \doi{10.1109/WSC.2005.1574234} \end{APACrefDOI}
\PrintBackRefs{\CurrentBib}

\bibitem [\protect \citeauthoryear {%
McCaffrey%
, Rhodes%
\BCBL {}\ \BBA {} Stidham%
}{%
McCaffrey%
\ \protect \BOthers {.}}{%
{\protect \APACyear {2015}}%
}]{%
McCaffreyWildfireevacuationits2015}
\APACinsertmetastar {%
McCaffreyWildfireevacuationits2015}%
\begin{APACrefauthors}%
McCaffrey, S\BPBI M.%
, Rhodes, A.%
\BCBL {}\ \BBA {} Stidham, M.%
\end{APACrefauthors}%
\unskip\
\newblock
\APACrefYearMonthDay{2015}{{\APACmonth{04}}}{}.
\newblock
{\BBOQ}\APACrefatitle {Wildfire Evacuation and Its Alternatives: Perspectives
  from Four {{United States}}' Communities} {Wildfire evacuation and its
  alternatives: Perspectives from four {{United States}}' communities}.{\BBCQ}
\newblock
\APACjournalVolNumPages{International Journal of Wildland
  Fire}{24}{2}{170-178}.
\newblock
\begin{APACrefDOI} \doi{10.1071/WF13050} \end{APACrefDOI}
\PrintBackRefs{\CurrentBib}

\bibitem [\protect \citeauthoryear {%
McCaffrey%
, Stidham%
, Toman%
\BCBL {}\ \BBA {} Shindler%
}{%
McCaffrey%
\ \protect \BOthers {.}}{%
{\protect \APACyear {2011}}%
}]{%
McCaffreyOutreachProgramsPeer2011}
\APACinsertmetastar {%
McCaffreyOutreachProgramsPeer2011}%
\begin{APACrefauthors}%
McCaffrey, S\BPBI M.%
, Stidham, M.%
, Toman, E.%
\BCBL {}\ \BBA {} Shindler, B.%
\end{APACrefauthors}%
\unskip\
\newblock
\APACrefYearMonthDay{2011}{{\APACmonth{09}}}{}.
\newblock
{\BBOQ}\APACrefatitle {Outreach {{Programs}}, {{Peer Pressure}}, and {{Common
  Sense}}: {{What Motivates Homeowners}} to {{Mitigate Wildfire Risk}}?}
  {Outreach {{Programs}}, {{Peer Pressure}}, and {{Common Sense}}: {{What
  Motivates Homeowners}} to {{Mitigate Wildfire Risk}}?}{\BBCQ}
\newblock
\APACjournalVolNumPages{Environmental Management}{48}{3}{475-488}.
\newblock
\begin{APACrefDOI} \doi{10.1007/s00267-011-9704-6} \end{APACrefDOI}
\PrintBackRefs{\CurrentBib}

\bibitem [\protect \citeauthoryear {%
McClelland%
}{%
McClelland%
}{%
{\protect \APACyear {1988}}%
}]{%
McClellandConnectionistmodelspsychological1988}
\APACinsertmetastar {%
McClellandConnectionistmodelspsychological1988}%
\begin{APACrefauthors}%
McClelland, J\BPBI L.%
\end{APACrefauthors}%
\unskip\
\newblock
\APACrefYearMonthDay{1988}{}{}.
\newblock
{\BBOQ}\APACrefatitle {Connectionist Models and Psychological Evidence}
  {Connectionist models and psychological evidence}.{\BBCQ}
\newblock
\APACjournalVolNumPages{Journal of Memory and Language}{27}{2}{107-123}.
\newblock
\begin{APACrefDOI} \doi{10.1016/0749-596X(88)90069-1} \end{APACrefDOI}
\PrintBackRefs{\CurrentBib}

\bibitem [\protect \citeauthoryear {%
McLennan%
, Elliott%
\BCBL {}\ \BBA {} Omodei%
}{%
McLennan%
\ \protect \BOthers {.}}{%
{\protect \APACyear {2012}}%
}]{%
McLennanHouseholderdecisionmakingimminent2012}
\APACinsertmetastar {%
McLennanHouseholderdecisionmakingimminent2012}%
\begin{APACrefauthors}%
McLennan, J.%
, Elliott, G.%
\BCBL {}\ \BBA {} Omodei, M.%
\end{APACrefauthors}%
\unskip\
\newblock
\APACrefYearMonthDay{2012}{}{}.
\newblock
{\BBOQ}\APACrefatitle {Householder Decision-Making under Imminent Wildfire
  Threat: Stay and Defend or Leave?} {Householder decision-making under
  imminent wildfire threat: Stay and defend or leave?}{\BBCQ}
\newblock
\APACjournalVolNumPages{International Journal of Wildland Fire}{21}{7}{915}.
\newblock
\begin{APACrefDOI} \doi{10.1071/WF11061} \end{APACrefDOI}
\PrintBackRefs{\CurrentBib}

\bibitem [\protect \citeauthoryear {%
Medsker%
}{%
Medsker%
}{%
{\protect \APACyear {1994}}%
}]{%
MedskerDesigndevelopmenthybrid1994}
\APACinsertmetastar {%
MedskerDesigndevelopmenthybrid1994}%
\begin{APACrefauthors}%
Medsker, L.%
\end{APACrefauthors}%
\unskip\
\newblock
\APACrefYearMonthDay{1994}{}{}.
\newblock
{\BBOQ}\APACrefatitle {Design and Development of Hybrid Neural Network and
  Expert Systems} {Design and development of hybrid neural network and expert
  systems}.{\BBCQ}
\newblock
\BIn{} \APACrefbtitle {Proceedings of 1994 {{IEEE International Conference}} on
  {{Neural Networks}} ({{ICNN}}'94)} {Proceedings of 1994 {{IEEE International
  Conference}} on {{Neural Networks}} ({{ICNN}}'94)}\ (\BVOL~3,
  \BPG~1470-1474).
\newblock
\APACaddressPublisher{Orlando, FL, USA}{{IEEE}}.
\newblock
\begin{APACrefDOI} \doi{10.1109/ICNN.1994.374503} \end{APACrefDOI}
\PrintBackRefs{\CurrentBib}

\bibitem [\protect \citeauthoryear {%
Miller%
\ \BBA {} Page%
}{%
Miller%
\ \BBA {} Page%
}{%
{\protect \APACyear {2007}}%
}]{%
MillerComplexAdaptiveSystems2007}
\APACinsertmetastar {%
MillerComplexAdaptiveSystems2007}%
\begin{APACrefauthors}%
Miller, J.%
\BCBT {}\ \BBA {} Page, S.%
\end{APACrefauthors}%
\unskip\
\newblock
\APACrefYear{2007}.
\newblock
\APACrefbtitle {Complex {{Adaptive Systems}}: {{An Introduction}} to
  {{Computational Models}} of {{Social Life}}} {Complex {{Adaptive Systems}}:
  {{An Introduction}} to {{Computational Models}} of {{Social Life}}}.
\newblock
\APACaddressPublisher{USA}{{Princeton University Press}}.
\PrintBackRefs{\CurrentBib}

\bibitem [\protect \citeauthoryear {%
Moeckel%
, Spiekermann%
\BCBL {}\ \BBA {} Wegener%
}{%
Moeckel%
\ \protect \BOthers {.}}{%
{\protect \APACyear {2003}}%
}]{%
MoeckelCreatingSyntheticPopulation2003}
\APACinsertmetastar {%
MoeckelCreatingSyntheticPopulation2003}%
\begin{APACrefauthors}%
Moeckel, R.%
, Spiekermann, K.%
\BCBL {}\ \BBA {} Wegener, M.%
\end{APACrefauthors}%
\unskip\
\newblock
\APACrefYearMonthDay{2003}{{\APACmonth{05}}}{}.
\newblock
{\BBOQ}\APACrefatitle {Creating a {{Synthetic Population}}} {Creating a
  {{Synthetic Population}}}.{\BBCQ}
\newblock
\BIn{} \APACrefbtitle {Computers in {{Urban Planning}} and {{Urban
  Management}}} {Computers in {{Urban Planning}} and {{Urban Management}}}\
  (\BPG~18).
\newblock
\APACaddressPublisher{Sendai, Japan}{}.
\PrintBackRefs{\CurrentBib}

\bibitem [\protect \citeauthoryear {%
Moore%
\ \BBA {} Newman%
}{%
Moore%
\ \BBA {} Newman%
}{%
{\protect \APACyear {2000}}%
}]{%
MooreEpidemicspercolationsmallworld2000}
\APACinsertmetastar {%
MooreEpidemicspercolationsmallworld2000}%
\begin{APACrefauthors}%
Moore, C.%
\BCBT {}\ \BBA {} Newman, M\BPBI E\BPBI J.%
\end{APACrefauthors}%
\unskip\
\newblock
\APACrefYearMonthDay{2000}{{\APACmonth{05}}}{}.
\newblock
{\BBOQ}\APACrefatitle {Epidemics and Percolation in Small-World Networks}
  {Epidemics and percolation in small-world networks}.{\BBCQ}
\newblock
\APACjournalVolNumPages{Physical Review E}{61}{5}{5678-5682}.
\newblock
\begin{APACrefDOI} \doi{10.1103/PhysRevE.61.5678} \end{APACrefDOI}
\PrintBackRefs{\CurrentBib}

\bibitem [\protect \citeauthoryear {%
Moritz%
\ \protect \BOthers {.}}{%
Moritz%
\ \protect \BOthers {.}}{%
{\protect \APACyear {2014}}%
}]{%
MoritzLearningcoexistwildfire2014}
\APACinsertmetastar {%
MoritzLearningcoexistwildfire2014}%
\begin{APACrefauthors}%
Moritz, M\BPBI A.%
, Batllori, E.%
, Bradstock, R\BPBI A.%
, Gill, A\BPBI M.%
, Handmer, J.%
, Hessburg, P\BPBI F.%
\BDBL {}Syphard, A\BPBI D.%
\end{APACrefauthors}%
\unskip\
\newblock
\APACrefYearMonthDay{2014}{{\APACmonth{11}}}{}.
\newblock
{\BBOQ}\APACrefatitle {Learning to Coexist with Wildfire} {Learning to coexist
  with wildfire}.{\BBCQ}
\newblock
\APACjournalVolNumPages{Nature}{515}{7525}{58-66}.
\newblock
\begin{APACrefDOI} \doi{10.1038/nature13946} \end{APACrefDOI}
\PrintBackRefs{\CurrentBib}

\bibitem [\protect \citeauthoryear {%
Norling%
, Sonenberg%
\BCBL {}\ \BBA {} R\"onnquist%
}{%
Norling%
\ \protect \BOthers {.}}{%
{\protect \APACyear {2000}}%
}]{%
NorlingEnhancingMultiAgentBased2000}
\APACinsertmetastar {%
NorlingEnhancingMultiAgentBased2000}%
\begin{APACrefauthors}%
Norling, E.%
, Sonenberg, L.%
\BCBL {}\ \BBA {} R\"onnquist, R.%
\end{APACrefauthors}%
\unskip\
\newblock
\APACrefYearMonthDay{2000}{{\APACmonth{07}}}{}.
\newblock
{\BBOQ}\APACrefatitle {Enhancing {{Multi}}-{{Agent Based Simulation}} with
  {{Human}}-{{Like Decision Making Strategies}}} {Enhancing {{Multi}}-{{Agent
  Based Simulation}} with {{Human}}-{{Like Decision Making
  Strategies}}}.{\BBCQ}
\newblock
\BIn{} \APACrefbtitle {Multi-{{Agent}}-{{Based Simulation}}}
  {Multi-{{Agent}}-{{Based Simulation}}}\ (\BPG~214-228).
\newblock
\APACaddressPublisher{}{{Springer, Berlin, Heidelberg}}.
\newblock
\begin{APACrefDOI} \doi{10.1007/3-540-44561-7-16} \end{APACrefDOI}
\PrintBackRefs{\CurrentBib}

\bibitem [\protect \citeauthoryear {%
Okaya%
\ \BBA {} Takahashi%
}{%
Okaya%
\ \BBA {} Takahashi%
}{%
{\protect \APACyear {2011}}%
}]{%
OkayaBDIAgentModel2011}
\APACinsertmetastar {%
OkayaBDIAgentModel2011}%
\begin{APACrefauthors}%
Okaya, M.%
\BCBT {}\ \BBA {} Takahashi, T.%
\end{APACrefauthors}%
\unskip\
\newblock
\APACrefYearMonthDay{2011}{}{}.
\newblock
{\BBOQ}\APACrefatitle {{{BDI Agent Model Based Evacuation Simulation}}} {{{BDI
  Agent Model Based Evacuation Simulation}}}.{\BBCQ}
\newblock
\BIn{} \APACrefbtitle {The 10th {{International Conference}} on {{Autonomous
  Agents}} and {{Multiagent Systems}} - {{Volume}} 3} {The 10th {{International
  Conference}} on {{Autonomous Agents}} and {{Multiagent Systems}} - {{Volume}}
  3}\ (\BPGS\ 1297--1298).
\newblock
\APACaddressPublisher{Richland, SC}{{International Foundation for Autonomous
  Agents and Multiagent Systems}}.
\PrintBackRefs{\CurrentBib}

\bibitem [\protect \citeauthoryear {%
Padgham%
, Nagel%
, Singh%
\BCBL {}\ \BBA {} Chen%
}{%
Padgham%
\ \protect \BOthers {.}}{%
{\protect \APACyear {2014}}%
}]{%
PadghamIntegratingBDIAgents2014}
\APACinsertmetastar {%
PadghamIntegratingBDIAgents2014}%
\begin{APACrefauthors}%
Padgham, L.%
, Nagel, K.%
, Singh, D.%
\BCBL {}\ \BBA {} Chen, Q.%
\end{APACrefauthors}%
\unskip\
\newblock
\APACrefYearMonthDay{2014}{}{}.
\newblock
{\BBOQ}\APACrefatitle {Integrating {{BDI Agents}} into a {{MATSim Simulation}}}
  {Integrating {{BDI Agents}} into a {{MATSim Simulation}}}.{\BBCQ}
\newblock
\BIn{} \APACrefbtitle {Proceedings of the {{Twenty}}-First {{European
  Conference}} on {{Artificial Intelligence}}} {Proceedings of the
  {{Twenty}}-first {{European Conference}} on {{Artificial Intelligence}}}\
  (\BPGS\ 681--686).
\newblock
\APACaddressPublisher{Amsterdam, The Netherlands, The Netherlands}{{IOS
  Press}}.
\newblock
\begin{APACrefDOI} \doi{10.3233/978-1-61499-419-0-681} \end{APACrefDOI}
\PrintBackRefs{\CurrentBib}

\bibitem [\protect \citeauthoryear {%
Padgham%
\ \BBA {} Singh%
}{%
Padgham%
\ \BBA {} Singh%
}{%
{\protect \APACyear {2016}}%
}]{%
PadghamMakingMATSimAgents2016}
\APACinsertmetastar {%
PadghamMakingMATSimAgents2016}%
\begin{APACrefauthors}%
Padgham, L.%
\BCBT {}\ \BBA {} Singh, D.%
\end{APACrefauthors}%
\unskip\
\newblock
\APACrefYearMonthDay{2016}{{\APACmonth{08}}}{}.
\newblock
{\BBOQ}\APACrefatitle {Making {{MATSim Agents Smarter}} with the
  {{Belief}}-{{Desire}}-{{Intention Framework}}} {Making {{MATSim Agents
  Smarter}} with the {{Belief}}-{{Desire}}-{{Intention Framework}}}.{\BBCQ}
\newblock
\BIn{} K.~Nagel, K.~Axhausen\BCBL {}\ \BBA {} A.~Horni\ (\BEDS), \APACrefbtitle
  {The {{Multi}}-{{Agent Transport Simulation MATSim}}} {The {{Multi}}-{{Agent
  Transport Simulation MATSim}}}\ (\BPG~3-8).
\newblock
\APACaddressPublisher{London}{{Ubiquity Press}}.
\PrintBackRefs{\CurrentBib}

\bibitem [\protect \citeauthoryear {%
Paveglio%
, Carroll%
\BCBL {}\ \BBA {} Jakes%
}{%
Paveglio%
\ \protect \BOthers {.}}{%
{\protect \APACyear {2008}}%
}]{%
PaveglioAlternativesEvacuationProtecting2008}
\APACinsertmetastar {%
PaveglioAlternativesEvacuationProtecting2008}%
\begin{APACrefauthors}%
Paveglio, T.%
, Carroll, M\BPBI S.%
\BCBL {}\ \BBA {} Jakes, P\BPBI J.%
\end{APACrefauthors}%
\unskip\
\newblock
\APACrefYearMonthDay{2008}{{\APACmonth{03}}}{}.
\newblock
{\BBOQ}\APACrefatitle {Alternatives to {{Evacuation}}\textemdash{{Protecting
  Public Safety}} during {{Wildland Fire}}} {Alternatives to
  {{Evacuation}}\textemdash{{Protecting Public Safety}} during {{Wildland
  Fire}}}.{\BBCQ}
\newblock
\APACjournalVolNumPages{Journal of Forestry}{106}{2}{65-70}.
\newblock
\begin{APACrefDOI} \doi{10.1093/jof/106.2.65} \end{APACrefDOI}
\PrintBackRefs{\CurrentBib}

\bibitem [\protect \citeauthoryear {%
Pel%
, Bliemer%
\BCBL {}\ \BBA {} Hoogendoorn%
}{%
Pel%
\ \protect \BOthers {.}}{%
{\protect \APACyear {2011}}%
}]{%
PelModellingTravellerBehaviour2011}
\APACinsertmetastar {%
PelModellingTravellerBehaviour2011}%
\begin{APACrefauthors}%
Pel, A.%
, Bliemer, M.%
\BCBL {}\ \BBA {} Hoogendoorn, S.%
\end{APACrefauthors}%
\unskip\
\newblock
\APACrefYearMonthDay{2011}{}{}.
\newblock
{\BBOQ}\APACrefatitle {Modelling {{Traveller Behaviour}} under {{Emergency
  Evacuation Conditions}}} {Modelling {{Traveller Behaviour}} under {{Emergency
  Evacuation Conditions}}}.{\BBCQ}
\newblock
\APACjournalVolNumPages{European Journal of Transport and Infrastructure
  Research}{11}{2}{166-193}.
\PrintBackRefs{\CurrentBib}

\bibitem [\protect \citeauthoryear {%
Pillac%
, Van~Hentenryck%
\BCBL {}\ \BBA {} Even%
}{%
Pillac%
\ \protect \BOthers {.}}{%
{\protect \APACyear {2016}}%
}]{%
Pillacconflictbasedpathgenerationheuristic2016}
\APACinsertmetastar {%
Pillacconflictbasedpathgenerationheuristic2016}%
\begin{APACrefauthors}%
Pillac, V.%
, Van~Hentenryck, P.%
\BCBL {}\ \BBA {} Even, C.%
\end{APACrefauthors}%
\unskip\
\newblock
\APACrefYearMonthDay{2016}{{\APACmonth{01}}}{}.
\newblock
{\BBOQ}\APACrefatitle {A Conflict-Based Path-Generation Heuristic for
  Evacuation Planning} {A conflict-based path-generation heuristic for
  evacuation planning}.{\BBCQ}
\newblock
\APACjournalVolNumPages{Transportation Research Part B:
  Methodological}{83}{}{136-150}.
\newblock
\begin{APACrefDOI} \doi{10.1016/j.trb.2015.09.008} \end{APACrefDOI}
\PrintBackRefs{\CurrentBib}

\bibitem [\protect \citeauthoryear {%
Popovici%
, Bucci%
, Wiegand%
\BCBL {}\ \BBA {} De~Jong%
}{%
Popovici%
\ \protect \BOthers {.}}{%
{\protect \APACyear {2012}}%
}]{%
PopoviciCoevolutionaryPrinciples2012}
\APACinsertmetastar {%
PopoviciCoevolutionaryPrinciples2012}%
\begin{APACrefauthors}%
Popovici, E.%
, Bucci, A.%
, Wiegand, R\BPBI P.%
\BCBL {}\ \BBA {} De~Jong, E\BPBI D.%
\end{APACrefauthors}%
\unskip\
\newblock
\APACrefYearMonthDay{2012}{}{}.
\newblock
{\BBOQ}\APACrefatitle {Coevolutionary {{Principles}}} {Coevolutionary
  {{Principles}}}.{\BBCQ}
\newblock
\BIn{} G.~Rozenberg, T.~B\"ack\BCBL {}\ \BBA {} J\BPBI N.~Kok\ (\BEDS),
  \APACrefbtitle {Handbook of {{Natural Computing}}} {Handbook of {{Natural
  Computing}}}\ (\BPG~987-1033).
\newblock
\APACaddressPublisher{Berlin, Heidelberg}{{Springer Berlin Heidelberg}}.
\newblock
\begin{APACrefDOI} \doi{10.1007/978-3-540-92910-9-31} \end{APACrefDOI}
\PrintBackRefs{\CurrentBib}

\bibitem [\protect \citeauthoryear {%
Rao%
}{%
Rao%
}{%
{\protect \APACyear {1996}}%
}]{%
RaoAgentSpeakBDIagents1996}
\APACinsertmetastar {%
RaoAgentSpeakBDIagents1996}%
\begin{APACrefauthors}%
Rao, A\BPBI S.%
\end{APACrefauthors}%
\unskip\
\newblock
\APACrefYearMonthDay{1996}{}{}.
\newblock
{\BBOQ}\APACrefatitle {{{AgentSpeak}}({{L}}): {{BDI}} Agents Speak out in a
  Logical Computable Language} {{{AgentSpeak}}({{L}}): {{BDI}} agents speak out
  in a logical computable language}.{\BBCQ}
\newblock
\BIn{} J\BPBI G.~Carbonell\ \BOthers {.}\ (\BEDS), \APACrefbtitle {Agents
  {{Breaking Away}}} {Agents {{Breaking Away}}}\ (\BVOL\ 1038, \BPG~42-55).
\newblock
\APACaddressPublisher{Berlin, Heidelberg}{{Springer Berlin Heidelberg}}.
\newblock
\begin{APACrefDOI} \doi{10.1007/BFb0031845} \end{APACrefDOI}
\PrintBackRefs{\CurrentBib}

\bibitem [\protect \citeauthoryear {%
Rao%
\ \BBA {} Georgeff%
}{%
Rao%
\ \BBA {} Georgeff%
}{%
{\protect \APACyear {1991}}%
}]{%
RaoModelingRationalAgents1991}
\APACinsertmetastar {%
RaoModelingRationalAgents1991}%
\begin{APACrefauthors}%
Rao, A\BPBI S.%
\BCBT {}\ \BBA {} Georgeff, M\BPBI P.%
\end{APACrefauthors}%
\unskip\
\newblock
\APACrefYearMonthDay{1991}{}{}.
\newblock
{\BBOQ}\APACrefatitle {Modeling {{Rational Agents}} within a
  {{BDI}}-{{Architecture}}} {Modeling {{Rational Agents}} within a
  {{BDI}}-{{Architecture}}}.{\BBCQ}
\newblock
\BIn{} J.~Allen, E.~Sandewall\BCBL {}\ \BBA {} R.~Fikes\ (\BEDS),
  \APACrefbtitle {2nd {{International Conference}} on {{Principles}} of
  {{Knowledge Representation}} and {{Reasoning}}} {2nd {{International
  Conference}} on {{Principles}} of {{Knowledge Representation}} and
  {{Reasoning}}}\ (\BPG~473-484).
\newblock
\APACaddressPublisher{San Mateo}{{Morgan Kaufmann Publishers, Inc.}}
\PrintBackRefs{\CurrentBib}

\bibitem [\protect \citeauthoryear {%
Rao%
\ \BBA {} Georgeff%
}{%
Rao%
\ \BBA {} Georgeff%
}{%
{\protect \APACyear {1995}}%
}]{%
RaoBDIAgentsTheory1995}
\APACinsertmetastar {%
RaoBDIAgentsTheory1995}%
\begin{APACrefauthors}%
Rao, A\BPBI S.%
\BCBT {}\ \BBA {} Georgeff, M\BPBI P.%
\end{APACrefauthors}%
\unskip\
\newblock
\APACrefYearMonthDay{1995}{}{}.
\newblock
{\BBOQ}\APACrefatitle {{{BDI Agents}}: {{From Theory}} to {{Practice}}} {{{BDI
  Agents}}: {{From Theory}} to {{Practice}}}.{\BBCQ}
\newblock
\BIn{} \APACrefbtitle {In {{Proceedings}} of the {{First International
  Conference}} on {{Multi}}-{{Agent Systems}} (Icmas-95} {In {{Proceedings}} of
  the {{First International Conference}} on {{Multi}}-{{Agent Systems}}
  (icmas-95}\ (\BPGS\ 312--319).
\PrintBackRefs{\CurrentBib}

\bibitem [\protect \citeauthoryear {%
Reid%
\ \BBA {} Beilin%
}{%
Reid%
\ \BBA {} Beilin%
}{%
{\protect \APACyear {2014}}%
}]{%
ReidWhereFireCoConstructing2014}
\APACinsertmetastar {%
ReidWhereFireCoConstructing2014}%
\begin{APACrefauthors}%
Reid, K.%
\BCBT {}\ \BBA {} Beilin, R.%
\end{APACrefauthors}%
\unskip\
\newblock
\APACrefYearMonthDay{2014}{{\APACmonth{02}}}{}.
\newblock
{\BBOQ}\APACrefatitle {Where's the {{Fire}}? {{Co}}-{{Constructing Bushfire}}
  in the {{Everyday Landscape}}} {Where's the {{Fire}}? {{Co}}-{{Constructing
  Bushfire}} in the {{Everyday Landscape}}}.{\BBCQ}
\newblock
\APACjournalVolNumPages{Society \& Natural Resources}{27}{2}{140-154}.
\newblock
\begin{APACrefDOI} \doi{10.1080/08941920.2013.840815} \end{APACrefDOI}
\PrintBackRefs{\CurrentBib}

\bibitem [\protect \citeauthoryear {%
Reynolds%
}{%
Reynolds%
}{%
{\protect \APACyear {2017}}%
}]{%
ReynoldsHistoryPrepareStay2017}
\APACinsertmetastar {%
ReynoldsHistoryPrepareStay2017}%
\begin{APACrefauthors}%
Reynolds, B\BPBI T.%
\end{APACrefauthors}%
\unskip\
\newblock
\APACrefYearMonthDay{2017}{}{}.
\newblock
{\BBOQ}\APACrefatitle {A {{History}} of the {{Prepare}}, {{Stay}} and
  {{Defend}} or {{Leave Early Policy}} in {{Victoria}}} {A {{History}} of the
  {{Prepare}}, {{Stay}} and {{Defend}} or {{Leave Early Policy}} in
  {{Victoria}}}.{\BBCQ}
\newblock
\APACjournalVolNumPages{}{}{}{312}.
\PrintBackRefs{\CurrentBib}

\bibitem [\protect \citeauthoryear {%
Rust%
, Miller%
\BCBL {}\ \BBA {} Palmer%
}{%
Rust%
\ \protect \BOthers {.}}{%
{\protect \APACyear {1994}}%
}]{%
RustCharacterizingeffectivetrading1994}
\APACinsertmetastar {%
RustCharacterizingeffectivetrading1994}%
\begin{APACrefauthors}%
Rust, J.%
, Miller, J\BPBI H.%
\BCBL {}\ \BBA {} Palmer, R.%
\end{APACrefauthors}%
\unskip\
\newblock
\APACrefYearMonthDay{1994}{{\APACmonth{01}}}{}.
\newblock
{\BBOQ}\APACrefatitle {Characterizing Effective Trading Strategies:
  {{Insights}} from a Computerized Double Auction Tournament} {Characterizing
  effective trading strategies: {{Insights}} from a computerized double auction
  tournament}.{\BBCQ}
\newblock
\APACjournalVolNumPages{Journal of Economic Dynamics and
  Control}{18}{1}{61-96}.
\newblock
\begin{APACrefDOI} \doi{10.1016/0165-1889(94)90069-8} \end{APACrefDOI}
\PrintBackRefs{\CurrentBib}

\bibitem [\protect \citeauthoryear {%
{Sadri Arif Mohaimin}%
, {Ukkusuri Satish V.}%
, {Murray-Tuite Pamela}%
\BCBL {}\ \BBA {} {Gladwin Hugh}%
}{%
{Sadri Arif Mohaimin}%
\ \protect \BOthers {.}}{%
{\protect \APACyear {2014}}%
}]{%
SadriArifMohaiminHowEvacuateModel2014}
\APACinsertmetastar {%
SadriArifMohaiminHowEvacuateModel2014}%
\begin{APACrefauthors}%
{Sadri Arif Mohaimin}%
, {Ukkusuri Satish V.}%
, {Murray-Tuite Pamela}%
\BCBL {}\ \BBA {} {Gladwin Hugh}.%
\end{APACrefauthors}%
\unskip\
\newblock
\APACrefYearMonthDay{2014}{{\APACmonth{01}}}{}.
\newblock
{\BBOQ}\APACrefatitle {How to {{Evacuate}}: {{Model}} for {{Understanding}} the
  {{Routing Strategies}} during {{Hurricane Evacuation}}} {How to {{Evacuate}}:
  {{Model}} for {{Understanding}} the {{Routing Strategies}} during {{Hurricane
  Evacuation}}}.{\BBCQ}
\newblock
\APACjournalVolNumPages{Journal of Transportation Engineering}{140}{1}{61-69}.
\newblock
\begin{APACrefDOI} \doi{10.1061/(ASCE)TE.1943-5436.0000613} \end{APACrefDOI}
\PrintBackRefs{\CurrentBib}

\bibitem [\protect \citeauthoryear {%
Scerri%
\ \protect \BOthers {.}}{%
Scerri%
\ \protect \BOthers {.}}{%
{\protect \APACyear {2010}}%
}]{%
ScerriBushfireBLOCKSModular2010}
\APACinsertmetastar {%
ScerriBushfireBLOCKSModular2010}%
\begin{APACrefauthors}%
Scerri, D.%
, Gouw, F.%
, Hickmott, S.%
, Yehuda, I.%
, Zambetta, F.%
\BCBL {}\ \BBA {} Padgham, L.%
\end{APACrefauthors}%
\unskip\
\newblock
\APACrefYearMonthDay{2010}{}{}.
\newblock
{\BBOQ}\APACrefatitle {Bushfire {{BLOCKS}}: {{A Modular Agent}}-Based
  {{Simulation}}} {Bushfire {{BLOCKS}}: {{A Modular Agent}}-based
  {{Simulation}}}.{\BBCQ}
\newblock
\BIn{} \APACrefbtitle {Proceedings of the 9th {{International Conference}} on
  {{Autonomous Agents}} and {{Multiagent Systems}}: {{Volume}} 1 - {{Volume}}
  1} {Proceedings of the 9th {{International Conference}} on {{Autonomous
  Agents}} and {{Multiagent Systems}}: {{Volume}} 1 - {{Volume}} 1}\ (\BPGS\
  1643--1644).
\newblock
\APACaddressPublisher{Richland, SC}{{International Foundation for Autonomous
  Agents and Multiagent Systems}}.
\PrintBackRefs{\CurrentBib}

\bibitem [\protect \citeauthoryear {%
Shahparvari%
, Chhetri%
, Abareshi%
\BCBL {}\ \BBA {} Abbasi%
}{%
Shahparvari%
\ \protect \BOthers {.}}{%
{\protect \APACyear {2015}}%
}]{%
ShahparvariMultiObjectiveDecisionAnalytics2015}
\APACinsertmetastar {%
ShahparvariMultiObjectiveDecisionAnalytics2015}%
\begin{APACrefauthors}%
Shahparvari, S.%
, Chhetri, P.%
, Abareshi, A.%
\BCBL {}\ \BBA {} Abbasi, B.%
\end{APACrefauthors}%
\unskip\
\newblock
\APACrefYearMonthDay{2015}{{\APACmonth{09}}}{}.
\newblock
{\BBOQ}\APACrefatitle {Multi-{{Objective Decision Analytics}} for
  {{Short}}-{{Notice Bushfire Evacuation}}: {{An Australian Case Study}}}
  {Multi-{{Objective Decision Analytics}} for {{Short}}-{{Notice Bushfire
  Evacuation}}: {{An Australian Case Study}}}.{\BBCQ}
\newblock
\APACjournalVolNumPages{Australasian Journal of Information Systems}{19}{0}{}.
\newblock
\begin{APACrefDOI} \doi{10.3127/ajis.v19i0.1181} \end{APACrefDOI}
\PrintBackRefs{\CurrentBib}

\bibitem [\protect \citeauthoryear {%
Singh%
\ \BBA {} Padgham%
}{%
Singh%
\ \BBA {} Padgham%
}{%
{\protect \APACyear {2017}}%
}]{%
SinghEmergencyEvacuationSimulator2017}
\APACinsertmetastar {%
SinghEmergencyEvacuationSimulator2017}%
\begin{APACrefauthors}%
Singh, D.%
\BCBT {}\ \BBA {} Padgham, L.%
\end{APACrefauthors}%
\unskip\
\newblock
\APACrefYearMonthDay{2017}{{\APACmonth{08}}}{}.
\newblock
{\BBOQ}\APACrefatitle {Emergency {{Evacuation Simulator}} ({{EES}}) - a
  {{Tool}} for {{Planning Community Evacuations}} in {{Australia}}} {Emergency
  {{Evacuation Simulator}} ({{EES}}) - a {{Tool}} for {{Planning Community
  Evacuations}} in {{Australia}}}.{\BBCQ}
\newblock
\BIn{} \APACrefbtitle {Proceedings of the {{Twenty}}-{{Sixth International
  Joint Conference}} on {{Artificial Intelligence}}} {Proceedings of the
  {{Twenty}}-{{Sixth International Joint Conference}} on {{Artificial
  Intelligence}}}\ (\BPG~5249-5251).
\newblock
\APACaddressPublisher{Melbourne, Australia}{{International Joint Conferences on
  Artificial Intelligence Organization}}.
\newblock
\begin{APACrefDOI} \doi{10.24963/ijcai.2017/780} \end{APACrefDOI}
\PrintBackRefs{\CurrentBib}

\bibitem [\protect \citeauthoryear {%
Singh%
, Padgham%
\BCBL {}\ \BBA {} Logan%
}{%
Singh%
\ \protect \BOthers {.}}{%
{\protect \APACyear {2016}}%
}]{%
SinghIntegratingBDIAgents2016}
\APACinsertmetastar {%
SinghIntegratingBDIAgents2016}%
\begin{APACrefauthors}%
Singh, D.%
, Padgham, L.%
\BCBL {}\ \BBA {} Logan, B.%
\end{APACrefauthors}%
\unskip\
\newblock
\APACrefYearMonthDay{2016}{{\APACmonth{11}}}{}.
\newblock
{\BBOQ}\APACrefatitle {Integrating {{BDI Agents}} with {{Agent}}-{{Based
  Simulation Platforms}}} {Integrating {{BDI Agents}} with {{Agent}}-{{Based
  Simulation Platforms}}}.{\BBCQ}
\newblock
\APACjournalVolNumPages{Autonomous Agents and Multi-Agent
  Systems}{30}{6}{1050-1071}.
\newblock
\begin{APACrefDOI} \doi{10.1007/s10458-016-9332-x} \end{APACrefDOI}
\PrintBackRefs{\CurrentBib}

\bibitem [\protect \citeauthoryear {%
Song%
, Koren%
, Wang%
\BCBL {}\ \BBA {} Barab\'asi%
}{%
Song%
\ \protect \BOthers {.}}{%
{\protect \APACyear {2010}}%
}]{%
SongModellingscalingproperties2010}
\APACinsertmetastar {%
SongModellingscalingproperties2010}%
\begin{APACrefauthors}%
Song, C.%
, Koren, T.%
, Wang, P.%
\BCBL {}\ \BBA {} Barab\'asi, A\BHBI L.%
\end{APACrefauthors}%
\unskip\
\newblock
\APACrefYearMonthDay{2010}{{\APACmonth{10}}}{}.
\newblock
{\BBOQ}\APACrefatitle {Modelling the Scaling Properties of Human Mobility}
  {Modelling the scaling properties of human mobility}.{\BBCQ}
\newblock
\APACjournalVolNumPages{Nature Physics}{6}{10}{818-823}.
\newblock
\begin{APACrefDOI} \doi{10.1038/nphys1760} \end{APACrefDOI}
\PrintBackRefs{\CurrentBib}

\bibitem [\protect \citeauthoryear {%
Strahan%
, Whittaker%
\BCBL {}\ \BBA {} Handmer%
}{%
Strahan%
\ \protect \BOthers {.}}{%
{\protect \APACyear {2018}}%
}]{%
StrahanSelfevacuationarchetypesAustralian2018}
\APACinsertmetastar {%
StrahanSelfevacuationarchetypesAustralian2018}%
\begin{APACrefauthors}%
Strahan, K.%
, Whittaker, J.%
\BCBL {}\ \BBA {} Handmer, J.%
\end{APACrefauthors}%
\unskip\
\newblock
\APACrefYearMonthDay{2018}{{\APACmonth{03}}}{}.
\newblock
{\BBOQ}\APACrefatitle {Self-Evacuation Archetypes in {{Australian}} Bushfire}
  {Self-evacuation archetypes in {{Australian}} bushfire}.{\BBCQ}
\newblock
\APACjournalVolNumPages{International Journal of Disaster Risk
  Reduction}{27}{}{307-316}.
\newblock
\begin{APACrefDOI} \doi{10.1016/j.ijdrr.2017.10.016} \end{APACrefDOI}
\PrintBackRefs{\CurrentBib}

\bibitem [\protect \citeauthoryear {%
Thompson%
, Haigh%
\BCBL {}\ \BBA {} Smith%
}{%
Thompson%
\ \protect \BOthers {.}}{%
{\protect \APACyear {2018}}%
}]{%
ThompsonPlannedultimateactions2018}
\APACinsertmetastar {%
ThompsonPlannedultimateactions2018}%
\begin{APACrefauthors}%
Thompson, K\BPBI R.%
, Haigh, L.%
\BCBL {}\ \BBA {} Smith, B\BPBI P.%
\end{APACrefauthors}%
\unskip\
\newblock
\APACrefYearMonthDay{2018}{{\APACmonth{03}}}{}.
\newblock
{\BBOQ}\APACrefatitle {Planned and Ultimate Actions of Horse Owners Facing a
  Bushfire Threat: {{Implications}} for Natural Disaster Preparedness and
  Survivability} {Planned and ultimate actions of horse owners facing a
  bushfire threat: {{Implications}} for natural disaster preparedness and
  survivability}.{\BBCQ}
\newblock
\APACjournalVolNumPages{International Journal of Disaster Risk
  Reduction}{27}{}{490-498}.
\newblock
\begin{APACrefDOI} \doi{10.1016/j.ijdrr.2017.11.013} \end{APACrefDOI}
\PrintBackRefs{\CurrentBib}

\bibitem [\protect \citeauthoryear {%
Tibbits%
\ \BBA {} Whittaker%
}{%
Tibbits%
\ \BBA {} Whittaker%
}{%
{\protect \APACyear {2007}}%
}]{%
TibbitsStaydefendleave2007}
\APACinsertmetastar {%
TibbitsStaydefendleave2007}%
\begin{APACrefauthors}%
Tibbits, A.%
\BCBT {}\ \BBA {} Whittaker, J.%
\end{APACrefauthors}%
\unskip\
\newblock
\APACrefYearMonthDay{2007}{{\APACmonth{01}}}{}.
\newblock
{\BBOQ}\APACrefatitle {Stay and Defend or Leave Early: {{Policy}} Problems and
  Experiences during the 2003 {{Victorian}} Bushfires} {Stay and defend or
  leave early: {{Policy}} problems and experiences during the 2003
  {{Victorian}} bushfires}.{\BBCQ}
\newblock
\APACjournalVolNumPages{Environmental Hazards}{7}{4}{283-290}.
\newblock
\begin{APACrefDOI} \doi{10.1016/j.envhaz.2007.08.001} \end{APACrefDOI}
\PrintBackRefs{\CurrentBib}

\bibitem [\protect \citeauthoryear {%
Tu%
, Tamminga%
, Drolenga%
, {de Wit}%
\BCBL {}\ \BBA {} {van der Berg}%
}{%
Tu%
\ \protect \BOthers {.}}{%
{\protect \APACyear {2010}}%
}]{%
TuEvacuationplancity2010}
\APACinsertmetastar {%
TuEvacuationplancity2010}%
\begin{APACrefauthors}%
Tu, H.%
, Tamminga, G.%
, Drolenga, H.%
, {de Wit}, J.%
\BCBL {}\ \BBA {} {van der Berg}, W.%
\end{APACrefauthors}%
\unskip\
\newblock
\APACrefYearMonthDay{2010}{}{}.
\newblock
{\BBOQ}\APACrefatitle {Evacuation Plan of the City of Almere: Simulating the
  Impact of Driving Behavior on Evacuation Clearance Time} {Evacuation plan of
  the city of almere: Simulating the impact of driving behavior on evacuation
  clearance time}.{\BBCQ}
\newblock
\APACjournalVolNumPages{Procedia Engineering}{3}{}{67-75}.
\newblock
\begin{APACrefDOI} \doi{10.1016/j.proeng.2010.07.008} \end{APACrefDOI}
\PrintBackRefs{\CurrentBib}

\bibitem [\protect \citeauthoryear {%
Waddell%
}{%
Waddell%
}{%
{\protect \APACyear {2002}}%
}]{%
WaddellUrbanSimModelingurban2002}
\APACinsertmetastar {%
WaddellUrbanSimModelingurban2002}%
\begin{APACrefauthors}%
Waddell, P.%
\end{APACrefauthors}%
\unskip\
\newblock
\APACrefYearMonthDay{2002}{}{}.
\newblock
{\BBOQ}\APACrefatitle {{{UrbanSim}}: {{Modeling}} Urban Development for Land
  Use, Transportation, and Environmental Planning} {{{UrbanSim}}: {{Modeling}}
  urban development for land use, transportation, and environmental
  planning}.{\BBCQ}
\newblock
\APACjournalVolNumPages{American Planning Association. Journal of the American
  Planning Association; Chicago}{68}{3}{297-314}.
\PrintBackRefs{\CurrentBib}

\bibitem [\protect \citeauthoryear {%
Whittaker%
\ \BBA {} Handmer%
}{%
Whittaker%
\ \BBA {} Handmer%
}{%
{\protect \APACyear {2010}}%
}]{%
WhittakerCommunityBushfireSafety2010}
\APACinsertmetastar {%
WhittakerCommunityBushfireSafety2010}%
\begin{APACrefauthors}%
Whittaker, J.%
\BCBT {}\ \BBA {} Handmer, J.%
\end{APACrefauthors}%
\unskip\
\newblock
\APACrefYearMonthDay{2010}{{\APACmonth{10}}}{}.
\newblock
{\BBOQ}\APACrefatitle {Community {{Bushfire Safety}}: {{A Review}} of
  {{Post}}-Black {{Saturday Research}}} {Community {{Bushfire Safety}}: {{A
  Review}} of {{Post}}-black {{Saturday Research}}}.{\BBCQ}
\newblock
\APACjournalVolNumPages{Australian Journal of Emergency Management,
  The}{25}{4}{7}.
\PrintBackRefs{\CurrentBib}

\bibitem [\protect \citeauthoryear {%
Whittaker%
, Haynes%
, Handmer%
\BCBL {}\ \BBA {} McLennan%
}{%
Whittaker%
\ \protect \BOthers {.}}{%
{\protect \APACyear {2013}}%
}]{%
WhittakerCommunitysafety20092013}
\APACinsertmetastar {%
WhittakerCommunitysafety20092013}%
\begin{APACrefauthors}%
Whittaker, J.%
, Haynes, K.%
, Handmer, J.%
\BCBL {}\ \BBA {} McLennan, J.%
\end{APACrefauthors}%
\unskip\
\newblock
\APACrefYearMonthDay{2013}{{\APACmonth{09}}}{}.
\newblock
{\BBOQ}\APACrefatitle {Community Safety during the 2009 {{Australian}} `{{Black
  Saturday}}' Bushfires: An Analysis of Household Preparedness and Response}
  {Community safety during the 2009 {{Australian}} `{{Black Saturday}}'
  bushfires: An analysis of household preparedness and response}.{\BBCQ}
\newblock
\APACjournalVolNumPages{International Journal of Wildland
  Fire}{22}{6}{841-849}.
\newblock
\begin{APACrefDOI} \doi{10.1071/WF12010} \end{APACrefDOI}
\PrintBackRefs{\CurrentBib}

\bibitem [\protect \citeauthoryear {%
Wickramasinghe%
, Singh%
\BCBL {}\ \BBA {} Padgham%
}{%
Wickramasinghe%
\ \protect \BOthers {.}}{%
{\protect \APACyear {2017}}%
}]{%
WickramasingheHeuristicDataMerging2017}
\APACinsertmetastar {%
WickramasingheHeuristicDataMerging2017}%
\begin{APACrefauthors}%
Wickramasinghe, B\BPBI N.%
, Singh, D.%
\BCBL {}\ \BBA {} Padgham, L.%
\end{APACrefauthors}%
\unskip\
\newblock
\APACrefYearMonthDay{2017}{}{}.
\newblock
{\BBOQ}\APACrefatitle {Heuristic {{Data Merging}} for {{Constructing Initial
  Agent Populations}}} {Heuristic {{Data Merging}} for {{Constructing Initial
  Agent Populations}}}.{\BBCQ}
\newblock
\BIn{} G.~Sukthankar\ \BBA {} J\BPBI A.~{Rodriguez-Aguilar}\ (\BEDS),
  \APACrefbtitle {Autonomous {{Agents}} and {{Multiagent Systems}}} {Autonomous
  {{Agents}} and {{Multiagent Systems}}}\ (\BPG~174-193).
\newblock
\APACaddressPublisher{}{{Springer International Publishing}}.
\PrintBackRefs{\CurrentBib}

\bibitem [\protect \citeauthoryear {%
Yuan%
\ \protect \BOthers {.}}{%
Yuan%
\ \protect \BOthers {.}}{%
{\protect \APACyear {2017}}%
}]{%
YuanTrafficevacuationsimulation2017}
\APACinsertmetastar {%
YuanTrafficevacuationsimulation2017}%
\begin{APACrefauthors}%
Yuan, S.%
, Chun, S\BPBI A.%
, Spinelli, B.%
, Liu, Y.%
, Zhang, H.%
\BCBL {}\ \BBA {} Adam, N\BPBI R.%
\end{APACrefauthors}%
\unskip\
\newblock
\APACrefYearMonthDay{2017}{{\APACmonth{05}}}{}.
\newblock
{\BBOQ}\APACrefatitle {Traffic Evacuation Simulation Based on Multi-Level
  Driving Decision Model} {Traffic evacuation simulation based on multi-level
  driving decision model}.{\BBCQ}
\newblock
\APACjournalVolNumPages{Transportation Research Part C: Emerging
  Technologies}{78}{}{}.
\PrintBackRefs{\CurrentBib}

\end{thebibliography}
